\documentclass[12pt,preprint]{aastex}
\usepackage{amsmath}

\newcommand{\boom}{{\scshape\footnotesize{BOOMERanG}} }
\newcommand{\bnine}{{\scshape\footnotesize{B98~}}}
\newcommand{\bk}{{\scshape\footnotesize{B03~}}}
\newcommand{\wmap}{{\scshape\footnotesize{WMAP~}}}

\newcommand{\eecmb}{\ensuremath{\langle{EE}\rangle} }
\newcommand{\tecmb}{\ensuremath{\langle{TE}\rangle} }

\newcommand{\be}{\begin{equation}}
\newcommand{\ee}{\end{equation}}

\def\ltsima{$\; \buildrel < \over \sim \;$}
\def\simlt{\lower.5ex\hbox{\ltsima}}
\def\gtsima{$\; \buildrel > \over \sim \;$}
\def\simgt{\lower.5ex\hbox{\gtsima}}

\shorttitle{\boom-\bk } \shortauthors{Silvia Masi et al.}

\begin{document}

\title{Instrument, Method, Brightness and Polarization Maps \\
       from the 2003 flight of BOOMERanG}

\author{S.Masi\altaffilmark{1}, P.Ade\altaffilmark{2}, 
J.Bock\altaffilmark{3}, J.Bond\altaffilmark{4},
J.Borrill\altaffilmark{5,}\altaffilmark{6},
A.Boscaleri\altaffilmark{7}, P.Cabella\altaffilmark{8},
C.Contaldi\altaffilmark{4,}\altaffilmark{9},
B.Crill\altaffilmark{10}, P.de Bernardis\altaffilmark{1}, G.De
Gasperis\altaffilmark{8}, A.de Oliveira-Costa\altaffilmark{11},
G.De Troia\altaffilmark{1}, G.Di Stefano\altaffilmark{12},
P.Ehlers\altaffilmark{13}, E.Hivon\altaffilmark{10},
V.Hristov\altaffilmark{14}, A.Iacoangeli\altaffilmark{1},
A.Jaffe\altaffilmark{9}, W.Jones\altaffilmark{14},
T.Kisner\altaffilmark{15,}\altaffilmark{16},
A.Lange\altaffilmark{14}, C.MacTavish\altaffilmark{17},
C.Marini-Bettolo\altaffilmark{1}, P.Mason\altaffilmark{14},
P.Mauskopf\altaffilmark{2},  T.Montroy\altaffilmark{15},
F.Nati\altaffilmark{1}, L.Nati\altaffilmark{1},
P.Natoli\altaffilmark{8,}\altaffilmark{18},
C.Netterfield\altaffilmark{13,}\altaffilmark{17},
E.Pascale\altaffilmark{17}, F.Piacentini\altaffilmark{1},
D.Pogosyan\altaffilmark{4,}\altaffilmark{19},
G.Polenta\altaffilmark{1}, S.Prunet\altaffilmark{20},
S.Ricciardi\altaffilmark{1}, G.Romeo\altaffilmark{12},
J.Ruhl\altaffilmark{15}, P.Santini\altaffilmark{1},
M.Tegmark\altaffilmark{11}, E.Torbet\altaffilmark{16},
M.Veneziani\altaffilmark{1}, and
N.Vittorio\altaffilmark{8,}\altaffilmark{18} }

\altaffiltext{1}{Dipartimento di Fisica, Universita' di Roma La
Sapienza, Roma, Italy}

\altaffiltext{2}{School of Physics and Astronomy, Cardiff
University, Wales, UK}

\altaffiltext{3}{Jet Propulsion Lanoratory, Pasadena, CA, USA}

\altaffiltext{4}{Canadian Institute for Theoretical Astrophysics
(CITA), University of Toronto, ON, Canada}

\altaffiltext{5}{Computational Research Division, Lawrence
Berkeley National Laboratory, Berkeley, CA, USA}

\altaffiltext{6}{Space Sciences Laboratory, University of
California, Berkeley, CA, USA}

\altaffiltext{7}{IFAC-CNR, Firenze, Italy}

\altaffiltext{8}{Dipartimento di Fisica, Universita' di Roma Tor
Vergata, Roma, Italy }

\altaffiltext{9}{Department of Physics, Imperial College, London,
UK}

\altaffiltext{10}{Infrared Processing and Analysis Center,
California Institute of Technology, Pasadena, CA, USA}

\altaffiltext{11}{Department of Physics, Massachusetts Institute
of Technology, Cambridge, MA, USA}

\altaffiltext{12}{Istituto Nazionale di Geofisica e Vulcanologia,
Roma, Italy}

\altaffiltext{13}{Department of Astronomy and Astrophysics,
University of Toronto, ON, Canada}

\altaffiltext{14}{Department of Astronomy, California Institute of
Technology, Pasadena, CA, USA}

\altaffiltext{15}{Physics Department, Case Western Reserve
University, Cleveland, OH, USA}

\altaffiltext{16}{Department of Physics, University of California,
Santa Barbara, CA, USA}

\altaffiltext{17}{Department of Physics, University of Toronto,
ON, Canada}

\altaffiltext{18}{INFN, Sezione di Roma 2, Roma, Italy}

\altaffiltext{19}{Department of Physics, University of Alberta,
Edmonton, AB, Canada}

\altaffiltext{20}{Institut d' Astrophysique de Paris, Paris,
France}

\begin{abstract}
We present the \boom-03  experiment and maps of the Stokes
parameters I, Q, U of the microwave sky obtained during a 14 day
balloon flight in 2003. Three regions of the southern sky were
surveyed: a deep survey ($\sim$ 90 square degrees) and a shallow
survey ($\sim$ 750 square degrees) at high Galactic latitudes
(both centered at $RA \simeq 5.5 h$, $dec \simeq -45^o$ ) and a
survey of $\sim$ 300 square degrees across the Galactic plane at
$RA \simeq 9.1 h$, $dec \simeq -47^o$. All three surveys were
carried out in three wide frequency bands centered at 145, 245 and
345 GHz, with an angular resolution of $\sim~10^{\prime}$. The 145
GHz maps of Stokes I are dominated by Cosmic Microwave Background
(CMB) temperature anisotropy, which is mapped with high signal to
noise ratio. The measured anisotropy pattern is consistent with
the pattern measured in the same region by \boom-98 and by \wmap.
The 145 GHz maps of Stokes Q and U provide a robust statistical
detection of polarization of the CMB when subjected to a power
spectrum analysis. This amplitude of the polarization is
consistent with that of the CMB in the $\Lambda$CDM cosmological
scenario. At 145 GHz, in the CMB surveys, the intensity and
polarization of the astrophysical foregrounds are found to be
negligible with respect to the cosmological signal. At 245 and 345
GHz we detect ISD emission correlated to the 3000 GHz IRAS/DIRBE
maps, and give upper limits for any other non-CMB component. We
also present intensity maps of the surveyed section of the
Galactic plane. These are compared to monitors of different
interstellar components, showing that a variety of emission
mechanisms is present in that region.
\end{abstract}


\section{Introduction}

The Cosmic Microwave Background (CMB) is a remnant of the early
Universe. Its existence is one of the pillars of the current Hot
Big Bang model; its spectrum, temperature anisotropy, and
polarization carry information about the fundamental properties of
the Universe. The power spectrum of the temperature anisotropy of
the CMB, $\langle TT \rangle$, is characterized by a flat plateau
at scales larger than the horizon at recombination ($\theta >>
1^\circ$; $\ell << 200$), where primordial perturbations froze
early in the history of the Universe, and by a series of peaks and
dips at sub-horizon scales: the signatures of acoustic
oscillations of the primeval plasma. Measurements of angular power
spectrum have been very effective in constraining cosmological
parameters \cite[e.g.,][]{debernardis94, bond98, bond00,
dodelson00, tegmark00a, tegmark00b, bridle01, douspis01, lange01,
jaffe01, lewis02, netterfield02, ruhl03, spergel03, bennett03,
tegmark04}. However, the temperature anisotropy power spectrum is
degenerate in some of these parameters; independent cosmological
information is required to break the degeneracy
\cite[]{efsthatiou99}.  All these studies make specific
assumptions on the type of the initial conditions (adiabatic, or
isocurvature) and on the shape of the power spectrum of the
initial perturbations (power-law, scale-invariance, running index,
etc.). When such assumptions are relaxed, the determination of the
cosmological parameters becomes much more uncertain
\cite[e.g.,][]{bucher02}.

There is additional information encoded in the linear polarization
properties of the CMB. CMB photons are last scattered at z $\sim$
1100. In Thomson scattering, any local quadrupole anisotropy in
the incoming photons creates a degree of linear polarization in
the scattered photons. The main term of the local anisotropy due
to density (scalar) fluctuations is dipole, while
the quadrupole term is much smaller.  For this reason the expected 
polarization is quite weak
\cite[]{rees68,kaiser83,hu97,kamionkowski97,zaldarriaga03}. The
polarization field can be expanded into a curl-free component
(E-modes) and a curl component (B-modes). Six auto and cross power
spectra can be obtained from these components: $\langle TT
\rangle$, $\langle TE \rangle$, $ \langle EE \rangle$, $\langle BB
\rangle$, $\langle TB \rangle$, and $\langle EB \rangle$. Due to
the parity properties of these components, standard cosmological
models have $\langle TB \rangle=0$ and $\langle EB \rangle=0$.
Linear scalar (density) perturbations can only produce E-modes of
polarization \cite[e.g.,][]{seljak97}. In the concordance model,
$\langle EE \rangle \sim 0.01 \langle TT \rangle$, making $\langle
EE \rangle$  a very difficult observable to measure. Tensor
perturbations (gravitational waves) produce both E-modes and
B-modes. If inflation  happened \cite[see,
e.g.,][]{mukhanov81,guth82,linde83,kolb90}, it produced a weak
background of gravitational waves. The resulting level of the
B-modes depends on the energy scale of inflation, but is in
general very weak \cite[see, e.g.,][]{copeland93,turner93}.
Alternative scenarios, like the cyclic model of
\cite{steinhardt02}, do not produce B-modes at all
\cite[]{boyle04}.

Sensitive measurements of the polarization spectra will provide a
confirmation of the current scenario of acoustic oscillations in
the early universe and improve the determination of cosmological
parameters, in particular those related to the optical depth and
reionization (see e.g. {\cite{{kaplinghat03}}). They will help
also in detecting deviations from a simple power-law spectrum of
the initial perturbations. Moreover, they will allow study of the
detailed mix of adiabatic and isocurvature initial perturbations
\cite[e.g.,][]{gordon03, peiris03}. The detection of $\langle BB
\rangle$ will probe the gravitational lensing of E-modes
\cite[]{zaldarriaga98}, and, if present, the inflation generated
component \cite[e.g.][]{leach03,song03}.

Confusion by galactic foregrounds will ultimately limit the
precision with which $\langle EE \rangle$ and $\langle BB \rangle$
can measured. Not much is known about the galactic polarized
background at microwave frequencies. The two mechanisms producing
diffuse brightness of the interstellar medium are synchrotron
radiation from relativistic electrons and thermal emission from
dust. The former is sampled at low frequencies. Patchy high
latitude observations at frequencies between 0.408 and 1.411 GHz
are collected in \cite{brouw76}. New observations carried out with
the ATCA telescope at 1.4GHz \cite[]{bernardi03} and at 2.3 GHz
\cite[]{carretti05} in the same high galactic latitude region
observed by our experiment show that the polarized synchrotron
emission is very weak. A naive extrapolation to 145 GHz predicts a
polarization of 0.2 $\mu$K rms, small with respect to the $\sqrt{
\langle EE \rangle} \sim 4 \mu K$ expected in the concordance
model. Polarized emission of galactic dust has been detected in
the 353 GHz survey of Archeops in the Galactic Plane
\cite[]{benoit03} and at high galactic latitudes
\cite[]{ponthieu05}. There, $\langle TE \rangle$ has been detected
at a level of 2 $\sigma$, while only an upper limit was obtained
for $\langle EE \rangle$. The polarized dust emission $\sqrt{
\langle EE \rangle}$ extrapolated to 145 GHz is quite weak, less
than 1 $\mu$K rms.  While the foreground signals are expected to
be smaller than the CMB signal at 145GHz, they should not be
ignored for future, very precise measurements of CMB polarization.
To do this, multiband measurements will be mandatory.

After a long pioneering phase
\cite[]{caderni78,nanos79,lubin81,masi87,partridge88,netterfield95,wollak97},
the measurement of CMB polarization is today a rapidly growing
field; new interest has been sparked especially by the possibility
of detecting the $\langle BB \rangle$ signature of the
inflationary gravity wave background
~\cite[]{keating01,subrahmanyan00,hedman02,piccirillo02,
delabrouille02,masi02,villa02,kovac02,johnson03,keating03,kogut03,
farese04,leitch04,barkats04,readhead04,cortiglioni04,cartwright05}.
To date, statistically significant detections of CMB polarization
have been reported by the experiments DASI, CAPMAP, CBI and WMAP,
all using coherent techniques. DASI has detected $\langle TE
\rangle$ at 2.9 $\sigma$ and $\langle EE \rangle$ at 6.3 $\sigma$
\cite[]{leitch04}; CAPMAP \cite[]{barkats04} has detected $\langle
EE \rangle$ at $\sim 2 \sigma$; CBI \cite[]{readhead04} has
detected $\langle EE \rangle$ at $\simgt 7 \sigma$; \wmap has
detected $\langle TE \rangle$ at many $\sigma$ \cite[]{kogut03}.
The polarization power spectra measured by these experiments are
all consistent with the forecast from the ``concordance" model best
fitting the \wmap $\langle TT \rangle$ power spectrum. Their
precision, however, is not yet good enough to improve
significantly the constraints on the cosmological parameters. The
only exception is the $\langle TE \rangle$ measurement by \wmap at
large angular scales, which provides evidence for an early,
complex reionization of the universe
~\cite[]{kogut03,kaplighat03}. The detailed structures in the
$\langle EE \rangle$ spectrum are still to be confirmed, and we
are very far from the sensitivity required to constrain the
initial conditions or inflation.

The CMB polarization signals are so small with respect to the
noise of current experiments that systematic effects are of
particular concern. Consistent detection by experiments using very
different techniques is important. This has been achieved only
recently for CMB temperature anisotropy measurements, where the
data obtained by DASI, CBI and \wmap at frequencies $\simlt$ 100
GHz are perfectly consistent with the bolometric maps obtained by
\boom, MAXIMA, ACBAR and Archeops at 150 GHz
\cite[]{debernardis03,abroe03,kuo03,hamilton03}.

All detections of CMB polarization to date have been made using
coherent detectors at frequencies $\simlt$ 100 GHz.  In this paper
we describe a completely orthogonal experiment that has, for the
first time, detected the CMB polarization at frequencies $ > $ 100
GHz.  The experiment is a modification of the BOOMERanG experiment
that produced the first resolved images of the CMB
\cite[]{debernardis00} and allowed the first detailed extraction
of cosmological parameters from the CMB \cite[]{lange01}. The
modified experiment, flown in January 2003 and hereafter referred
to as \bk, is sensitive to polarization in three bands centered at
145, 245 and 345 GHz. We present here the measurement method, the
instrument, and the maps of the Stokes parameters I, Q, U of the
CMB detected by \bk in the 2003 campaign.

Maps of CMB anisotropy and polarization are an important step in
compressing the cosmological information into power spectra, but
they are also important on their own. Maps are essential for
understanding systematic effects in the measurement and the level
of foreground contamination, and can be used to test the
Gaussianity of the CMB fluctuations (see e.g.
\cite{polenta02,detroia03,komatsu03,aliaga03,savage04}).

Estimates of the power spectra $\langle TT \rangle$, $\langle
TE \rangle$ and $\langle EE \rangle$ from \bk are described in
three companion papers ~\cite[]{jones05,piacentini05,montroy05},
and the resulting constraints on cosmological parameters in a
further paper ~\cite[]{mactavish05}.

\section{The \boom-03 Instrument}

\subsection{Generality \label{subs:general}}

This instrument derives directly from the \boom payload flown in
1997 ~\cite[]{piacentini02} and in 1998 ~\cite[]{crill03}. That
instrument provided the first high signal-to-noise maps of the CMB
anisotropy with sub-horizon resolution
\cite[]{debernardis00,netterfield02,ruhl03}, and identified three
peaks in the angular power spectrum of the CMB
\cite[]{debernardis02,ruhl03}. After the 1998/1999 flight, the
instrument was recovered and modified to make it sensitive to
polarization and to improve the attitude reconstruction hardware.
In this section we describe the different subsystems, with focus
on the new ones.

\bk is a scanning polarimeter, composed of an off-axis, 1.3m
diameter mm-wave telescope, a cryogenic multi-band bolometric
receiver, and an attitude control system. The latter is able to
control the azimuth and elevation of the telescope while the
payload is floating in the stratosphere, at an altitude of
$\simgt$ 30 km, under a long duration stratospheric balloon.

We use the sky scan to modulate the signal. We map the anisotropy
of the linear polarization by means of two separate bolometers,
$B_1$ and $B_2$, that observe the sky through the same feed
structure but are sensitive to orthogonal polarization directions.
This device is called a Polarization Sensitive Bolometer (PSB).
Each bolometer signal is processed and amplified separately.

In principle, we can then difference the two signals to obtain the
Stokes parameter Q of linear polarization. The U parameter is
measured by means of an identical PSB, containing bolometers $B_3$
and $B_4$, rotated by $\pi/4$ in the focal plane with respect to
the first one. In this minimal set of four bolometers, the
principal axis of each sensor is rotated with respect to the focal
plane by an angle $\alpha_k$. For the first PSB $\alpha_1=0$ and
$\alpha_2=\pi/2$; for the second one $\alpha_3=\pi/4$ and
$\alpha_4=3\pi/4$.

In practice, our sky scan strategy uses repeated scans over the
same sky pixel $p$. At different times $t_i$ during the
survey, the focal plane rotates with respect to the sky by an
angle $\gamma_i$. Information on Q and U in each sky pixel thus
comes from all the bolometers present in the focal plane, according
to the relation
\begin{equation}
V^p_{i,k} = {1 \over 2} \mathcal{S}_k \left[ I_p + Q_p \cos [2
(\alpha_k+\gamma_i)] + U_p \sin[2 (\alpha_k+\gamma_i)] \right] +
n_{i,k} ~~~~~~~. \label{eq:polar}
\end{equation}
Here $V^p_{i,k}$ is the signal measured by bolometer $k$ at time
$t_i$; $\mathcal{S}_k$ is the responsivity of bolometer $k$;
$I_p$, $Q_p$, $U_p$ are the Stokes parameters of pixel $p$ in the
chosen celestial coordinates, and $n_{i,k}$ is the noise
contribution to the $i$-th measurement on that pixel. This system
of equations can be inverted and the Stokes parameters estimated
if a sufficient number of measurements over a range of angles
$\alpha_k+\gamma_i$ is taken.

Other ways to modulate the polarization involve the use of a
modulating analyzer to extract the polarized component by
synchronous demodulation. Rotating wire grids, half wave plates,
K-mirrors, Faraday rotators, Fresnel rombs, have been used or
proposed as polarization analyzers
\cite[e.g.][]{keating02,battistelli02,hanany03,gervasi03,gundersen03,catalano04}.
While all of these techniques can in principle provide a valuable
means of reducing requirements on the stability of detector gains
and offsets, they come at a cost in both complexity and bandwidth.
Correlation polarimeters have so far been implemented only with
coherent detectors \cite[e.g.][]{carretti01,padin02}.

The polarization measurement strategy defined by
eq.(\ref{eq:polar}) is prone to leakage of the unpolarized
component $I$ into the polarized ones $Q$ and $U$, if the
responsivities $\mathcal{S}_k$ are not known exactly. Similarly,
errors in the principal axes angles $\alpha_k$ mix $Q$ and $U$
into each other. The polarimetric calibration consists of
measuring all $\mathcal{S}_k$ and $\alpha_k$. The precision
required to obtain the common-mode rejection needed in our case
can be estimated as follows.

For simplicity we choose the reference frame to have $\gamma=0$,
and we consider the pair of detectors with $\alpha_1=0,
\alpha_2=\pi/2$. In this case we can't recover all the parameters
but only $I$ and $Q$. Equation~(\ref{eq:polar}) becomes
\begin{eqnarray}
V_1 &=& {\mathcal{S}_1 \over 2} \left[I +  Q \right] ={\mathcal{S} \over 2} \left[I +  Q \right] \\
V_2 &=& {\mathcal{S}_2 \over 2} \left[I - Q \right] = {\mathcal{ S
R } \over 2} \left[I - Q \right].
\end{eqnarray}
 where we have expressed the
calibration constants in terms of an absolute calibration
$\mathcal S = \mathcal S_1$ and a relative calibration $\mathcal R
= \mathcal S_2/\mathcal S_1$. The solution is
\begin{eqnarray}
I &=& \frac 1{\mathcal S} \left[V_1 + \frac{V_2}{\mathcal R}
\right]
 \\
Q &=& \frac 1{\mathcal S} \left(V_1 - \frac{V_2}{\mathcal
R}\right)
\end{eqnarray}
Uncertainties on the calibration constant $\sigma_{\mathcal S}$
and on the relative calibration $\sigma_{\mathcal R}$ are
propagated in the error on $Q$ by
\begin{equation}
\sigma_Q^2 =\left| \frac{dQ}{d\mathcal S}\right|^2
\sigma_{\mathcal S}^2 + \left| \frac{dQ}{d\mathcal R}\right|^2
\sigma_{\mathcal R}^2
\end{equation}
Using $I \simeq 2V / {\mathcal S} $ (true if $Q \ll I$)
\begin{equation}
\frac{\sigma_Q^2}{Q^2} = \frac{\sigma_{\mathcal S}^2}{\mathcal
S^2}+ \left(\frac 12 \frac IQ \right)^2\frac{\sigma_{\mathcal
R}^2}{\mathcal R^2}
\end{equation}
For $E$-mode polarization we expect to have a factor $I/Q$ of the
order of 20. If we want to achieve a $\sim 10 \%$ accuracy in the
determination of Q and U, we need to have $\sigma_{\mathcal
R}/\mathcal R \simlt 1\%$ for each detector. We show below that
relative calibration constants can be measured within an error of
$\simlt$ 2\% by comparing the CMB temperature anisotropy measured
in different detectors (see \S \ref{subs:speccal}). The absolute
calibration $\mathcal S$ has to be determined with a relative
error $\sigma_{\mathcal S}/{\mathcal S} \simlt 5\%$.

From equation~(\ref{eq:polar}) we can also estimate the acceptable
principal axis angle uncertainty. Again we choose the reference
frame to have $\gamma=0$, and we consider the couple of detectors
with $\alpha_1=0, \alpha_2=\pi/2$, with errors $\sigma_{\alpha_1}$
and $\sigma_{\alpha_2}$. To first order the measured  Stokes
parameter $Q_m$ will be $Q_m =  Q + 2U
\sqrt{\sigma_{\alpha_1}^2+\sigma_{\alpha_2}^2}$, i.e. $(Q_m - Q)/U
\simeq 2\sqrt{\sigma_{\alpha_1}^2+\sigma_{\alpha_2}^2}$. From this
equation we see that $\sigma_{\alpha_i} \lesssim 2^o$ produces an
error in $Q$ (and $U$) $\simlt 10\%$. Mixing Q and U also mixes E
and B-mode signals, so errors in alpha may affect the level at
which one can set an upper limit on the B-mode polarization
anisotropy signal.

Non-ideal polarized detectors have a residual sensitivity to the
polarization component orthogonal to the principal axis
(cross-polar response). As we rotate a perfect polarizing grid in
front of a non ideal polarization sensitive detector, the detector
response is given by a modified Malus law:
\begin{equation}
V= \mathcal{S} I_0 (1-\xi \sin^2(\alpha_{det}-\alpha_{grid})),
\label{eq:poleff}
\end{equation}
where ${\mathcal S}$ is the responsivity, $I_0$ is the incident
power and $\xi$ is the polarization efficiency. $\alpha_{det}$ is
the angle of the main axis of the detector with respect to the x
axis of the reference frame, and $\alpha_{grid}$ is the position
of the principal axis of the polarizing grid.

This can be rewritten in terms of the cross-polar response
coefficient $\epsilon = V_\perp / V_\parallel$ defining the
response of the detector to radiation polarized orthogonally to
its main axis, as a fraction of the response to radiation
polarized parallel to its main axis. We have $\epsilon = 1 - \xi$.
Using an operative definition of responsivity $\mathcal{S^\prime}=
\mathcal{S} (1 + \epsilon) /2 $ we rewrite
\begin{equation}
\label{eqn:shape} V=\mathcal{S^\prime}I_0\left[ 1 +
\frac{1-\epsilon}{1+\epsilon} \cos(2(\alpha_{det} -
\alpha_{grid})) \right]
\end{equation}.

Taking into account cross-polar response, and dropping the
$^\prime$ for simplicity, equation (\ref{eq:polar}) becomes
\begin{equation}
\label{eqn:vcross} V_k = {\mathcal{S}_k \over 2} \left[ I + Q
\frac{1-\epsilon_k}{1+\epsilon_k} \cos(2\alpha) + U
\frac{1-\epsilon_k}{1+\epsilon_k} \sin(2\alpha) \right]+n_k
\end{equation}

This system can be inverted to measure $Q$, $U$, and $I$ if the
cross-polar response coefficients $\epsilon_k$ are known. In the
same simplified case considered above the system becomes
\begin{eqnarray}
\label{eqn:vcross1} V_1 &=& \mathcal{S} \left[I +
\frac{1-\epsilon_1}{1+\epsilon_1} Q \right] \\
 V_2 &=&
\mathcal{R S} \left[I - \frac{1-\epsilon_2}{1+\epsilon_2} Q
\right]
\end{eqnarray}
Assuming that $\epsilon_1$ and $\epsilon_2$ are
known, this is solved by
\begin{eqnarray}
I&=& \frac 1{2\mathcal S} \left[V_1 \left(
\frac{1-\epsilon_2}{1+\epsilon_2} \right) + \frac{V_2}{\mathcal R}
\left( \frac{1-\epsilon_1}{1+\epsilon_1} \right)\right]
(1+\epsilon_1+\epsilon_2) \\ Q&=& \frac 1{2\mathcal S} \left(V_1 -
\frac{V_2}{\mathcal R}\right) (1+\epsilon_1 + \epsilon_2) .
\end{eqnarray} where the terms of the second order in $\epsilon$
have been neglected. Proceeding as above, we find
\begin{equation} \frac{\sigma_Q^2}{Q^2} =
\frac{\sigma_{\mathcal S}^2}{\mathcal S^2}+ \left(\frac 12 \frac
IQ \right)^2\frac{\sigma_{\mathcal R}^2}{\mathcal R^2}
+\sigma_{\epsilon_1}^2 + \sigma_{\epsilon_2}^2
\end{equation} We conclude that in our case
the uncertainty on the cross-polar response parameters can be
$\simlt 5\%$. All the conclusions obtained in this section have
been verified by means of numerical simulations with realistic
parameters \cite[]{masi02}.

Scanning the sky with AC-coupled detectors is an effective way to
measure the intensity and polarization of the sky at the angular
scales of interest for the CMB. This strategy has been used by
\boom-98 and Archeops, is in use in QuAD, and will be used by
Planck-HFI and LFI.

In the case of \bk, we scan the sky rotating the full payload in
azimuth, so that atmospheric emission is almost constant along the
scan. Working with an AC coupled amplifier, our system is
insensitive to constant signals.

We alternate forward and reverse scans to map the low foreground
region already observed by \boom, centered in the constellations
Caelum and Horologium, at $RA \simeq 5.5 h$, $dec \simeq -45^o$
($b \simlt -30^o$). During Antarctic LDB flights, the average
latitude of the payload is $\sim -78^o$. In one day, due to sky
rotation, this procedure produces a highly cross-linked scan
pattern (see \cite{crill03}, Fig.9), which is important for map
making.

\subsection{Observation strategy optimization}

Our sky coverage is optimized to reduce the errors on the CMB
power spectra given the constraints imposed by our telescope's
hardware and the presence of bright celestial sources (the Sun and
Galaxy). With the assumption of uniform sky coverage and
uncorrelated noise from pixel to pixel and between T,Q and U, the
errors on \eecmb are approximated by \cite[]{zaldarriaga97}

\begin{equation}
\sigma_{E,\ell}^2=\frac{2}{(2\ell+1)f_{{sky}}}\left[{C_{E,\ell}+\frac{4{\pi}f_{{sky}}{N_E}^2}{\tau
B_\ell^2} }\right]^2
\end{equation}
and the errors on \tecmb are given by
\begin{equation}
\sigma_{X,\ell}^2=\frac{2}{(2\ell+1)f_{{sky}}}\left[{C_{X,\ell}^2+\left({C_{T,\ell}+\frac{4{\pi}f_{{sky}}{N_T}^2}{\tau
B_\ell^2}}\right)\left({C_{E,\ell}+\frac{4{\pi}f_{{sky}}{N_E}^2}{\tau
B_\ell^2}}\right)} \right]
\end{equation}
where $N_T$ and $N_E$ are the effective noise equivalent
temperatures in $T$ and $E$ of the combined set of detectors;
$\tau$ is the integration time spent uniformly covering the
$f_{\text{sky}}$ fraction of the sky, and $B_\ell^2$ is the beam
response in multipole space. For the case of \bk,
$N_E^2=2{\times}N_T^2$ since there are eight linear detectors
arranged in four orthogonal pairs, each with a different
orientation on the sky (see Figure \ref{fig:focalplane}). If we
use an estimate of our detector sensitivity and flight time (both
based on the previous flights of \boom ), and a set of power
spectra based on the best models to date, then we can calculate
the uniform sky coverage needed to maximize the signal to noise
ratio near the first peaks of \eecmb and \tecmb. The signal to
noise ratios for those band-powers are a fairly flat function of
the sky coverage. The signal to noise ratio for \eecmb peaks near
$4 \pi f_{{sky}}\approx{70}$ square degrees. For \tecmb, the peak
is near $4 \pi f_{{sky}}\approx{1600}$ square degrees (see Figures
\ref{fig:sn_col7.eps} and \ref{fig:sn_col7_TE.eps}).

\begin{figure}[p]
\plotone{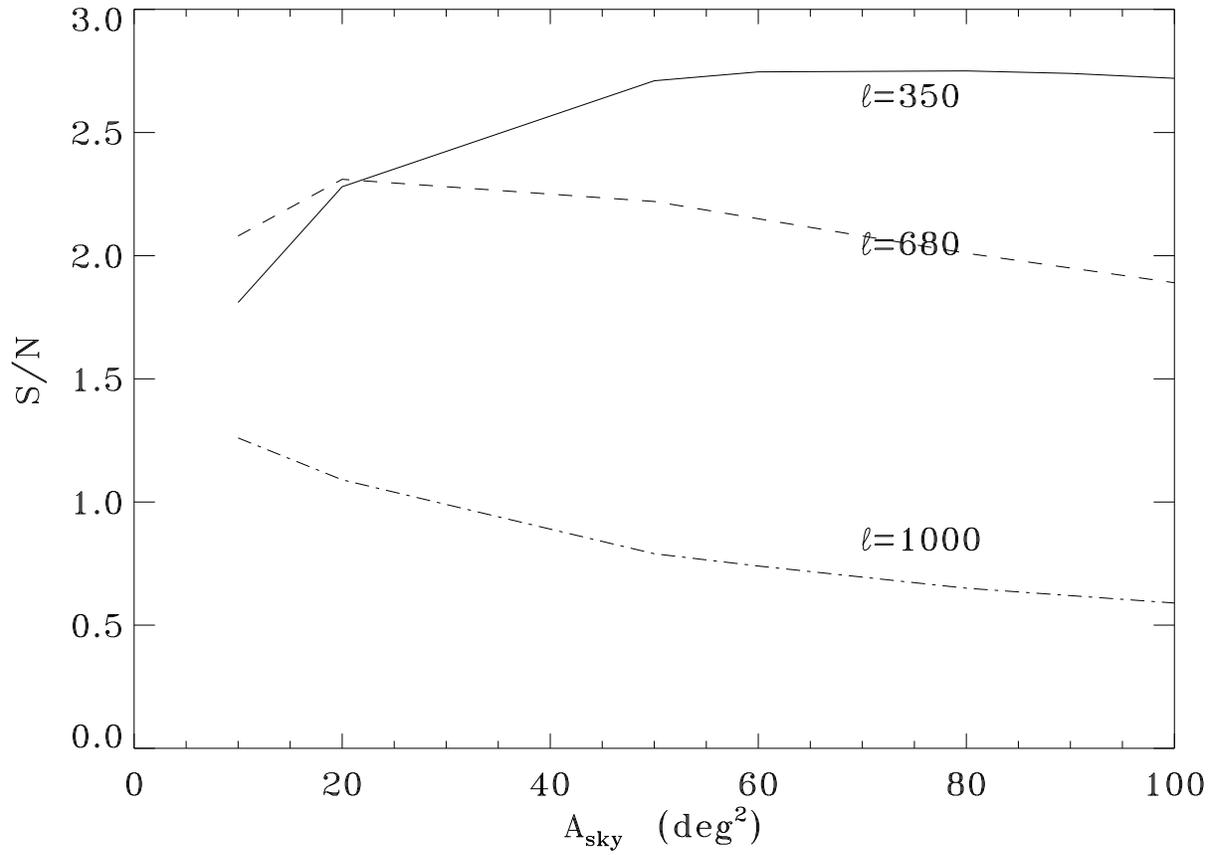} \caption{The signal to noise ratio vs.\ sky
coverage  for the \eecmb computed using the best estimate of
$N_E$. The three curves refer to different $\Delta \ell = 50$
bandpowers, and are labelled using the center multipole. These
multipoles correspond to peaks in the "concordance" \eecmb power
spectrum. \label{fig:sn_col7.eps}}
\end{figure}

\begin{figure}[p]
\plotone{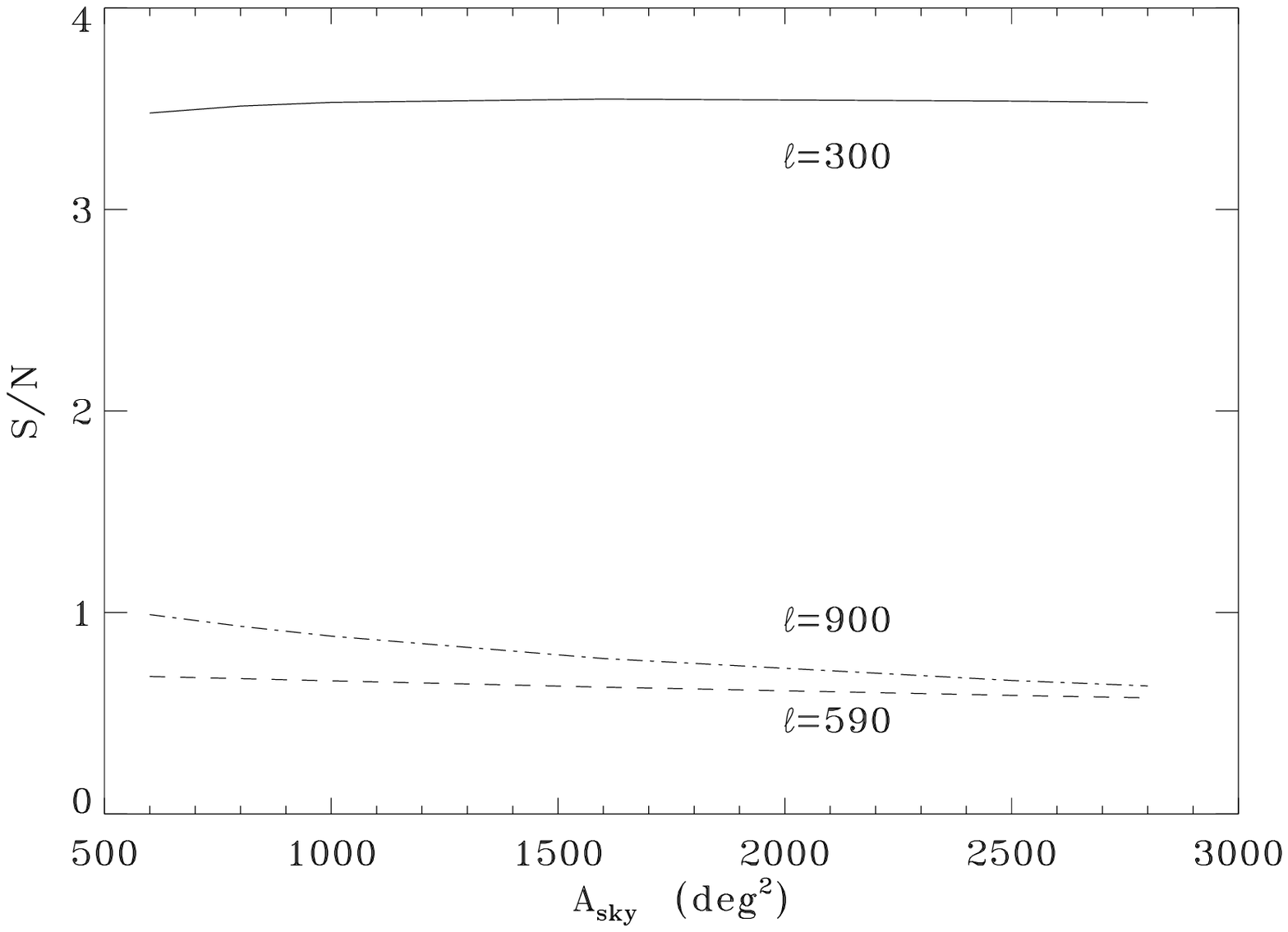} \caption{The signal to noise ratio vs.\ sky
coverage for \tecmb computed using the best estimate of $N_T$ and
$N_E$. The three curves refer to different $\Delta \ell = 50$
bandpowers, and are labelled using the center multipole. The
selected multipoles correspond to extremals in the "concordance"
\tecmb power spectrum. \label{fig:sn_col7_TE.eps}}
\end{figure}

The \boom telescope has a useable elevation range of 35 to 55
degrees.  The telescope is designed to scan no more than
60$^{\circ}$ from the anti-sun direction.  Exceeding this range
could cause heating of the telescope baffles by sunlight.  There
are also constraints on the acceptable scan periods of the
telescope. From the 1998 flight of \boom , we know that certain
scan periods excite pendulations in the balloon-gondola system.
The scan speed is restricted by the thermal time constants of the
detectors, the mechanics of the telescope control systems, and the
stability of our readout electronics.

Given these constraints, we created a scan strategy that came as
close as possible to producing uniform coverage over both a
``deep'' region (for sensitivity to \eecmb) and a larger
``shallow'' region (for sensitivity to \tecmb). Because each
change in pointing elevation perturbs the telescope, we decided to
adjust the elevation no more than once per hour. With this
restriction, we found that the smallest reasonable size that we
could achieve for the ``deep'' region was $\approx{100}$ square
degrees.  The size of our ``shallow'' region ($\approx{800}$
square degrees) was bounded on one side by the galaxy and on the
other side by the distance from the anti-sun direction.

When determining the details of the scan strategy, we simulated
the scanning of the telescope based on the same ``schedule file''
used to actually control the telescope during the flight.  These
simulations produced a coverage map for a given schedule file.
Since this coverage was non-uniform, we approximated the spectral
errors by the sum of the error contributions from each pixel
\begin{eqnarray}
\sigma_{E,\ell}^2 & = & \sum\limits_{p}^{}\;\frac{2}{(2\ell+1)f_{{p}}}\left[{C_{E,\ell}+\frac{4{\pi}f_{{p}}{n_{p,E}}^2}{\tau}}\right]^2\\
\sigma_{X,\ell}^2 & = &
\sum\limits_{p}^{}\;\frac{2}{(2\ell+1)f_{{p}}}\left[{C_{X,\ell}^2+\left({C_{T,\ell}+\frac{4{\pi}f_{{p}}{n_{p,T}}^2}{\tau}}\right)\left({C_{E,\ell}+\frac{4{\pi}f_{{p}}{n_{p,E}}^2}{\tau}}\right)}\right]
\end{eqnarray}
where $f_{{p}}$ is the sky fraction of a typical pixel and
$n_{p,T}$ and $n_{p,E}$ are the noise in a given pixel computed
from the $NET$ of the detectors and the integration time on the
pixel.  We decided to spend the first four days of the flight
scanning over the ``shallow'' region, and to spend the remainder
of the flight on the ``deep'' region.  After confirming that the
scanning schedule produced the desired sky coverage (and signal to
noise), we adjusted the scan speed so that it did not coincide
with any pendulation modes discovered in \bnine. Because of these
adjustments, \bk uses a range of many different azimuth scan
speeds: between 0.2 and 0.6 deg/s for the deep survey, and between
0.6 and 1 deg/s in the shallow survey.

Twice a day, when the scans over the deep and shallow regions
would be at nearly constant declination and therefore give minimal
cross-linking, the telescope was directed to scan a region
spanning the galactic plane, including RCW38 and other galactic
sources useful for calibration (see section \ref{subs:galaxy}).

\subsection{Detectors}

Cryogenic bolometers are the most sensitive detectors for
continuum mm-wave radiation (see e.g. \cite{richards94}).

The focal plane of \bk consists of eight optically active
bolometric receivers, two dark bolometers, and a fixed resistor,
all of which operate from a 270 mK base temperature.  The
optically inactive channels are read out using identical
electronics as a check against microphonics, RFI and baseplate
temperature fluctuations. The focal plane is split equally between
two types of receivers: four (dual-polarized) Polarization
Sensitive Bolometer (PSB) pixels operating at 145 GHz, and four
(single-polarization) two-color photometers using spider-web
bolometers operating at 245 and 345 GHz.

Polarization Sensitive Bolometers consist of a pair of co-located
silicon nitride micromesh absorbers which couple aniostropically
to linearly polarized radiation through a corrugated waveguide
structure \cite[]{jones03,jones05a}. The system allows
simultaneous background limited measurements of the Stokes I and Q
parameters over $\simgt 30\%$ bandwidths. The absorbers, separated
by $\simeq 60\mu$m, are electrically and thermally isolated from
one another. The devices used in \bk are shown in Fig.
\ref{fig:PSB}.
\begin{figure}[p]
\plotone{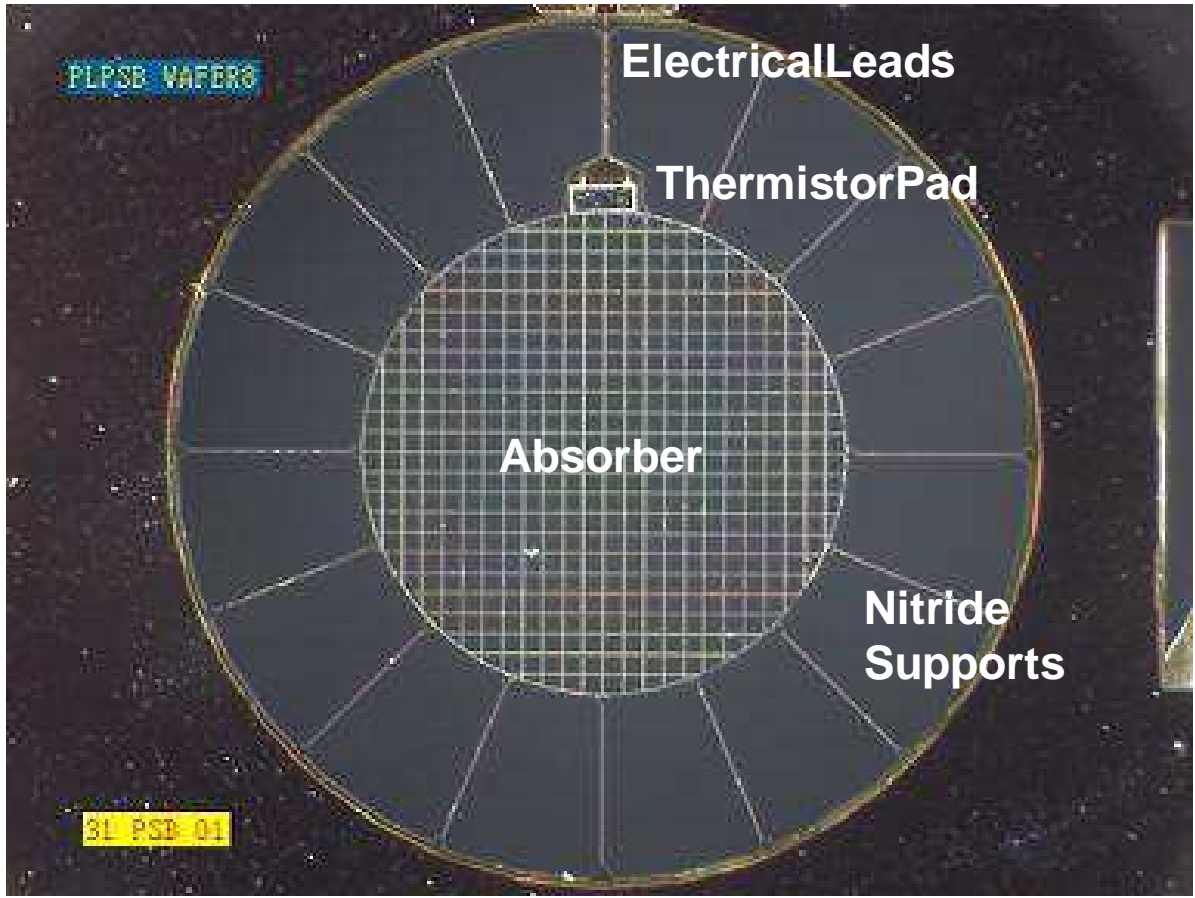} \caption{Photograph of the radiation
absorber of a Polarization Sensitive Bolometer used in \bk. The
$Si^3N^4$ micromesh absorbing grid has a diameter of 2.6mm, the
absorber leg spacing is 108 $\mu m$, and each leg is 3 $\mu m$
wide. Only the vertical wires are metalized. The horizontal wires
are used for structural support and do not absorb radiation.
\label{fig:PSB}}
\end{figure}
The common thermal and radiative environment resulting from the
physical proximity of the two detectors provides gain stability
and rejection of differential temperature fluctuations which are
limited only by differences in the properties of the NTD Ge
thermistors and the electrical leads, which determine the thermal
conductance to the bath. Both linear polarizations propagate
through a shared waveguide structure and set of optical filters
and couple to the telescope through a cryogenic corrugated feed,
ensuring an identical electromagnetic bandpass and highly
symmetric beams.  PSBs are fabricated using the same proven
photolithographic techniques used to make spider-web
bolometers~\cite[]{yun03}, and enjoy the same benefits of reduced
heat capacity, low cross section to cosmic rays, and reduced
susceptibility to microphonic response relative to monolithic
bolometers.

The two-color photometer is an evolutionary development of the
photometers originally designed for MAX~\cite[]{fischer92},  and
used subsequently by the SuZIE~\cite[]{holzapfel97}, the FIRP
instrument on the IRTS~\cite[]{lange94}, and the \boom 98
~\cite[]{piacentini02,crill03} CMB experiments. The \bk photometer
has been optimized for only two frequencies, and has been made
polarization sensitive by the fitting of a polarizing grid to the
feed aperture. The detectors are all similar, if not identical, to
the detectors flown on \boom 98. The \bk feed design, consisting
of a multi-mode back-to-back profiled corrugated horn, is
significantly advanced relative to earlier versions of the
photometer.  This system achieved high efficiencies and symmetric
beam patterns over the full 200-420 GHz bandwidth.

The radiation is coupled from the photometer feed through a 420
GHz metal mesh low-pass filter into the 12.7 mm diameter
photometer body.  A dichroic filter, oriented at $45^\circ$ with
respect to the optical axis, directs radiation at frequencies
above 295 GHz to the 345 GHz detector module while passing the
lower frequencies to the 245 GHz detector module.  The detector
modules are thermally isolated from the photometer body, which is
held near 2K, by a $\sim 5$-mm gap.  The photometer sub-Kelvin
feeds are smooth walled with an exit aperture matched to the
geometric area of the absorber.  Corrugated feeds are not
necessary, as the polarization discrimination and beam forming is
determined by the 2-Kelvin feed antenna.

The two configurations used in the focal plane are presented in
Fig. \ref{fig:photometers}. The performance of the receivers as
integrated in the \boom focal plane is reported in Section
\ref{sect:pre_flight_cal}.

\begin{figure}[p]
\plotone{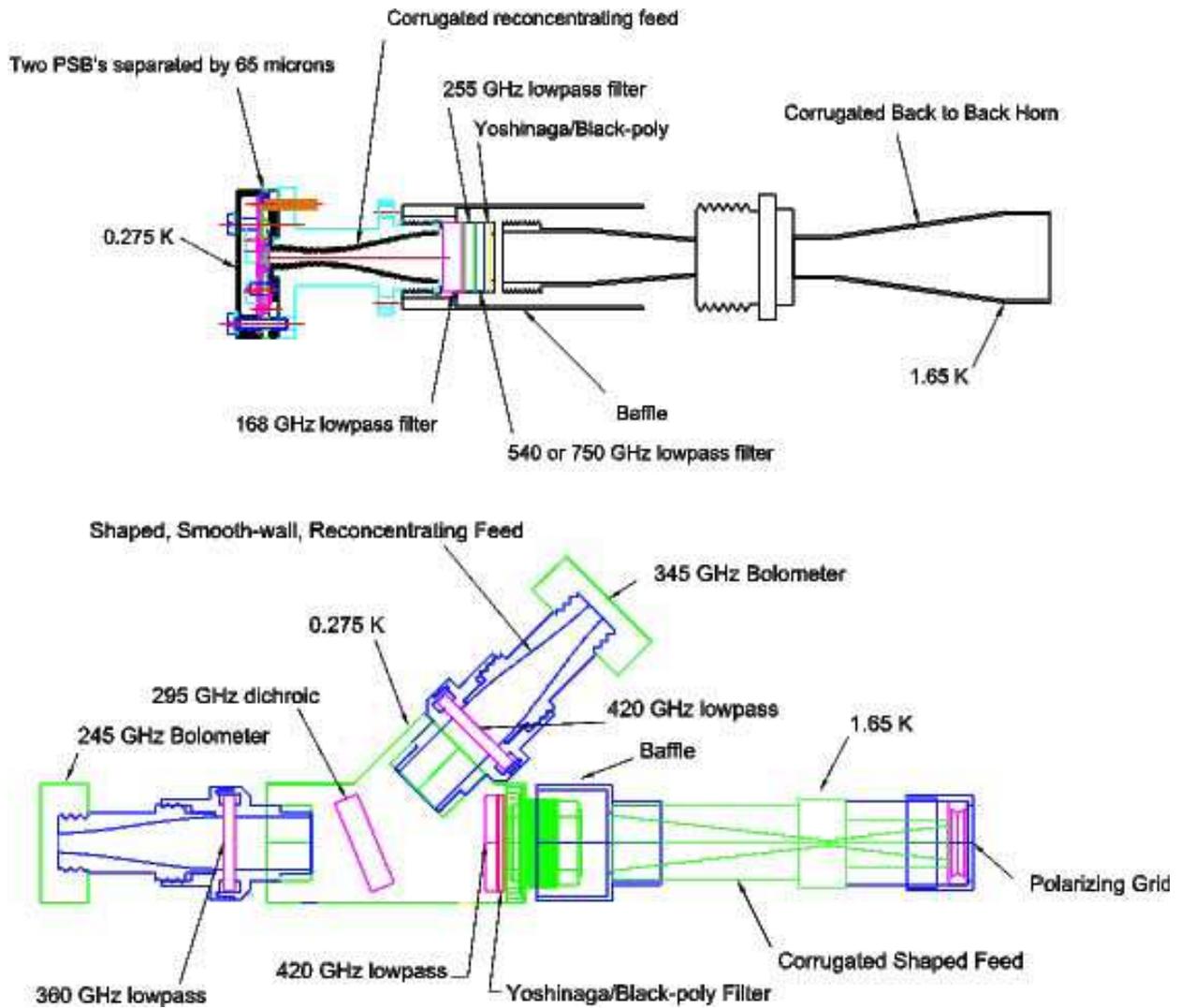} \caption{Schematic of the 2-color
photometers (bottom) and of the PSB polarimeters (top) used in
\bk. \label{fig:photometers}}
\end{figure}

\subsection{Optical filtering}

Optical filtering is of critical importance to a bolometric
receiver; in addition to defining the optical pass-band, care must
be taken to ensure that the detector is shielded from out-of-band
radiation originating from within the cryostat. These filters also
play a significant role in the determination of the end-to-end
optical efficiency of the system.  Finally, the optical filtering
must reduce the radiative loading on the various stages of the
cryogenic system to acceptable levels.

The window of the \bk cryostat consists of a $50-\mu$m
polypropylene film stretched on an elliptical aluminum frame.  The
window has excellent transmission at all three wavelengths but is
exceedingly fragile, requiring replacement after each cycle the
cryostat.

\begin{table}[p]
\begin{center}
\begin{tabular}{c|c|c|c}
\multicolumn{4}{c}{\large Optical Filters} \\
\hline\hline
Temp. & 145 GHz & 245 GHz & 345 GHz \\
\hline 77 K & \multicolumn{3}{c}{$\sim 100\mu$m IR blocker} \\
\cline{2-4} \vdots  & \multicolumn{3}{c}{540 GHz AR-coated LPF} \\
\cline{2-4} 2 K & \multicolumn{3}{c}{$\sim 100\mu$m IR blocker} \\
\cline{2-4} \vdots  & \multicolumn{3}{c}{450 GHz AR-coated LPF} \\
\cline{2-4} \vdots  & \multicolumn{3}{c}{Removable NDF$^\dagger$}
\\ \cline{2-4} \vdots  & &\multicolumn{2}{c}{Polarizing Grid} \\
\cline{3-4} \vdots  & &\multicolumn{2}{c}{420 GHz LPF} \\
\cline{3-4} \vdots  & &\multicolumn{2}{c}{$180\mu$m BP/Yosh
LPF$^\ddagger$} \\ \cline{3-4} \vdots  & & 295 GHz dichroic LPF &
295 GHz dichroic HPF \\ \cline{2-4} 0.3 K   & $180\mu$m BP/Yosh
LPF$^\ddagger$ & 360 GHz LPF & 410 GHz LPF \\ \cline{2-4} \vdots
& 255 GHz LPF &\multicolumn{2}{c}{ }\\ \cline{2-2} \vdots  & 540
GHz LPF &\multicolumn{2}{c}{ }\\ \cline{2-2} \vdots  & 168 GHz LPF
&\multicolumn{2}{c}{ }\\ \cline{2-2} \hline
\multicolumn{4}{l}{$^\dagger$ The neutral density filter can be
mechanically rotated in and out of the beam}\\
\multicolumn{4}{l}{\hspace{3mm}while the system is cooled down.}\\
\multicolumn{4}{l}{$^\ddagger$ A black-polyethilene (BP) Yoshinaga
filter is an absorptive filter consisting of a thallium} \\
\multicolumn{4}{l}{\hspace{3mm}salt deposited on a black
polyethylene substrate, the thickness of which is}\\
\multicolumn{4}{l}{\hspace{3mm}tuned to minimize reflections~\cite[]{yamada62}.}\\
\end{tabular}
\end{center}
\caption[Optical filters]{\small The optical filtering scheme
employed
  by \bk .  In order to take advantage of the low backgrounds
  available at float altitudes, much care must be taken to reduce the
  background originating from within the cryostat.  While the metal mesh
  filters, which consist of bonded layers of polyethylene, exhibit
  in-band emissivities at the percent level, the PTFE antireflection
  coating is several times more emissive. It is crucial that these
  filters remain well
  heat-sunk and protected from infrared emission from the warmer stages.}
\label{tbl:filters}
\end{table}

Most of the filters used in the \bk optical system consist of
layers of patterned meshes deposited on polypropylene
substrates~\cite[]{lee96}, with the gaps between layers filled by
polypropylene as well. The dichroic beam-splitter used in the
photometer body is the sole exception, being an air-gap the
inductive layers are deposited on a thin Mylar substrate and
stretched on an aluminum frame.  The polarizing grids used on the
photometer and in laboratory testing, and the neutral-density
filter, are made in a similar fashion.  Instead of inductive or
capacitive grids, a linear pattern is used for the polarizer,
while a uniform lossy coating is used for the NDF.

The layers of the filter are hot-pressed to form a single
self-supporting filter. Some of the thicker hot-pressed filters
are antireflection  coated with a tuned layer of PTFE (Teflon).
PTFE has high infrared emissivity, which initially resulted in
excessive heating of the \bk filters and a large thermal load on
the LN$_2$ and $^4$He stages.  In addition, the heating of the
filters led to a significant increase in background loading of the
detectors resulting from in-band thermal emission. To ameliorate
this problem, large-format, composite IR blockers were fitted in
front of the 77K and 2K filters.  These filters have high in-band
transmission and reflect radiation at wavelengths shortward of
$\sim 100 \mu$m.

\subsection{The focal plane}

To measure polarization, we combine information from spatially
separated pixels, as shown in eq.(\ref{eq:polar}). The focal plane
layout (Fig. \ref{fig:focalplane}) is designed to minimize spatial
separation between pixels, given the constraints of the existing
\boom optics and the size of the feed-horns. This allows for
maximal overlap of maps made by spatially separate pixels. The
wide focal plane of the \bk telescope is thus populated by 8
pixels with independent corrugated feed horn systems. Each pixel
contains two detectors. The four pairs of 145 GHz PSBs in the
lower row of detectors provide the best sensitivity for CMB
temperature and polarization anisotropy. We have introduced some
level of redundancy by using four independent PSB pairs, covering
with their principal axes the range $\alpha=[0,\pi]$ in $\pi/8$
steps. The four pixels in the upper row are 2-color photometers
operating at 245 and 345 GHz. These are included to provide a
lever arm for discriminating CMB from dusty foregrounds. Table
\ref{tab:rec} summarizes the properties of the \bk receiver.

\begin{table}[p]
\begin{center}
\begin{tabular}{|c|c|c|c|c|}
\hline
Freq & Bandwidth & $\#$detectors& Beam FWHM  & NET$_{CMB}$\\
\hline
145 GHz & 45 GHz & 8 & $9.95^{\prime}$  & 170 $\mu K \sqrt{s}$\\
245 GHz & 80 GHz & 4 & $6.22^{\prime}$ & 320 $\mu K \sqrt{s}$\\
345 GHz & 100 GHz & 4 & $6.90^{\prime}$ & 450 $\mu K \sqrt{s}$ \\
\hline
\end{tabular}
\end{center}
\begin{center}
\caption{\small Summary of the properties of the \bk receiver. The
noise reported in the last column is for a frequency of 1 Hz and
is the average noise of all the detectors at that frequency.
\label{tab:rec}}
\end{center}
\end{table}

\begin{figure}[p]
\plotone{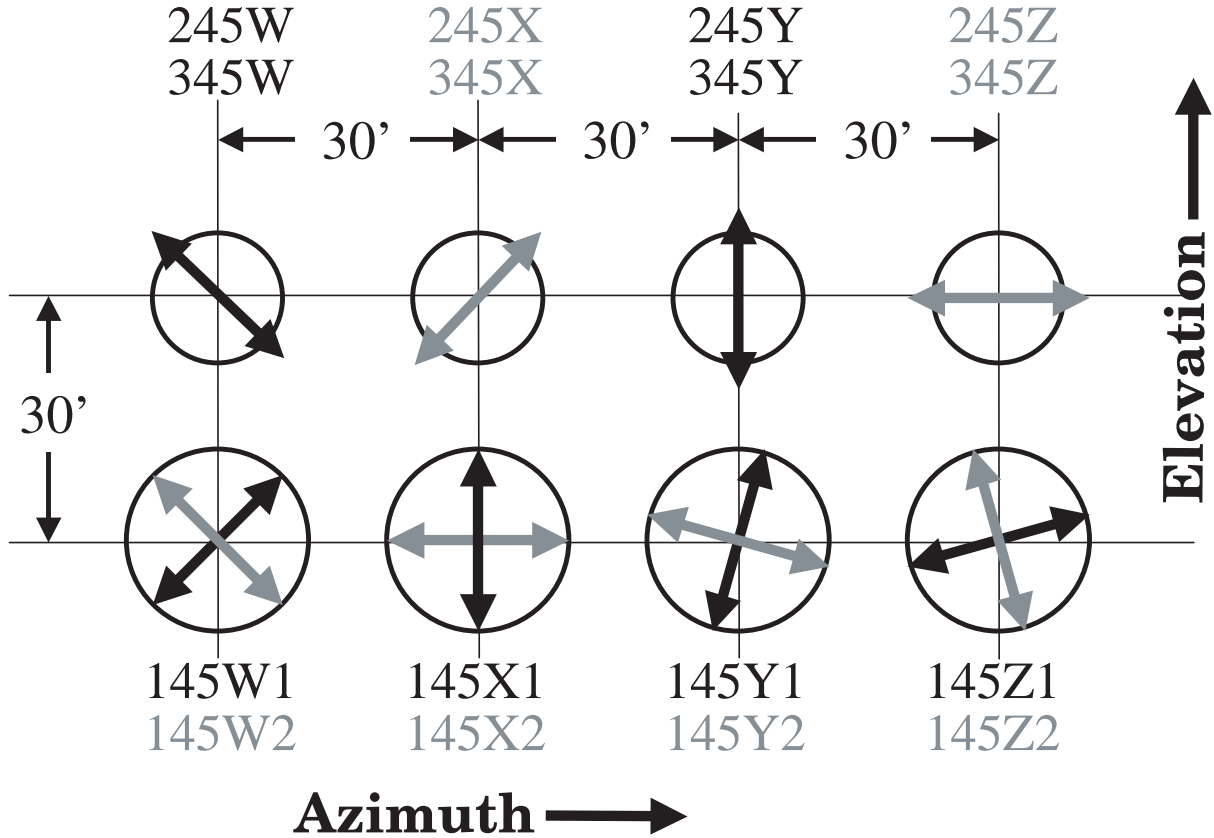} \caption{Focal Plane Schematic. 2-color
photometers with band centers at 245 GHz and 345 GHz populate the
upper row. Each photometer is only sensitive to one polarization.
The lower row has 4 pairs of PSB's. The elements in a PSB pair are
sensitive to orthogonal polarizations. The circles representing
the pixels show relative beams sizes: $\sim 7^{\prime}$ for both
photometer channels and $9.5^{\prime}$ for the PSB's. The arrows
through the circles show the orientation of the principal axis of
polarization. The photometer and PSB rows are separated by
$30^{\prime}$ in elevation, while the pixels in a row are
separated by $30^{\prime}$ in cross-elevation. The labels of the
two bolometers used for each pixel are also reported.
\label{fig:focalplane}}
\end{figure}

\subsection{Telescope and optics polarization properties \label{subs:telescope}}

The \boom telescope is an off-axis system, minimizing the
radiative loading on the detectors. Its configuration is close to
the Dragone condition \cite[]{dragone74}, which nulls the
cross-polar response in the center of the focal plane. However,
the need to accommodate a large number of detectors in the focal
plane drove our optimization towards a wide corrected focal plane
rather than nulling the cross-polar response only in the center.
We optimized the optics for diffraction limited performance at
1~mm over a $2^{\circ} \times 5^{\circ}$ field of view. Radiation
from the sky is reflected by the parabolic primary mirror (1.3~m
diameter, f=1280~mm, 45$^\circ$ off-axis) and enters the cryostat
through a thin (50 $\mu$m) polypropylene window near the prime
focus. Inside the cryostat, at 2K, the fast off-axis secondary
(elliptical) and tertiary (parabolic) mirrors re-image the prime
focus onto the detector focal plane. They are also configured to
form an image of the primary mirror at the 10~cm diameter tertiary
mirror, which is the Lyot-stop of the system. In the center of the
tertiary mirror a $\sim 1$ cm diameter hole hosts a thermal
calibration source (callamp) which is flashed at fixed intervals
during the flight (see \cite{crill03} for details). The size of
the tertiary mirror therefore limits the illumination pattern on
the primary mirror, which is under-filled by 50\% in area (85~cm
in diameter) to improve the rejection of side-lobes. This is
further improved by cold absorbing baffles surrounding the cold
mirrors and rejecting stray light. Detailed parameters of the
optics are described in \cite{piacentini02} and \cite{crill03}.

We have studied the polarization properties of this system by
means of the physical optics code BMAX \cite[]{jones05a}. The
total power beams $B(\theta, \phi)$ resulting from the system of
the feed-horns and telescope are presented in Fig.
\ref{fig:beams}.

\begin{figure}[p]
\plotone{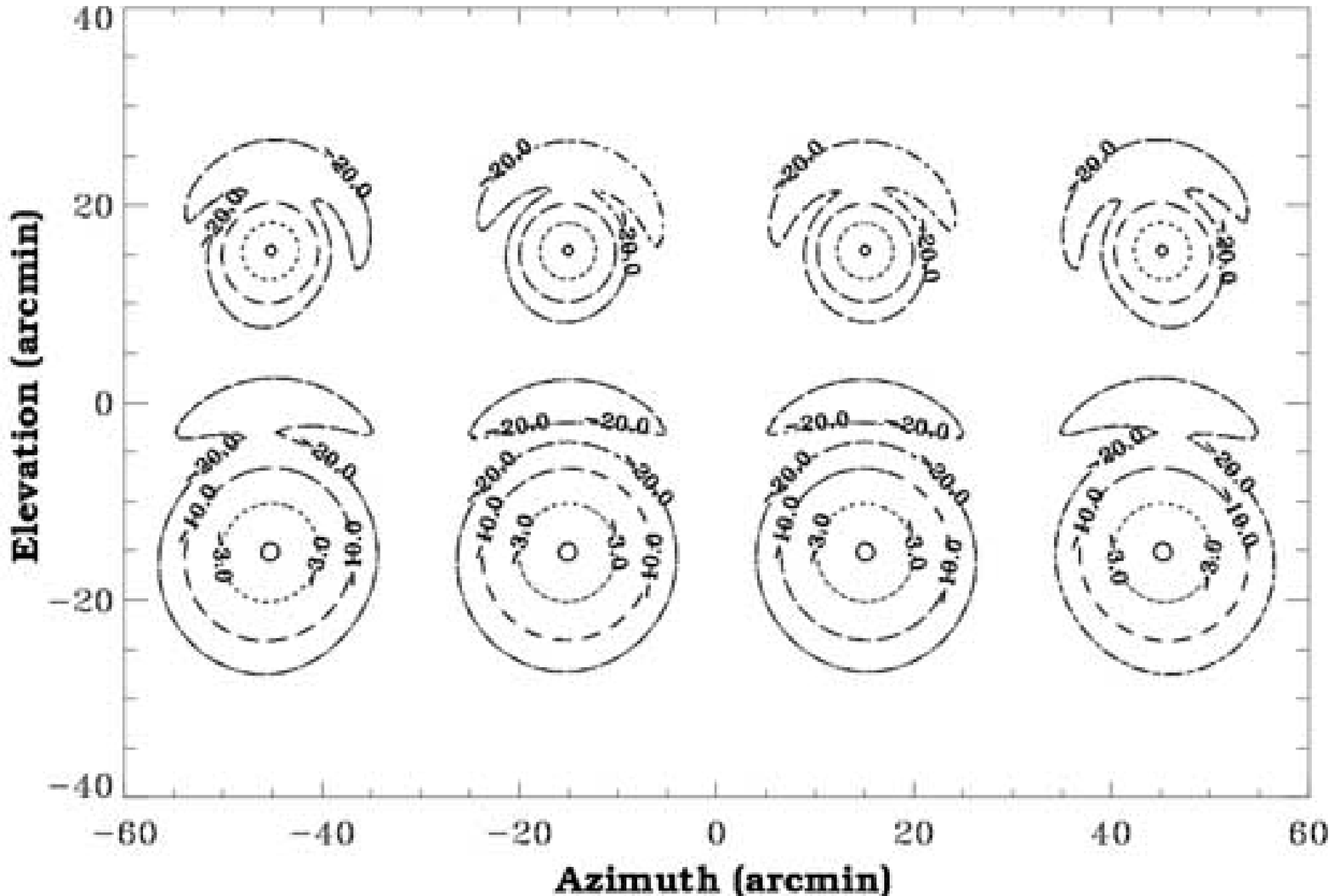} \caption{Total power beams $B(\theta,
\phi)$ computed for the \bk optical system using BMAX. The lower
row is populated by the 145 GHz PSBs, the upper row is populated
by the 245/345 GHz photometers. Contours are plotted for power
rejection levels of 0, -3, -10, -20 dB. \label{fig:beams}}
\end{figure}

In Fig. \ref{fig:xbeam}} we compare the cross-polar beam
(contours) to the co-polar beam (colors) computed for one of the
145 GHz channels. When integrated over $4 \pi$, the cross-polar
response is $\sim$ a few $\times 10^{-3}$ of the co-polar one. We
conclude that the cross-polar contribution due to the optics is
negligible with respect to the one intrinsic to the detectors.

\begin{figure}[p]
\plotone{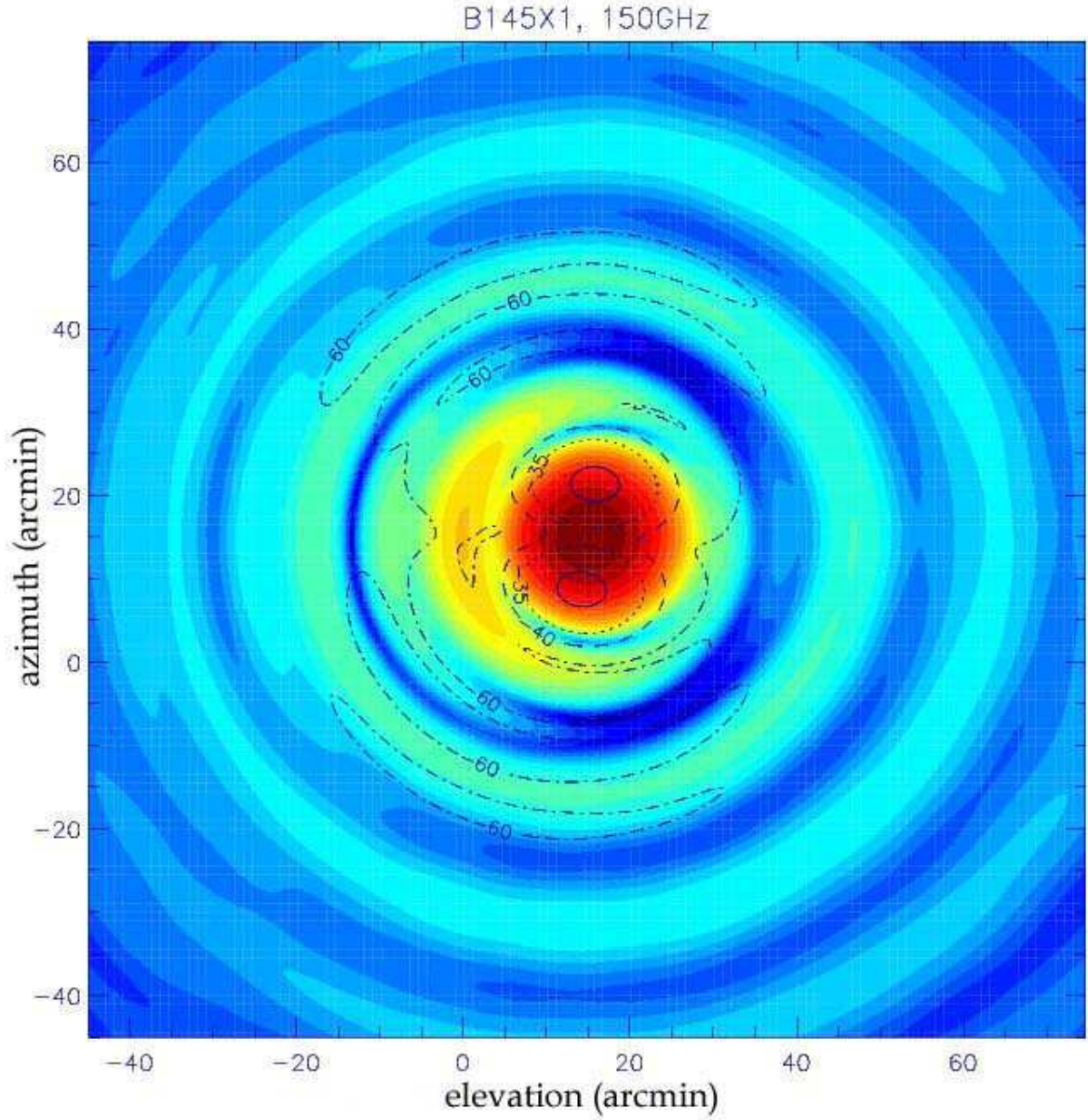} \caption{Comparison of the cross-polar
(contours) and co-polar (colors) beams for one of the 145 GHz
channels, as computed with the physical optics code BMAX.
\label{fig:xbeam}}
\end{figure}

\subsection{Readout and signal processing}

The bolometer readout unit is the same used in the previous
flights of \boom \cite[]{piacentini02,crill03}. The \bk bolometers
are AC biased with a differential sine wave at $\sim$ 140~Hz. The
AC voltage across the bolometer is modulated by the resistance
variations induced by changes in the absorbed microwave power. A
matched pair of low noise J-FETs (based on Infrared Laboratories
cryogenic JFET modules) inside the cryostat reduces the signal
impedance from $\sim 10 {\rm M}\Omega$ down to $\sim 1 {\rm
k}\Omega$; the signal is then amplified by a differential preamp
(AD624), band-pass filtered to remove noise outside the signal
bandwidth, and synchronously demodulated by a phase sensitive
detector (AD630). The output of the AD630 is proportional to the
instantaneous resistance of the bolometer. High frequencies (above
20~Hz, i.e. above the cutoff frequency of the bolometer) are
removed by means of a 4-pole order low-pass filter. Signal (and
noise) components below 5.6~mHz are attenuated using a single pole
high pass filter. A further 100$\times$ amplification is applied
to fill the dynamic range of the ADC. The total amplification of
the readout chain (from the bolometer to the ADC input) is 50000.
This ensures that bolometer plus readout noise is more than the
quantization noise of the ADC. The output signal is analog to
digital converted with 16~bit resolution at the sampling frequency
of 60.0~Hz. The warm readout circuit has a gain stability of $<
10$ ppm/$^\circ$C. The readout noise is flat down to a few mHz,
and is negligible with respect to the bolometer noise with NEP
$\sim 1.5 \times 10^{-17}$~W$/\sqrt{\rm Hz}$.

\subsection{Cryogenic System}

The cryogenic system is the same flown in 1997 and 1998. The main
cryostat \cite[]{masi99} has a 65 liter tank filled with
pressurized (1 atmosphere) liquid nitrogen, and a 60 liter liquid
helium tank, which is pumped during flight so the He is
superfluid. It maintains at 1.6K a large (60 liters) experimental
space containing the optical filters, a movable neutral density
filter, and the $^3$He refrigerator \cite[]{masi98} cooling the
bolometers and feed horn systems. The hold time of the system is
around 20 days. While for the 1998 flight we used two separate
entrance windows for the left and right sides of the focal plane,
for the \bk flight we used a single vacuum window, roughly
elliptical (100 mm $\times$ 65 mm). The material is the same $50
\mu m$ thick polypropylene used in 1997 and 1998.

\subsection{Attitude Control System}

Azimuthal attitude control is provided by a reaction wheel with a
moment of inertia approximately 0.2$\%$ of that of the complete
gondola, and by a second torque motor in the pivot which couples
the gondola to the flight train.  The torque applied to the
reaction wheel is proportional to the error in angular velocity,
and the torque applied to the pivot is proportional to the
rotation rate of the reaction wheel.  On short time scales torque
is from the reaction wheel motor,
while on long time scales torque is from the pivot motor. During a
$\sim 1$ dps scan, the reaction wheel has a peak rotation rate of
$\sim 100$ rpm. The motors are driven by a custom MOSFET bridge,
controlled by a 20KHz PWM signal from the attitude control
computer.  The back-emf of the reaction wheel motor is compensated
for in the motor control software, based on the reaction wheel
rotation rate.

Angular velocity readout is provided by KVH ECore2000 fibre optic
rate gyroscopes, which provide an angular random walk noise of
around 8'/$\sqrt{h}$ down to 0.01Hz.  Coarse absolute pointing is
provided by a TANS-VECTOR differential GPS array (well calibrated,
but with 6$^\prime$ drifts on $\sim$ 20 min time scales), and a
fixed sun sensor (sub-arc-minute precision, but difficult to
calibrate). The position of the inner frame relative to the outer
frame is determined using a 16 bit absolute encoder.  Of these
sensors, only the GPS array provides a complete measure of the
gondola orientation in Az, Pitch, and Roll, though with
significant uncertainty.

The gondola is scanned in azimuth with a rounded saw-tooth wave
form. At all times, feedback is to angular velocity from the
azimuthal fibre optic rate gyroscope.  During the linear part of
the scan, the request velocity is constant (0.3$^o$/s typical).
When the end of the linear scan is reached (as determined by an
absolute sensor, typically the differential GPS) the control
changes from fixed angular velocity to fixed angular acceleration,
until the velocity has reached the desired value in the opposite
direction. The absolute sensors are only used to define the
turnarounds.

The elevation is changed by moving the inner frame, which contains
the cryostat, optics, and receiver readout electronics, relative
to the outer frame using a geared dc motor driving a worm gear.
The elevation is only changed between observing modes.  During an
observation, the gondola is only scanned in azimuth.  No attempt
is made to remove pitch motions of the outer frame by controlling
the inner frame.

Two redundant, watchdog switched 80386 based computers take care
of the attitude control logic, including adherence to the
observation schedule file and in-flight commanding.

\subsection{Attitude Reconstruction Sensors}

Two pointed sensors, a pointed sun sensor and a pointed star
camera, which have excellent intrinsic calibration, were added for
the \bk flight.

The \bk tracking star camera (SC) consists of a video, {\it COHU}
brand (4920 series), monochrome, Peltier cooled, CCD camera
equipped with a Maksutov 500 mm focal length, f/5.6 telephoto
lens.  This setup yields approximately 4 arcsecond per pixel
resolution and $\sim$ 30 arcminute field of view. Affixed to the
lens is a seven ring baffle which is covered in aluminized mylar.
The interior of the baffle is painted with black water-based
theatre paint to prevent light scatter from entering the optics.
For the flight, a 715 nm high pass filter was installed and the
camera gain was set to minimum in an effort to lower the risk of
CCD saturation at float, in the daytime Antarctic sky.

The SC is attached to a yoke type equatorial mount. Motion control
of the two axis system is provided by two {\it Applied Motion
Products} high torque stepper motors. An encoder on each axis
provides position feedback for controlling the motion of the
mount.  Motor current is controlled via pulse width modulation and
a PID loop is used for the logic. Additional feedback from the
azimuth gyroscope is required to facilitate star tracking while
the telescope is scanning.  Video images are captured with a {\it
MATROX METEOR} frame grabber at a rate of 10 Hz.  The SC raw data
consists of readouts from the two encoders, star pixel location in
the camera field of view and star ID.  Star azimuth and elevation
relative to the gondola are reconstructed post-flight.

The first solar sensor (Pointed Sun Sensor, or PSS) is a motorized
two axis sensor \cite[]{romeo02} based on a four quadrant
photo-diode. Unbalance on the sensor activates the motors to keep
the sun spot centered on the sensor, so that the sensor accurately
tracks the sun. The angles of the sensor with respect to the
payload frame are measured by means of two 16 bit absolute
encoders (resolution $0.33^\prime$). The second Solar sensor
(Fixed Sun Sensor, or FSS) has no moving parts \cite[]{romeo02}
and is composed of two orthogonal digital meridians. In such a
device a linear slit is orthogonal to a linear CCD.  The sunlight
entering the slit excites different pixels for different angles of
incidence. The output of the sensor is the position of the center
of mass of light, which can be related to the sensor-sun angle
after appropriate calibration. Due to the variation of the
luminosity level, an in-flight calibration is needed.

\section{Pre-flight calibration \label{sect:pre_flight_cal}}

We extensively tested the instrument before the flight. This
allowed tuning of some of the parameters for maximum performance,
such as the bolometer bias frequency, preamplifier gains, and ACS
sensor gains. It also allowed measurement of instrument parameters
which are not expected to change from the lab to the flight, and
those that cannot be measured in flight. These are the time-domain
transfer function, the spectral response, the angular response,
the principal axes of the polarimeters, and the cross-polar
response. In the following subsections we describe the set of
measurements we performed before launch.

\subsection{Transfer function  \label{subs:labtransfer}}

In a scanning instrument, multipoles of the CMB
fluctuations are encoded at  frequencies $f$ in the
time-ordered detector data:
\begin{equation}
f \simeq {\dot{ \theta } \over \pi} \ell
\end{equation}
where $\dot{ \theta }$ is the sky scan speed. The frequency domain
response of the instrument has to be known in order to deconvolve
the signals obtained during the flight. In \bk we use azimuth scan
speeds between 0.2 and 0.5 deg/s: multipoles in the range $100 <
\ell < 1500$ produce bolometer signals at frequencies in the range
0.07 Hz $\simlt f \simlt$ 3 Hz.  The bolometers used in \bk have
thermal time constants in the $\sim$ 50 ms range. Their response
is closely described by a first order low-pass filter. The readout
electronics has a transfer function that matches this response and
cuts low frequencies in order to remove the DC level and low
frequency drifts.
\begin{figure}[p]
\begin{center}
\includegraphics[angle=90,width=14cm]{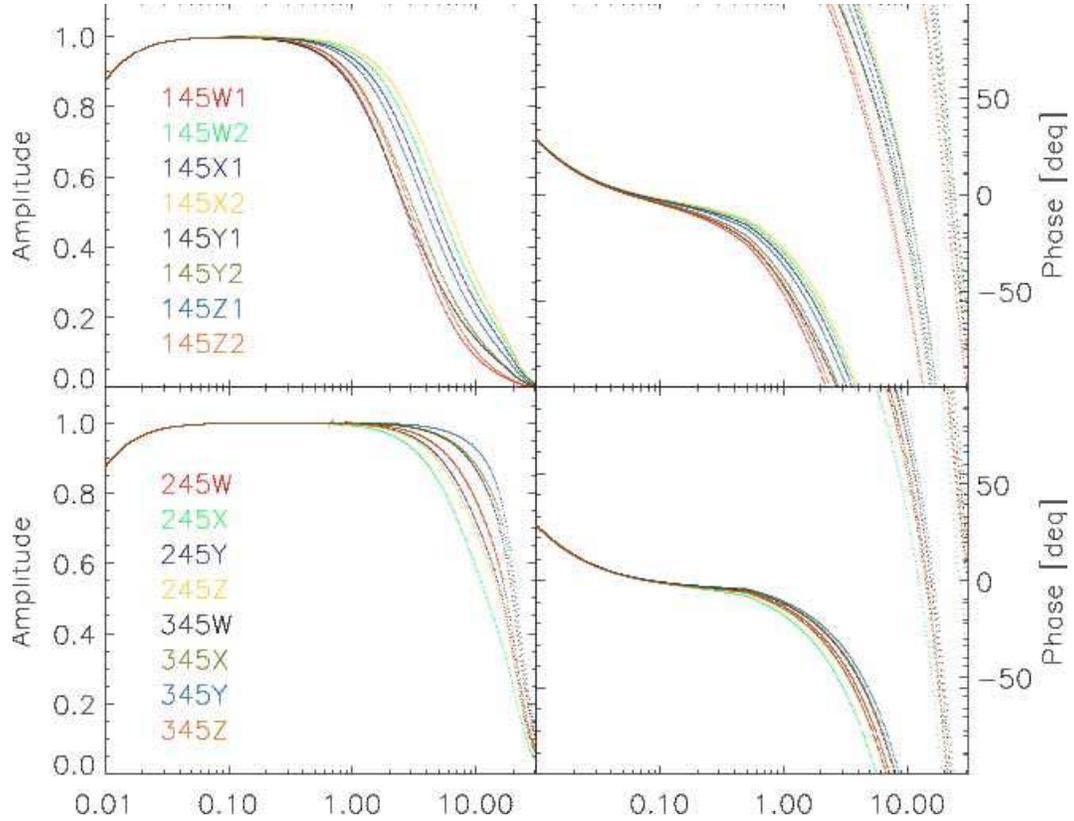}
\caption{Measured transfer function of all the \bk channels. The
CMB anisotropy and polarization signals of interest produce
detector signals in the range 0.1 Hz $\simlt f \simlt$ 5 Hz. The
low frequency response is due to the signal electronics chain.
The high frequency response is affected by the individual
bolometers' time constants;  hence the variations at high
frequencies from channel to channel.
 \label{fig:transfer}}
\end{center}
\end{figure}
The transfer functions of the detectors for the full system
(bolometers + readout) are measured with a 77K blackbody load,
with a cold (2K) Neutral Density Filter in the optical path
(transmission $\sim$ 1.5 \%). This closely simulates the in-flight
loading on the bolometers, as confirmed by the measurement of the
in-flight DC level across the detectors. The source is a modulated
signal obtained chopping a small 300K source against the 77K
blackbody background (filling about 5$\%$ of the beam solid
angle). The results are reported in Fig. \ref{fig:transfer}. The
high frequency response is limited by the time constant of the
bolometers. This is between 50 and 100 ms for the PSBs, and
between 5 and 20 ms for the 245 and 345 GHz bolometers (see table
\ref{tab:optical_efficiencies}). We estimate a $\simlt 10 \%$
uncertainty in applying these data to flight conditions.

\subsection{Spectral Calibration \label{subs:fts}}

The optical power absorbed by the bolometer is a function of the
incident optical power, the spectral response of the bolometer and
the optical efficiency of entire system.

For an unpolarized beam filling source and a detector sensing a
single polarization, the absorbed optical power is
\begin{equation}
W = {\eta \over 2} A\Omega \int B(\nu) e(\nu) d\nu,
\label{eq:popt}
\end{equation}
where $A$ is the aperture area, $\Omega$ is the beam solid angle,
$e(\nu)$ is the spectral response of the system (normalized to 1
in the maximum), $\eta$ is the spectral normalization (i.e. $\eta
e(\nu)$ is the actual spectral response), and $B(\nu)$ is the
spectrum of the input source.

The optical efficiency is then defined as the average of the
function $\eta e(\nu)$ over the bandwidth of the detector.

\begin{figure}[p]
\begin{center}
\includegraphics[angle=90,width=14cm]{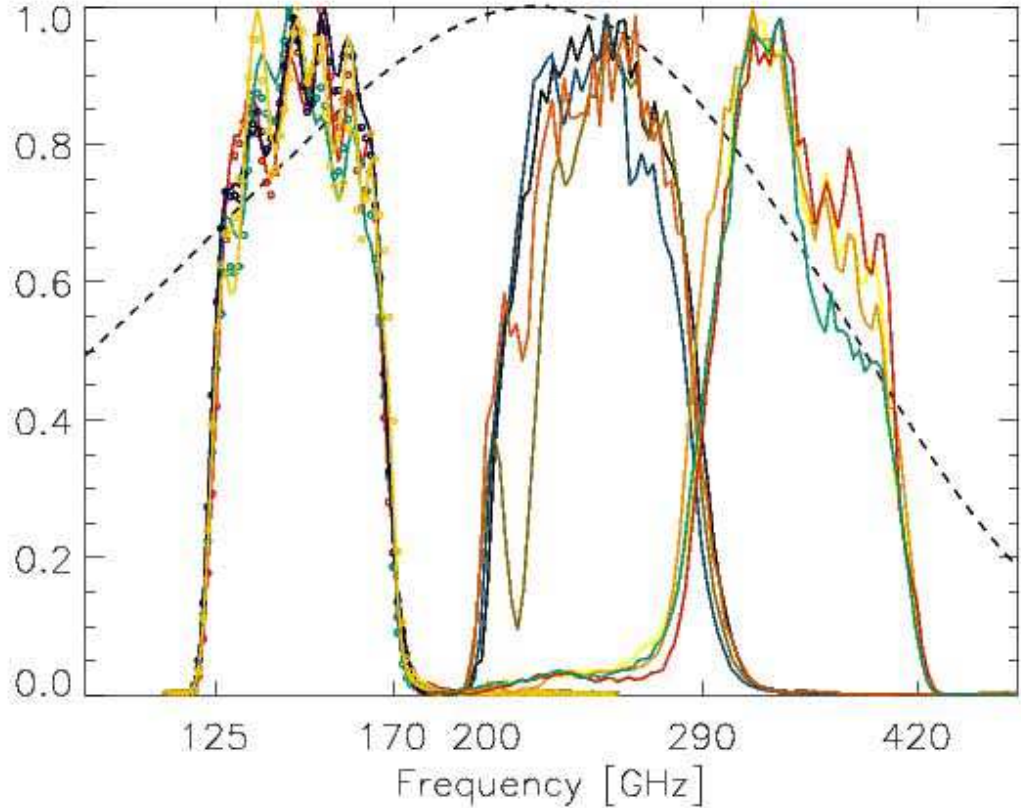}
\caption{Measured spectral response $e(\nu)$ of the 16 bolometers
used in \bk. The spectra have been measured before the flight by
means of a thermal blackbody source and a Martin-Puplett
interferometer. All the transmission measurements are normalized
for unit peak transmission. The efficiencies of the different
channels are reported in table \ref{tab:optical_efficiencies}. The
dashed line is the normalized brightness of CMB fluctuations,
normalized to the peak (217 GHz) brightness: our three bands
bracket the maximum of such a spectrum.
 \label{fig:specplot}}
\end{center}
\end{figure}

The spectral response of our detectors is measured using a
polarized Martin-Puplett Fourier Transform Spectrometer (FTS).
Figure \ref{fig:specplot} shows the details of the spectral
response $e(\nu)$ of the different bolometers, as measured before
the flight.

To normalize the spectral response, we use load curve data taken
with the bolometers viewing blackbody optical loads of 77~K
(liquid nitrogen) and 90~K (liquid oxygen) \cite[]{montroy03}. For
load curves taken with two different optical loadings, we can
assume that the incident power (optical plus electrical) on the
bolometer is the same when the bolometer resistance measured from
the two curves is equal
\begin{equation}
P_{elec}(I_{bias1}) + W^{load1} = P_{elec}(I_{bias2}) +W^{load2},
\end{equation}
By differencing the electrical power, we can calculated the
difference in optical power
 \begin{equation}
\Delta P = P_{elec}(I_{bias1}) - P_{elec}(I_{bias2}) = W^{load2} -
W^{load1}.
\end{equation}
We calculate the spectral normalization $\eta$ from this power
difference. Table \ref{tab:optical_efficiencies} shows the
spectral normalization and flat band optical efficiency calculated
from load curve differences.

\begin{table}[p]
\begin{center}
\begin{tabular}{|c|c|c|c|c|c|c|}
\hline
Channel & $\langle \nu \rangle$ (GHz) & $\sqrt{\langle (\Delta \nu)^2 \rangle}$ (GHz) & ${MJy/sr \over K_{CMB}}$ & Spec. Norm. & Opt. Eff. & $\tau$ (ms) \\
\hline
145W1 & 147.3 & 13.5 & 388 & 0.35 & 0.26 & 89\\
145W2 & 146.7 & 12.4 & 387 & 0.35 & 0.26 & 50\\
145X1 & 146.2 & 12.6 & 386 & 0.37 & 0.28 & 58\\
145X2 & 146.6 & 12.4 & 387 & 0.33 & 0.25 & 43\\
145Y1 & 147.0 & 12.8 & 388 & 0.38 & 0.29 & 97\\
145Y2 & 146.8 & 12.7 & 387 & 0.38 & 0.31 & 82\\
145Z1 & 147.0 & 12.6 & 388 & 0.26 & 0.19 & 66\\
145Z2 & 147.1 & 13.1 & 388 & 0.21 & 0.14 & 81\\
\hline
245W  & 248.0 & 26.0 & 461 & 0.38 & 0.33 & 12\\
245X  & 250.7 & 24.7 & 460 & 0.36 & 0.28 & 20\\
245Y  & 244.0 & 24.6 & 465 & 0.33 & 0.27 & 15\\
245Z  & 247.6 & 26.5 & 461 & 0.32 & 0.27 & 16\\
\hline
345W  & 340.3 & 38.6 & 321 & 0.81 & 0.70 & 7.9\\
345X  & 338.9 & 38.9 & 323 & 0.89 & 0.77 & 8.0\\
345Y  & 344.1 & 37.6 & 314 & 0.82 & 0.72 & 4.7\\
345Z  & 337.8 & 38.7 & 326 & 0.87 & 0.68 & 12\\
\hline
\end{tabular}
\end{center}
\caption{Spectral and time response calibrations: \small For all
channels (named in column 1) we report in the second column the
average frequency, and in the third column the optical bandwidth.
These are estimated in terms of the first and second moments of
the transmission spectra of Fig. \ref{fig:specplot}. The
conversion factor between Specific Brightness and CMB temperature
fluctuations, as computed from the transmission spectra of Fig.
\ref{fig:specplot}, is reported in the fourth column. Spectral
normalizations and flat band optical efficiencies are measured
using NDF-up load curve power differences. The spectral
normalization $\eta$ (fifth column) is calculated using the
spectral response from the FTS measurements. The optical
efficiency $\langle \eta e(\nu) \rangle_{band}$ (sixth column) is
calculated assuming a flat spectral response.
The detectors are assumed to be single-moded.  The high optical
efficiency of the 345~GHz channels is likely due to propagation of
multiple modes to the detector. The time constant $\tau$ is
reported in the seventh column, as measured in the laboratory with
loading conditions similar to the flight ones.
\label{tab:optical_efficiencies}}
\end{table}

The spectral calibration of the detectors allow us to compute
color corrections for all our photometric channels. In fact our
system is calibrated in flight using the CMB anisotropy as a
reference source (see \S \ref{subs:speccal}). For this reason the
responsivity must be corrected when observing sources with a
different spectrum, like, e.g. interstellar dust emission.
However, the spectral matching of PSBs is good enough that even
not applying any color correction, the spurious polarization
degree is $\simlt 5 \times 10^{-3}$ for any reasonable ISD
spectrum.

\subsection{Polarimetric Calibration}

To characterize the polarization properties of our system, we
measured the polarization efficiency, $\xi$ (or the cross-polar
response $\epsilon=1-\xi$), and the polarization angle of each
detector, $\alpha_{det}$ (see eqn. \ref{eq:poleff} and
\ref{eqn:shape}).

We made these measurements with the receiver alone and with the
whole system including the telescope.

\begin{figure}[p]
\includegraphics[angle=-90,width=18cm]{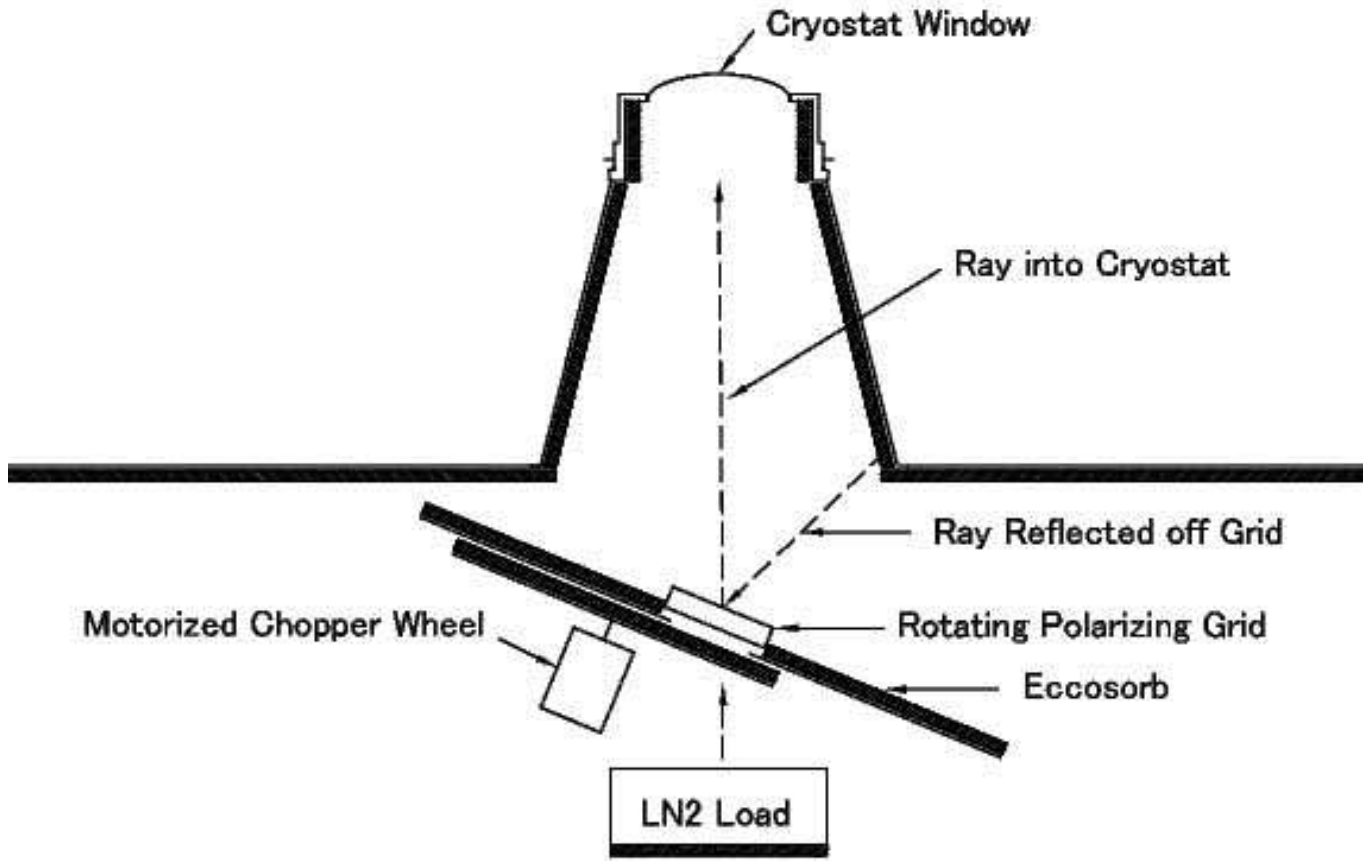}
\caption{The source used for the polarization calibration of the
receiver. The system is tilted at $22.5^{\circ}$ so that rays
reflecting off the grid go to 300 K eccosorb absorbers (shown as
black, thick lines in the drawing). The aperture of the system is
$\sim 5$ cm. The chopper wheel chops between 77 K and 300 K, while
a belt drive rotates the polarizing grid. The chopper wheel
rotates at 2~Hz, while the grid has a rotation period of about 10
minutes. \label{fig:rotator}}
\end{figure}

Figure \ref{fig:rotator} shows the configuration of the receiver
polarization calibration. A rotating polarizing grid is placed
directly beneath the cryostat. It sits above a cold load (liquid
nitrogen) which is modulated by a chopper wheel. The chopper wheel
is rotated at 2~Hz and the grid has a rotation period of
approximately 10 minutes. When the transmission axis of the grid
is aligned with the polarization axis of the detector, the
detector sees whatever is behind the grid (i.e. the chopper wheel
or the cold load). When the grids are $90^\circ$ out of alignment
the detector sees radiation reflected off the grid. The size of
the source aperture is such that the modulated signal fills $\sim
5 \%$ of the receiver solid angle. The best fit values of the
cross-polar response parameter $\epsilon$ when the source is on
the optical axis are reported in table~\ref{tab:crosspol}.

\begin{figure}[p]
\begin{center}
\includegraphics[angle=0,width=11.5cm]{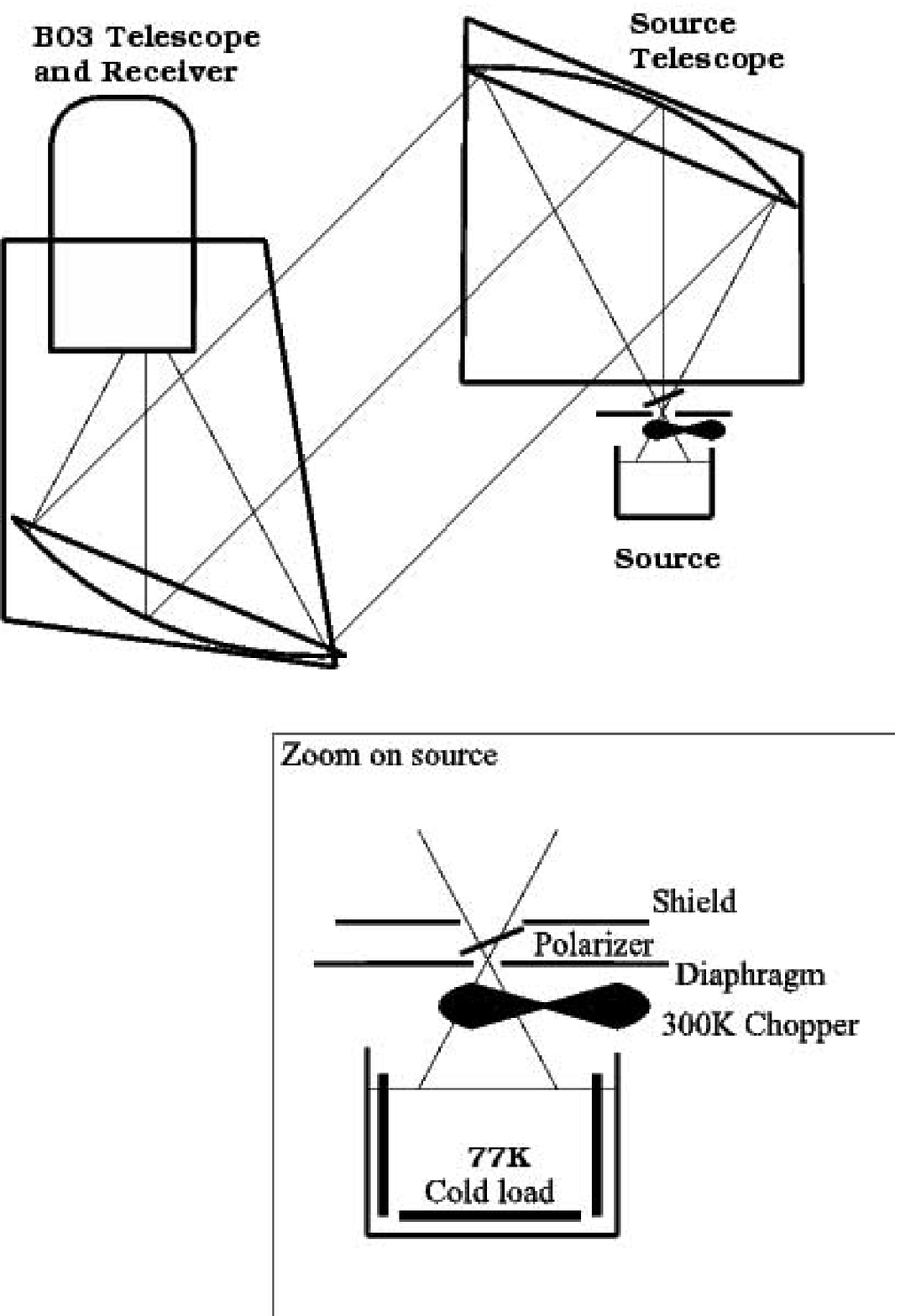}
\caption{The apparatus for measuring polarization angle and cross
polar response of the entire telescope. The cold load is kept at
77K by a liquid nitrogen bath. The source telescope is a 1.3 m off
axis paraboloid (the spare of the \bk mirror), and is kept at a
distance of $\sim 6$ m from the \bk telescope. The source is
scanned in Azimuth by rotating the \bk payload around its vertical
axis. With a small diaphragm in the source, scans are repeated at
different elevations to obtain a full map of the beam and of its
polarization properties. With a large diaphragm, producing a
beam-filling beam, we change azimuth and elevation of \bk to find
the maximum signal. \label{fig:paolorizer}}
\end{center}
\end{figure}
For the measurement of the full system we used a fully polarized
modulated plane wave filling the telescope aperture. This was
produced using a thermal source in the focus of the spare \boom
primary mirror. This inverted telescope system was placed in front
of the \bk telescope (see Figure~\ref{fig:paolorizer}). Analytical
calculations \citep[]{dijk74} and optical simulations show that
the cross-polar response introduced by the spare primary at our
frequency is $\simlt 1 \%$. The source at the focus is made with a
high efficiency wire grid (12.5 $\mu m$ diameter Tungsten wires,
25 $\mu m$ spacing) in front of a chopper alternating a 77K
blackbody and a 273K blackbody. The wire grid axis was tilted by
28$^o$ with respect to the optical axis of the system. The \bk
telescope was mounted on a rotating platform with absolute
encoders for angle measurements. Bolometer data and the chopper
reference signals were read by the Data Acquisition System of the
experiment. A software demodulator measuring amplitude and phase
of the synchronous signal was used to extract the polarized source
signal from the background. Data were taken for 18 positions of
the wire grid between 0 and 180 degrees. After correcting for the
tilt of the grid with respect to the optical path, the data for
all the bolometers and for all the incidence angles were fit with
equation (\ref{eqn:shape}).

\begin{figure}[p]
\begin{center}
\includegraphics[angle=0,width=6cm]{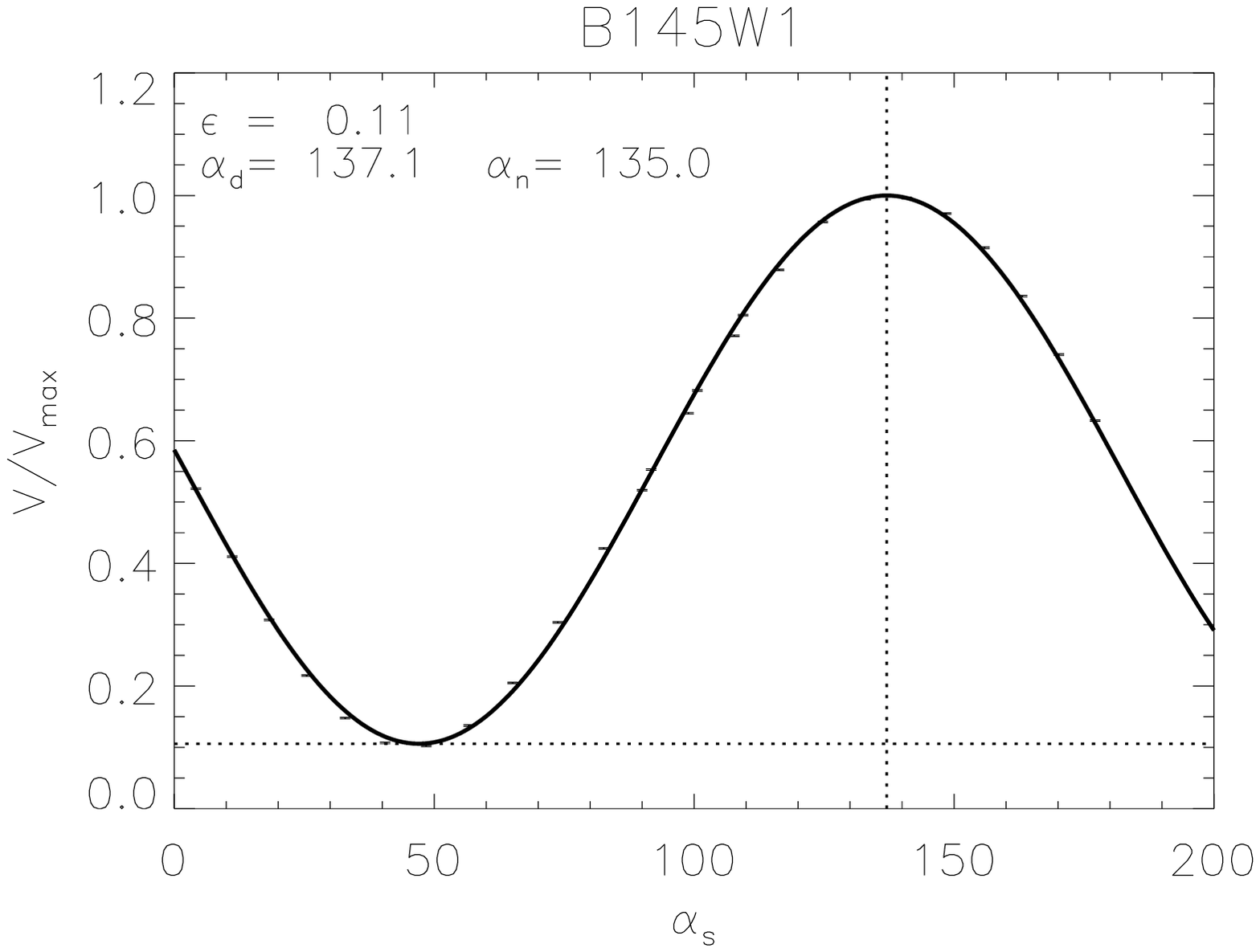}
\includegraphics[angle=0,width=6cm]{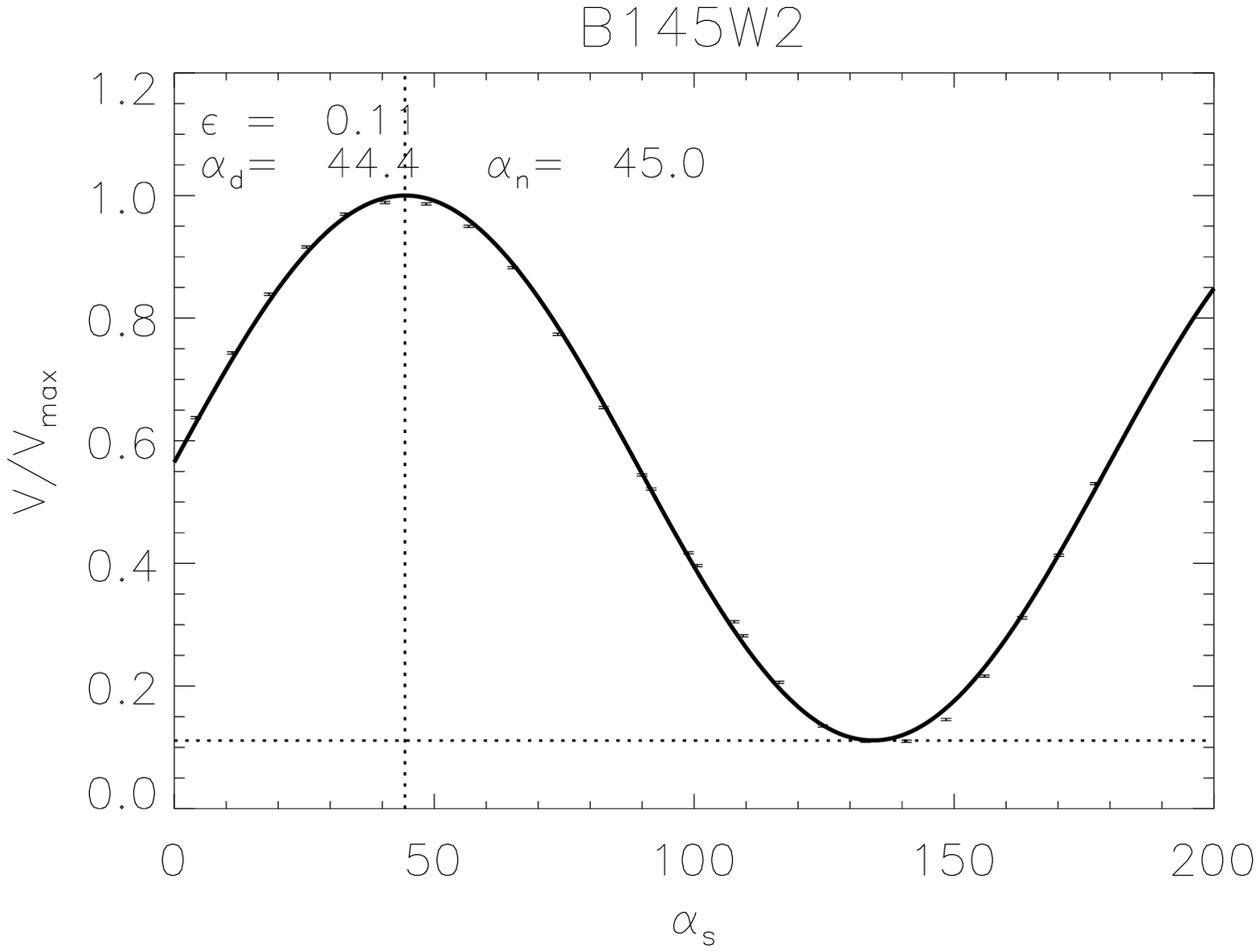}
\includegraphics[angle=0,width=6cm]{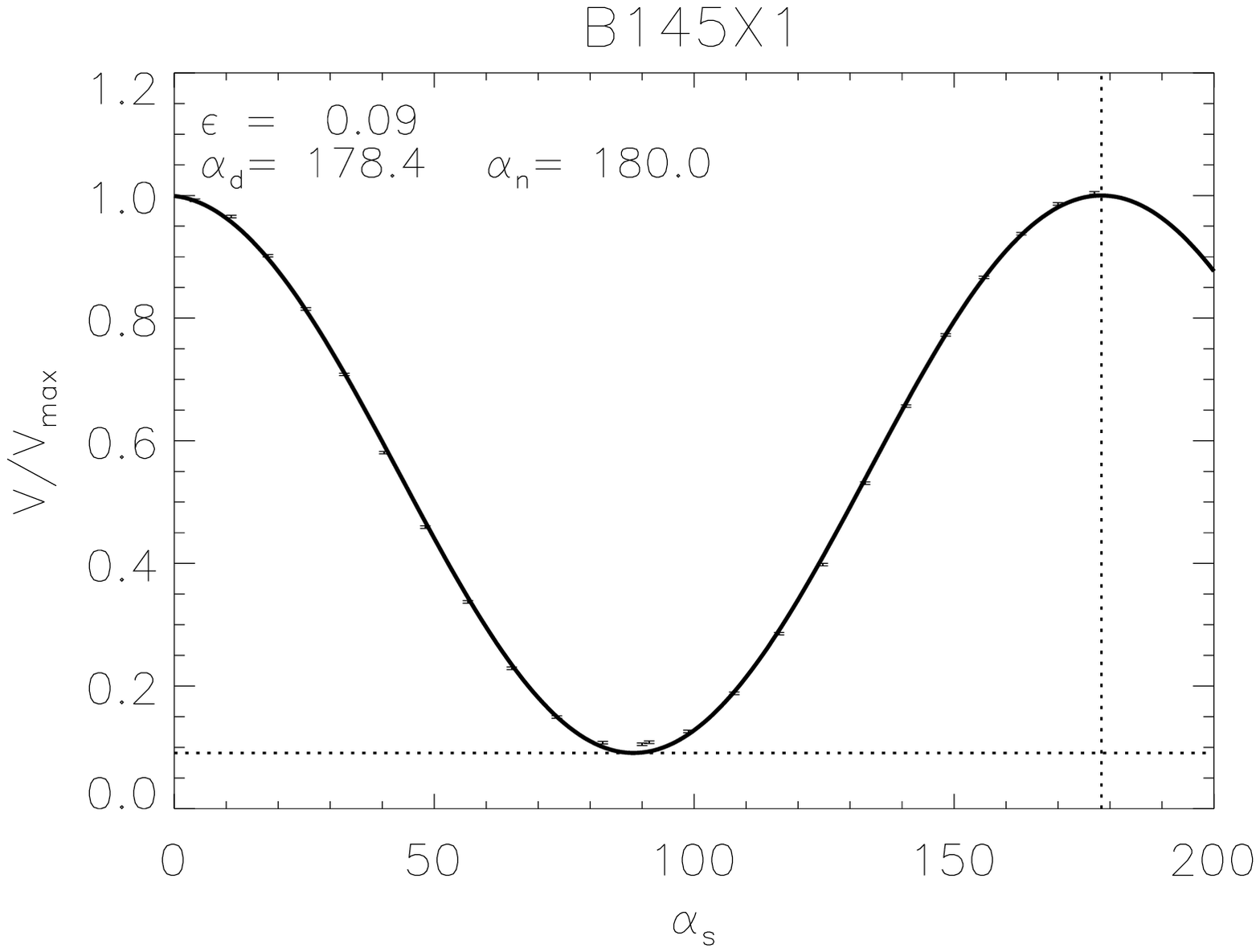}
\includegraphics[angle=0,width=6cm]{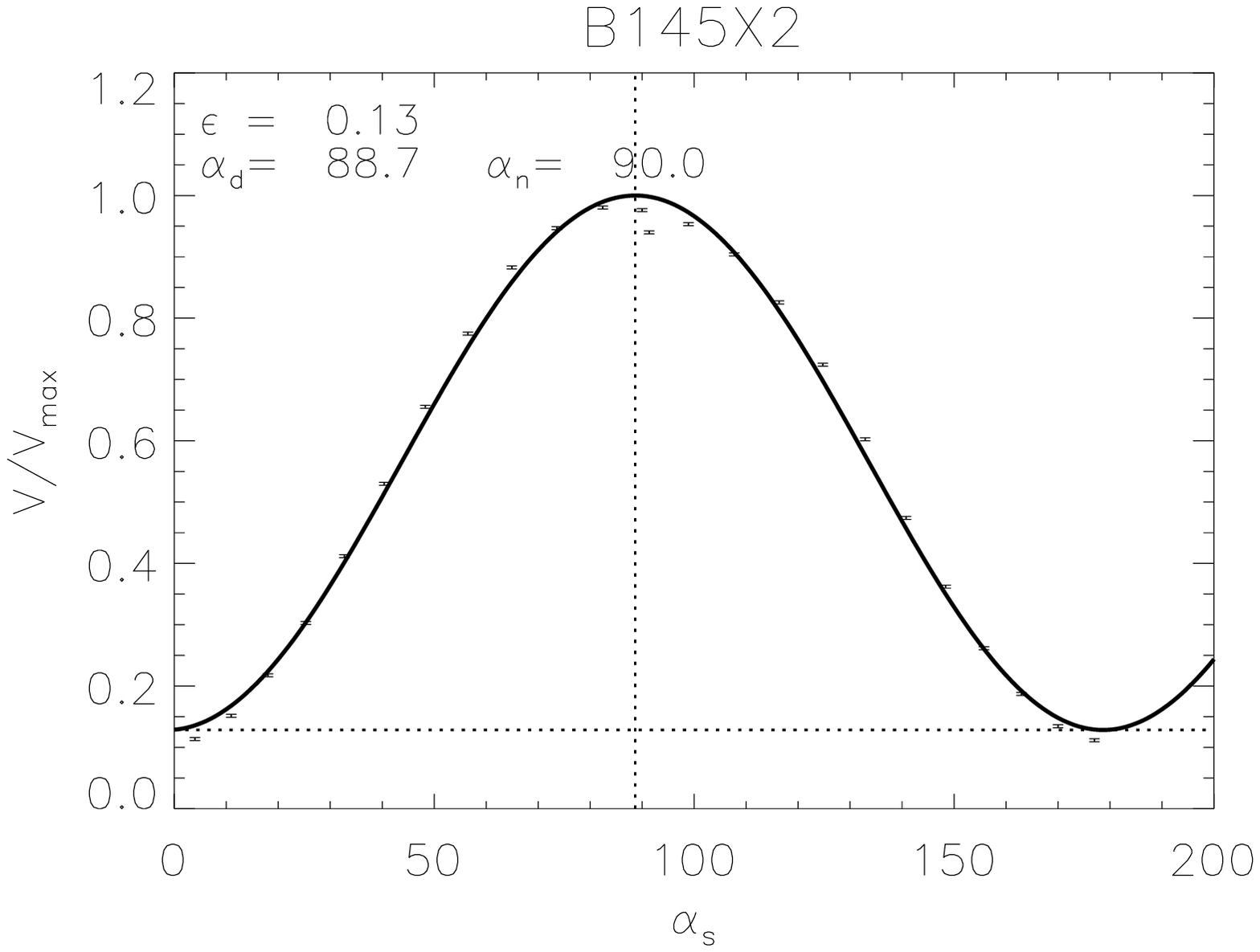}
\includegraphics[angle=0,width=6cm]{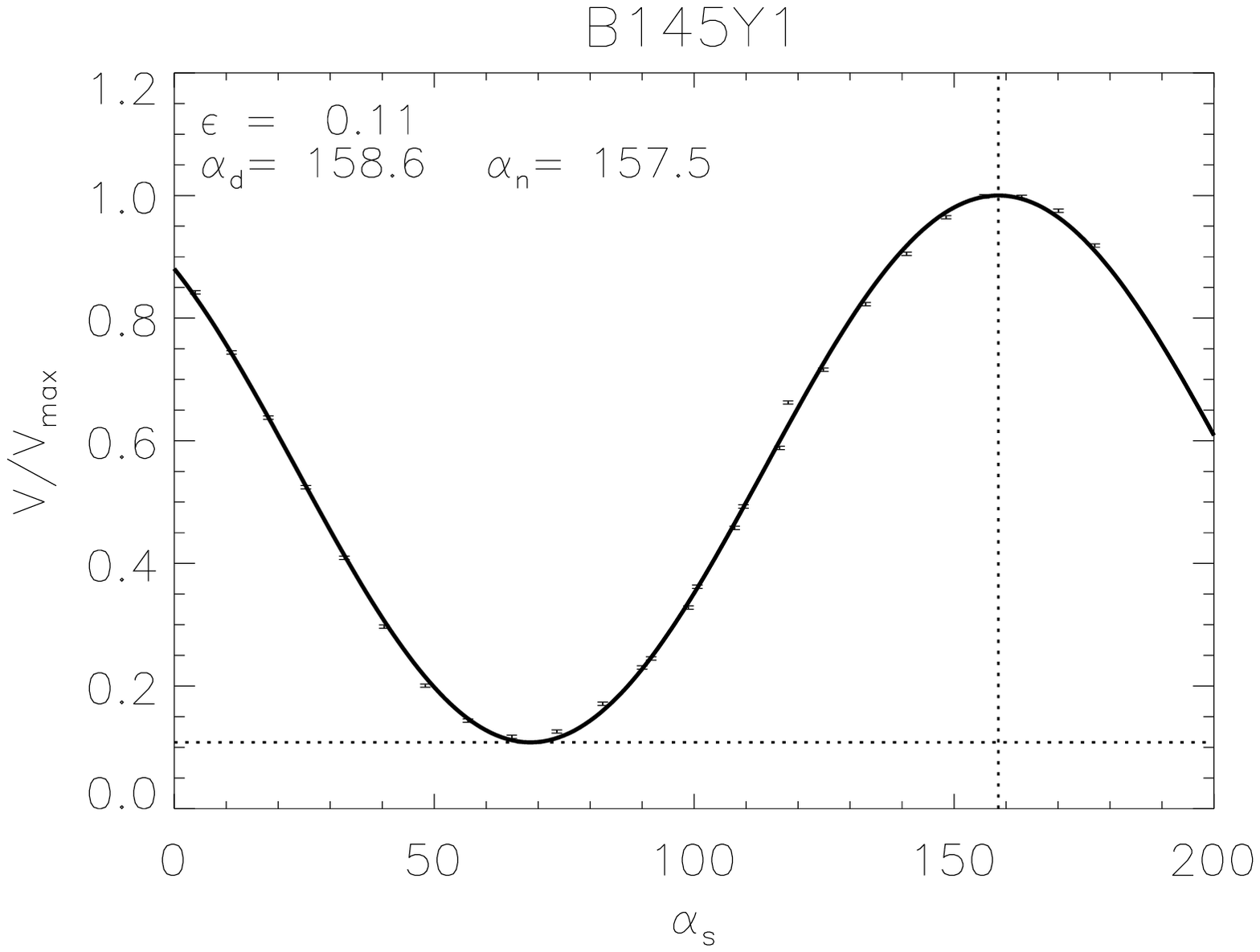}
\includegraphics[angle=0,width=6cm]{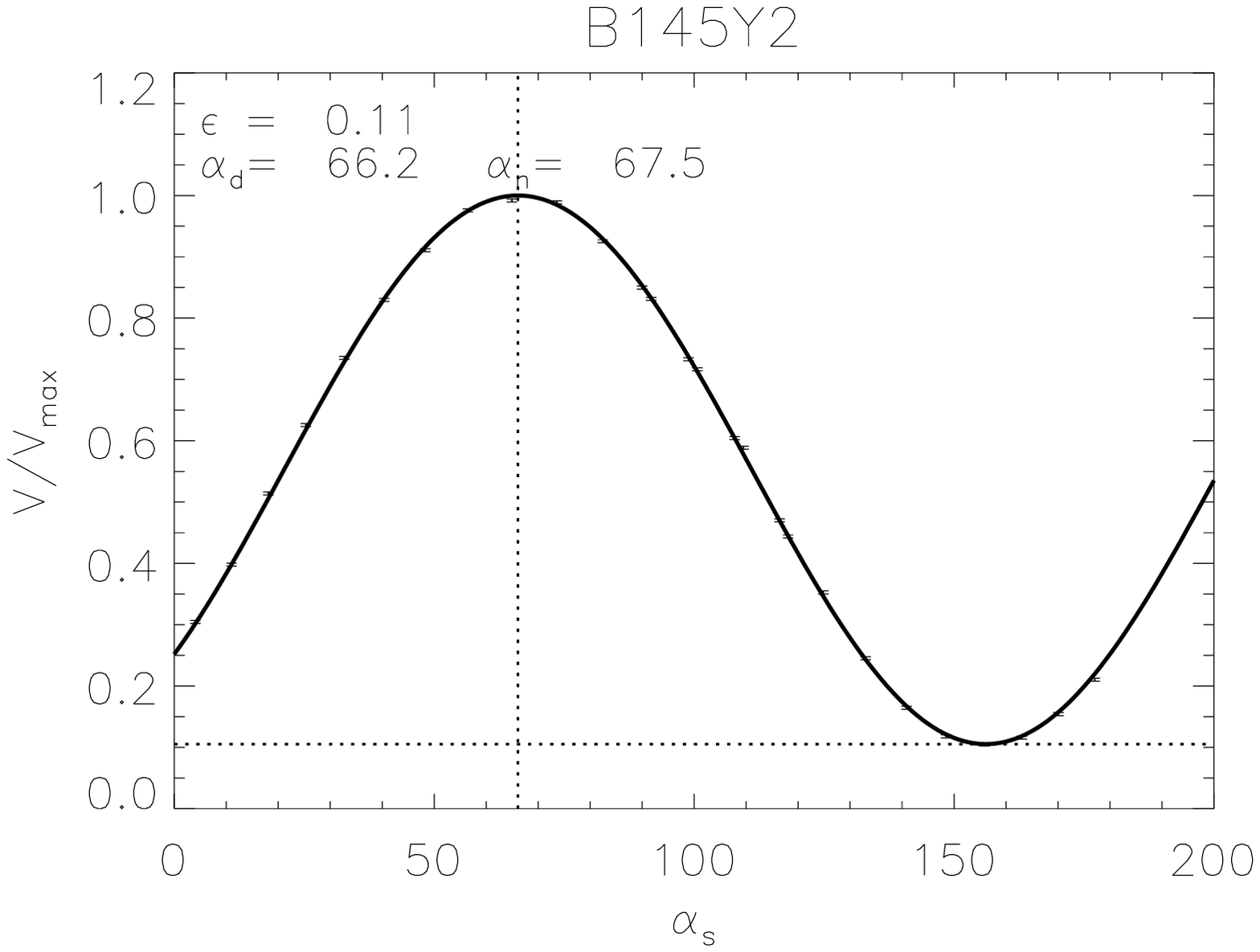}
\includegraphics[angle=0,width=6cm]{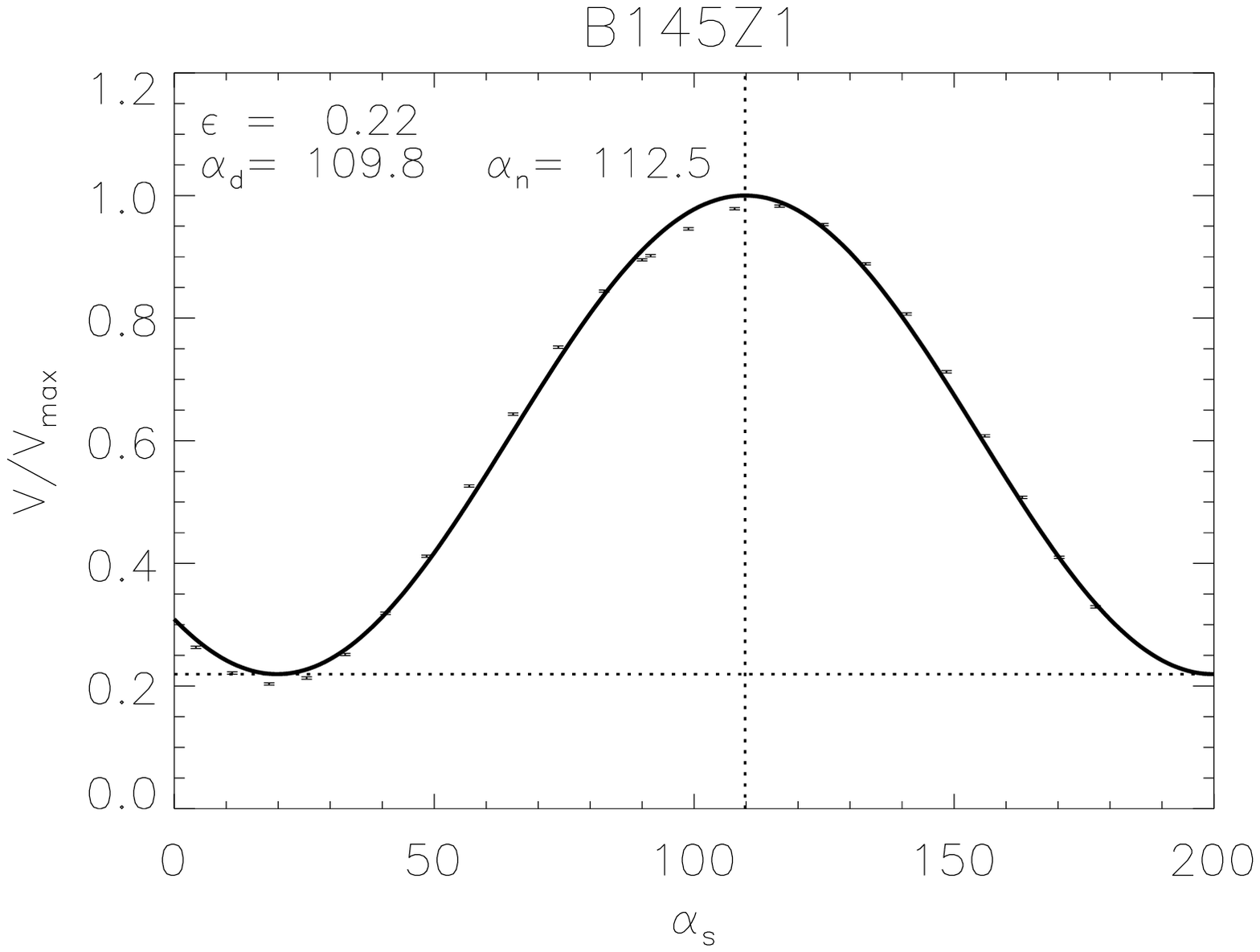}
\includegraphics[angle=0,width=6cm]{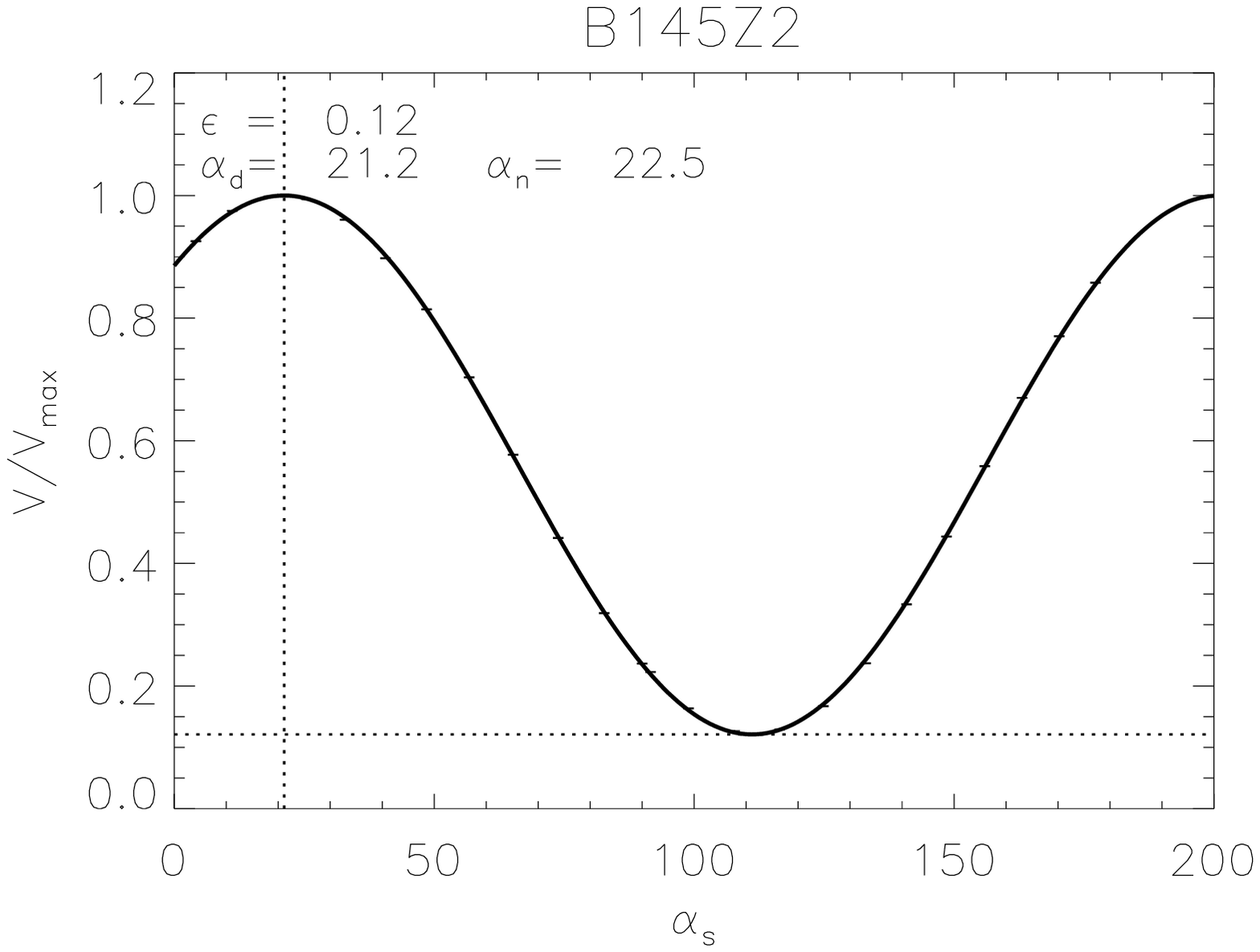}
\caption{Beam-integrated polarization response of the entire
instrument (PSB channels).} \label{fig:crosspsb}
\end{center}
\end{figure}

\begin{figure}[p]
\begin{center}
\includegraphics[angle=0,width=6cm]{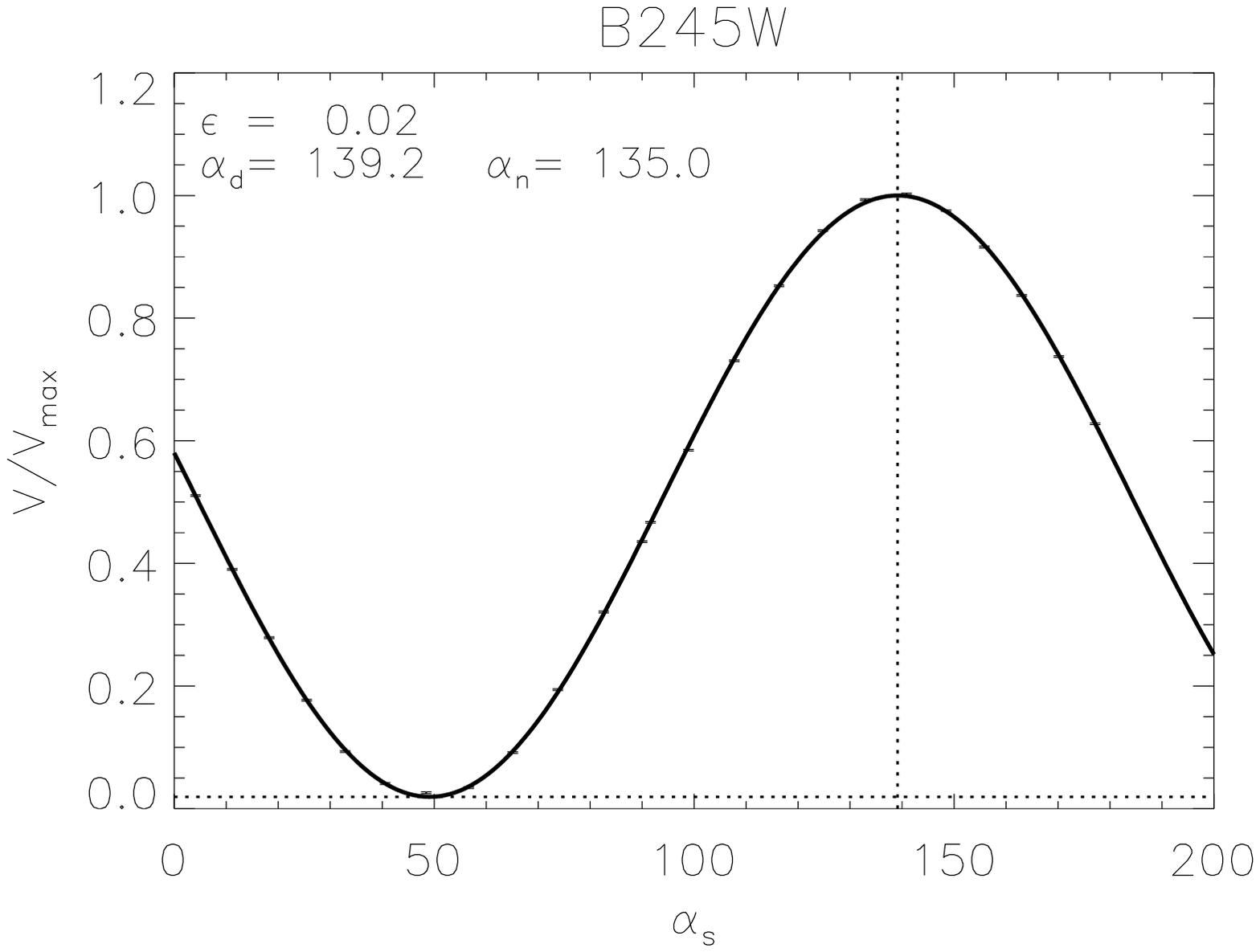}
\includegraphics[angle=0,width=6cm]{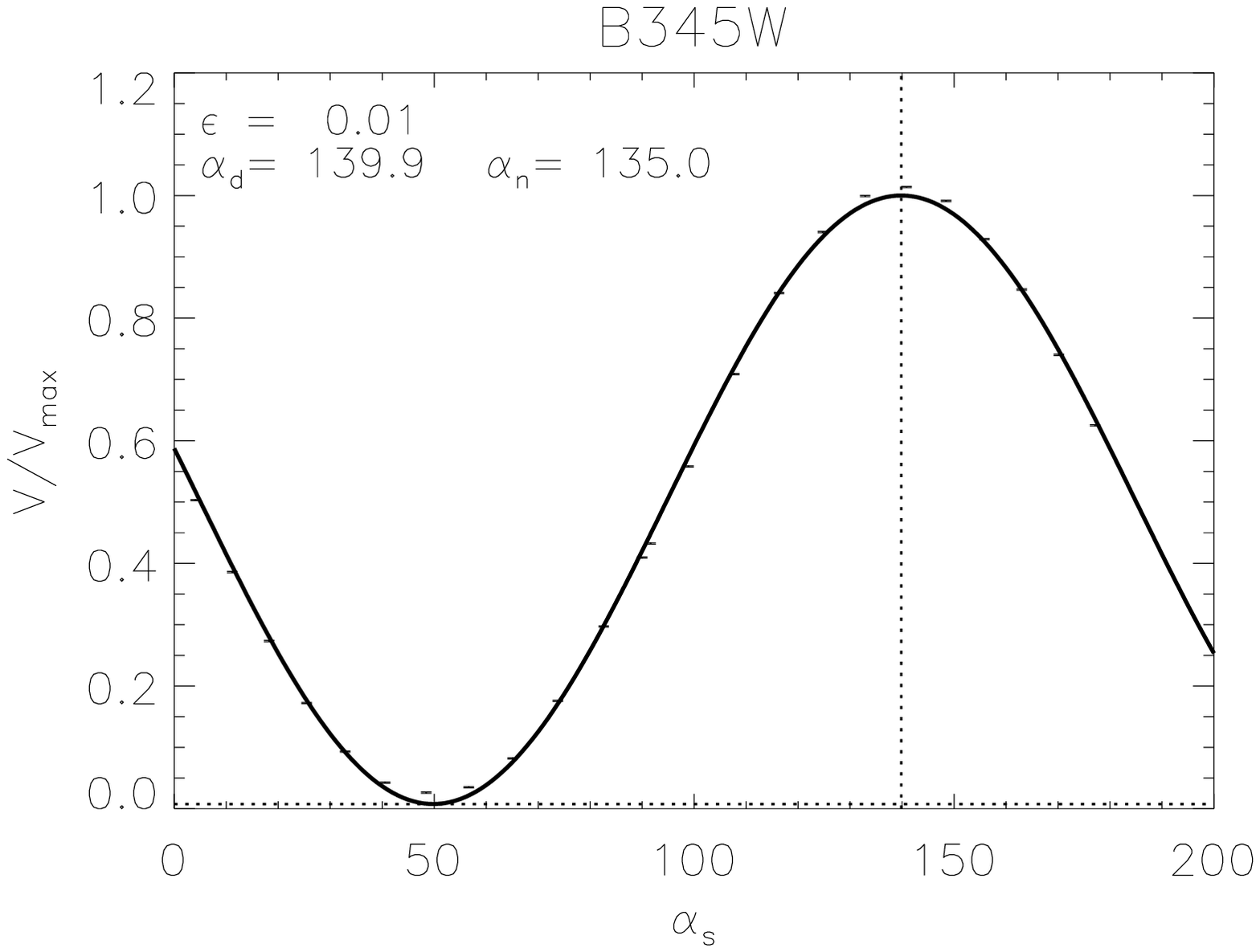}
\includegraphics[angle=0,width=6cm]{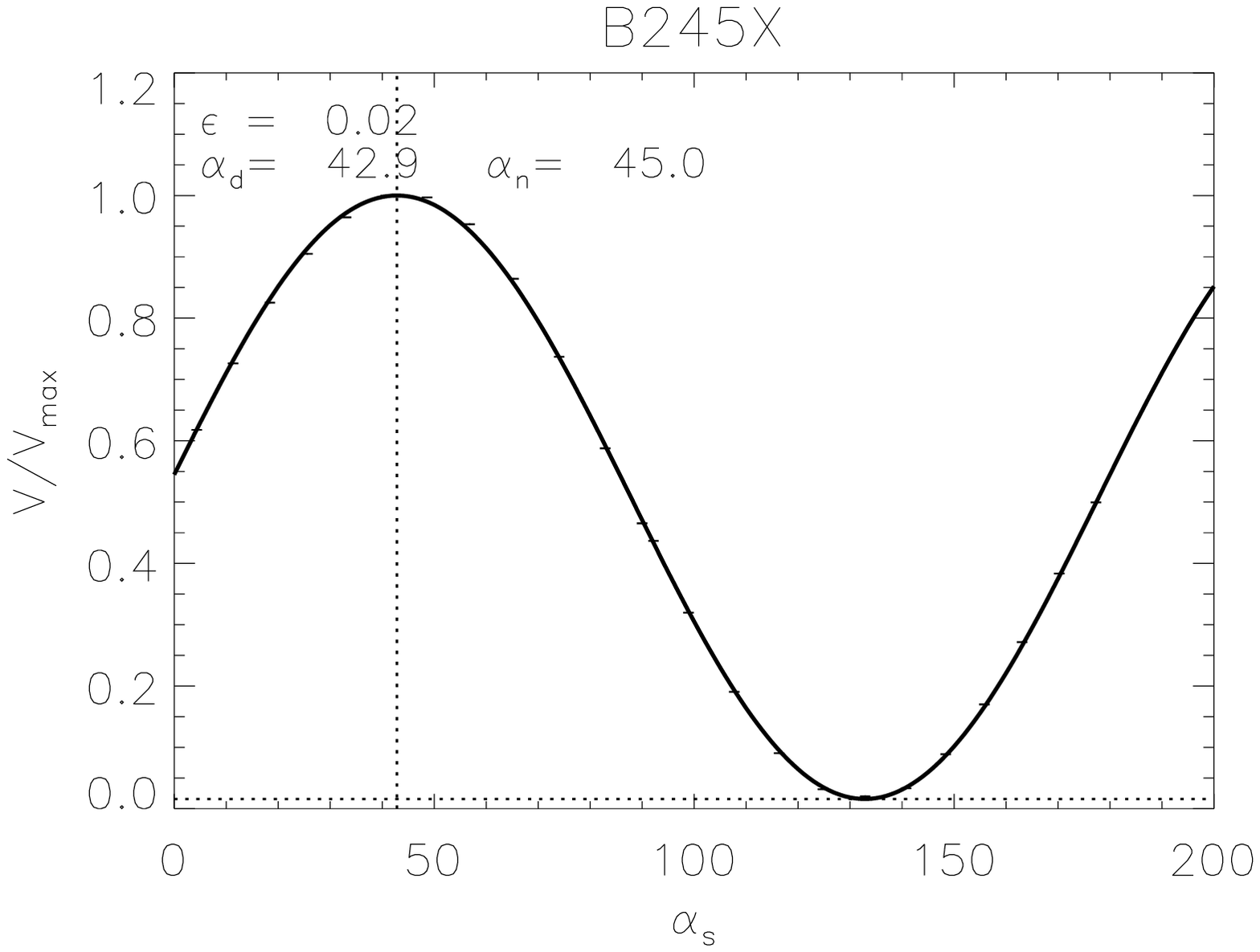}
\includegraphics[angle=0,width=6cm]{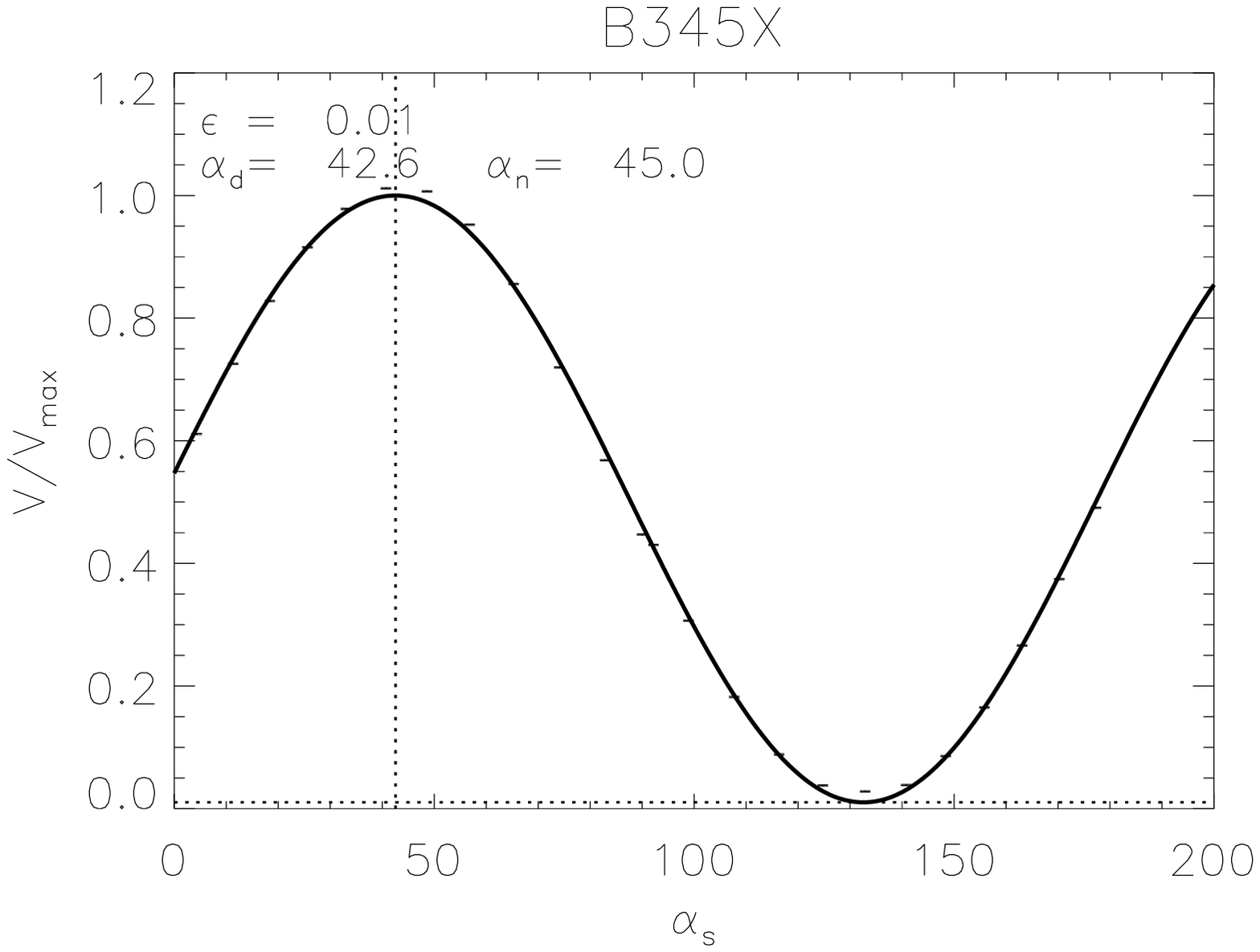}
\includegraphics[angle=0,width=6cm]{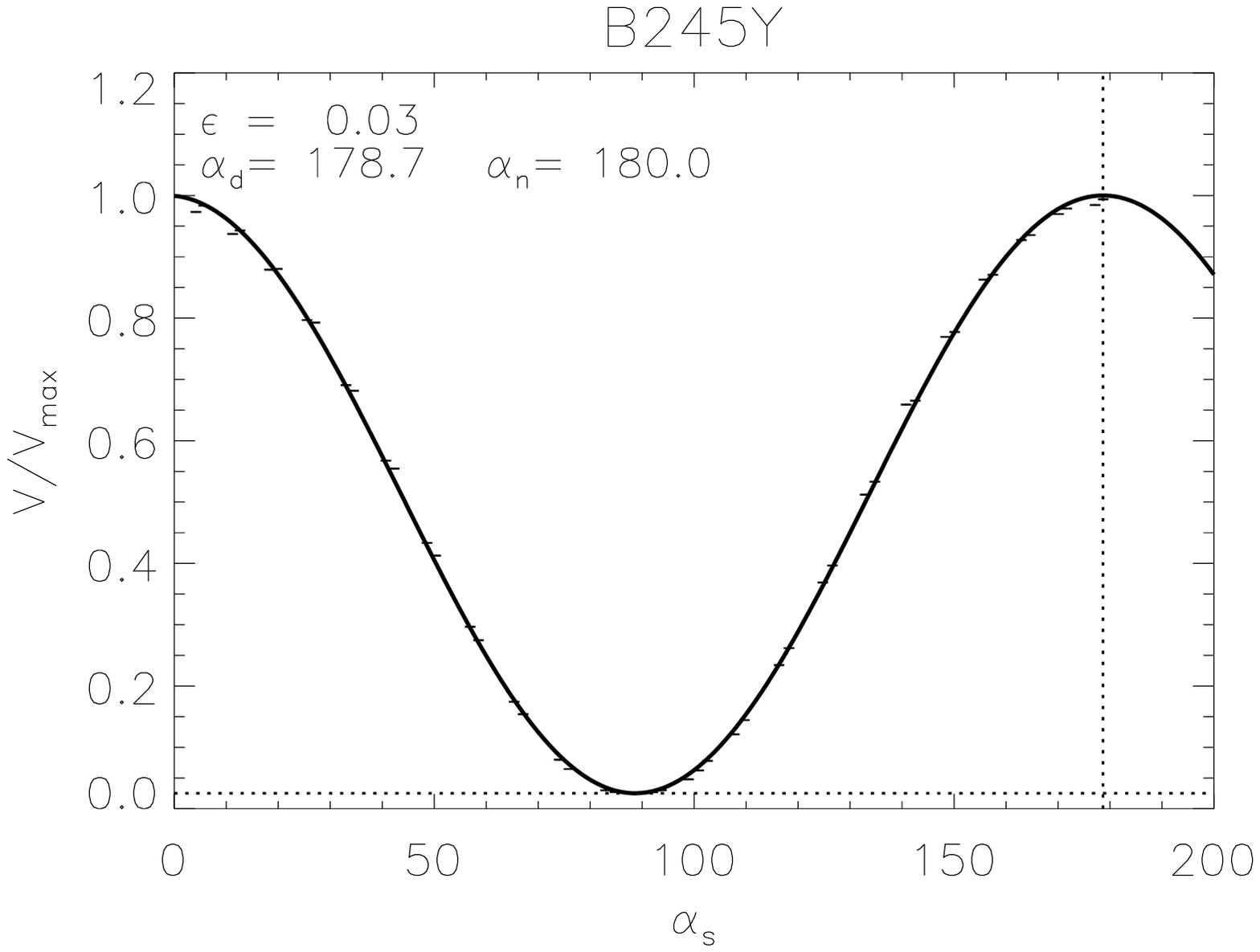}
\includegraphics[angle=0,width=6cm]{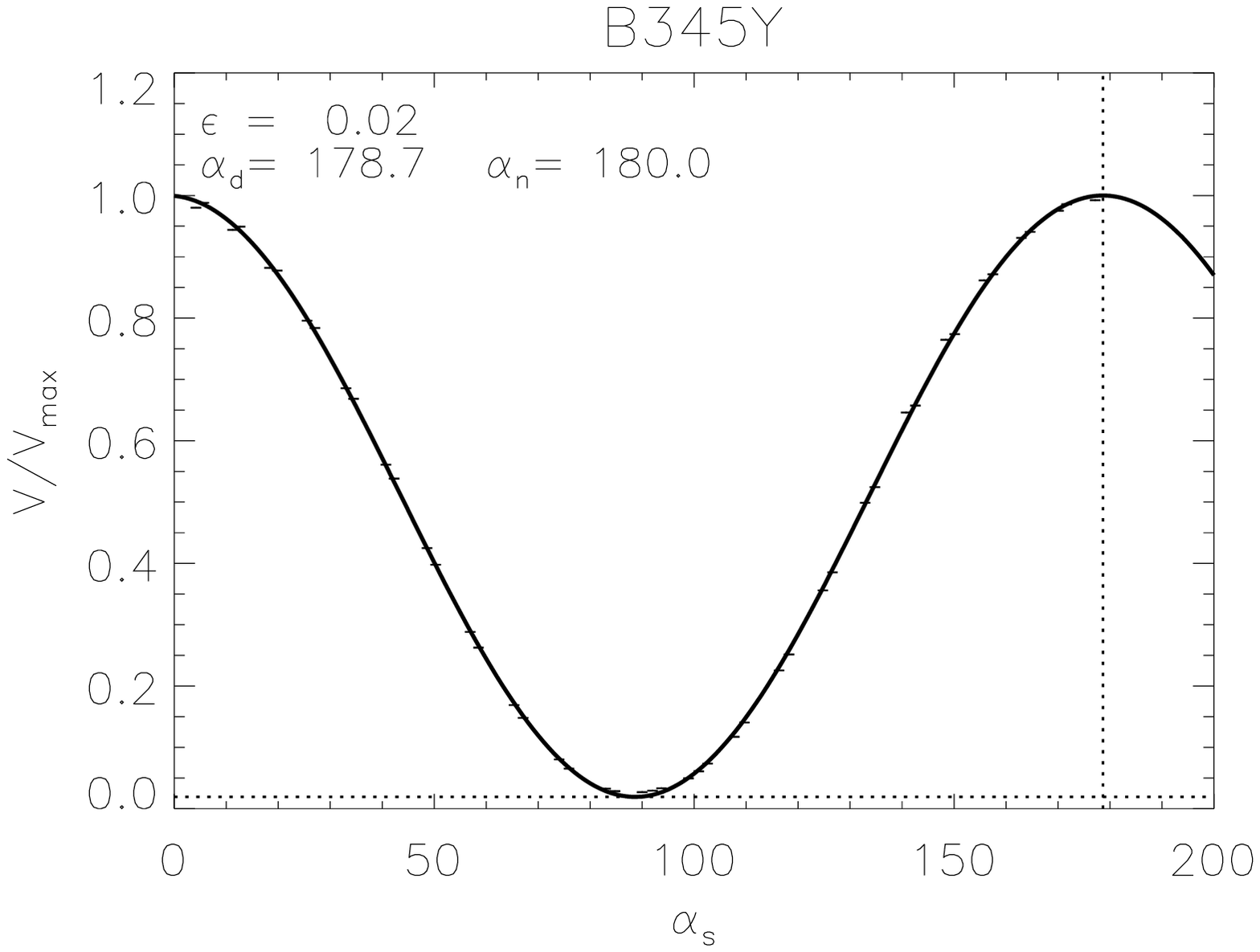}
\includegraphics[angle=0,width=6cm]{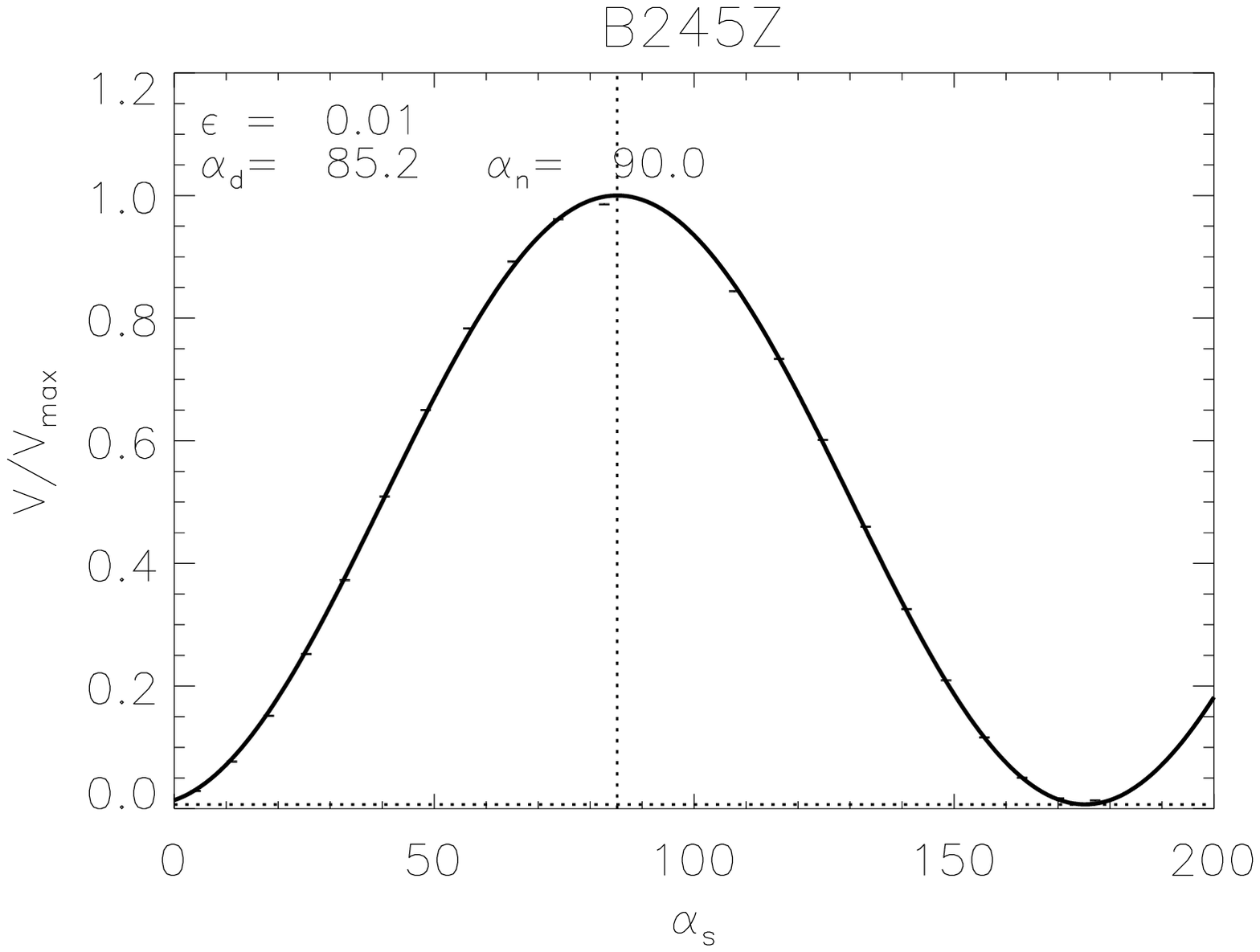}
\includegraphics[angle=0,width=6cm]{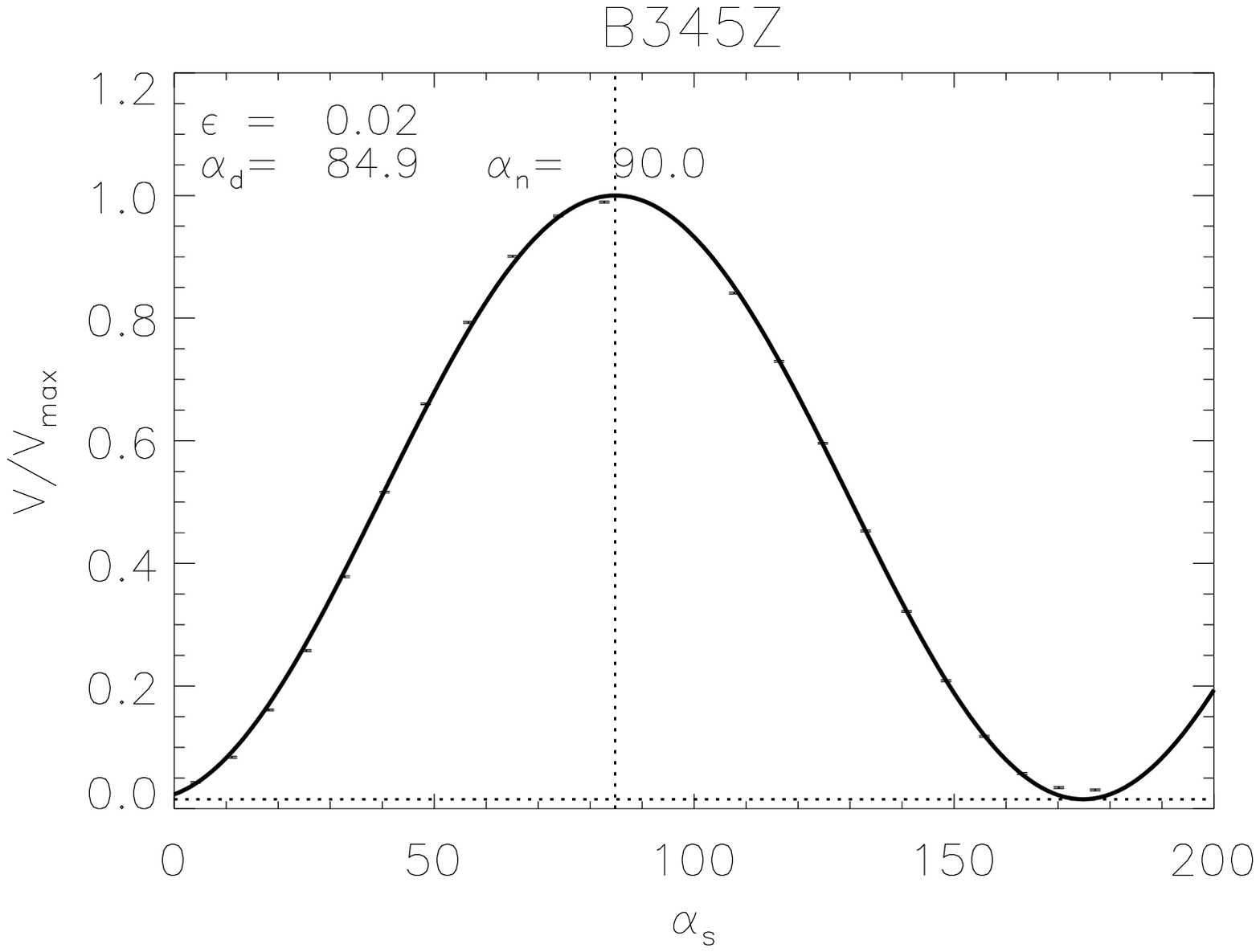}
\caption{Beam-integrated polarization response of the entire
instrument (two-colors photometers.)} \label{fig:crossfot}
\end{center}
\end{figure}

The measurement was carried out in two different configurations.
In the first we used a wide aperture on the cold source, so the
telescope beam was completely filled by a $\simgt 1^o$ source. The
source-telescope relative position was then optimized for maximum
signal. In this configuration we measured the integrated
polarization properties of the instrument. The measured data for
this configuration are shown in Figure~\ref{fig:crosspsb} for the
PSB channels, and in Figure~\ref{fig:crossfot} for the two-color
photometer channels. The best fit values of the parameters
$\epsilon$ and $\alpha_{det}$ are reported in
table~\ref{tab:crosspol}. The measured cross-polar response for
the PSB channels is of the order of 0.1. It is lower for the two
color photometer channels, because of the high efficiency of the
wire grids.

In the second configuration we used a small aperture ($\sim
3^\prime$ FWHM), in order to investigate the cross-polar response
properties as a function of the off-axis angle. In general the
on-axis cross-polar response is lower than the integral values of
table~\ref{tab:crosspol}, while off-axis is higher. This is in
agreement with the physical optics model of our system (shown in
Fig. \ref{fig:xbeam}) and with the receiver polarization
calibration described above.

\begin{table}[p]
\begin{center}
\begin{tabular}{lrrcc}
\hline \hline
Channel & \multicolumn{2}{c}{Polarization angle (deg)} & \multicolumn{2}{c}{Cross-polarization $\epsilon$}\\
     & nominal & measured & on-axis & beam-filling\\
\hline
145W1 & 135.0  & $137.1 \pm 2.0$ & $0.080 \pm 0.006$ & $0.11 \pm 0.02$ \\
145W2 & 45.0   & $44.4  \pm 2.0$ & $0.062 \pm 0.005$ & $0.11 \pm 0.02$ \\
145X1 & 180.0  & $178.4 \pm 2.0$ & $0.055 \pm 0.013$ & $0.09 \pm 0.02$ \\
145X2 & 90.0   & $88.7  \pm 2.0$ & $0.070 \pm 0.008$ & $0.13 \pm 0.02$ \\
145Y1 & 157.5  & $158.6 \pm 2.0$ & $0.051 \pm 0.012$ & $0.11 \pm 0.02$ \\
145Y2 & 67.5   & $66.2  \pm 2.0$ & $0.060 \pm 0.006$ & $0.11 \pm 0.02$ \\
145Z1 & 112.5  & $109.8 \pm 2.0$ & $0.182 \pm 0.004$ & $0.22 \pm 0.02$ \\
145Z2 & 22.5   & $21.2  \pm 2.0$ & $0.088 \pm 0.008$ & $0.12 \pm 0.02$ \\
\hline
245W  & 135.0  & $139.2 \pm 2.0$ & $0.007 \pm 0.005$ & $0.02 \pm 0.02$ \\
245X  & 45.0   & $42.9  \pm 2.0$ & $0.007 \pm 0.005$ & $0.02 \pm 0.02$ \\
245Y  & 180.0  & $178.7 \pm 2.0$ & $0.000 \pm 0.015$ & $0.03 \pm 0.02$ \\
245Z  & 90.0  &  $85.2  \pm 2.0$ & $0.014 \pm 0.013$ & $0.01 \pm 0.02$ \\
\hline
345W  & 135.0  & $139.9 \pm 2.0$ & $0.008 \pm 0.005$ & $0.01 \pm 0.02$ \\
345X  & 45.0   & $42.6  \pm 2.0$ & $0.008 \pm 0.001$ & $0.01 \pm 0.02$ \\
345Y  & 180.0  & $178.7 \pm 2.0$ & $0.004 \pm 0.014$ & $0.02 \pm 0.02$ \\
345Z  & 90.0   & $84.9  \pm 2.0$ & $0.019 \pm 0.013$ & $0.02 \pm 0.02$ \\
\hline
\end{tabular}
\end{center}
\caption{Cross-polarization and principal axis angle measurements
for all the channels. The polarization angle and the beam filling
values refer to measurements on the complete instrument obtained
with the apparatus described in Fig. \ref{fig:paolorizer}, while
the on-axis measurements refer to measurements on the receiver
alone, obtained with the apparatus described in Fig.
\ref{fig:rotator} and also in a test cryostat testing single
channels. } \label{tab:crosspol}
\end{table}

\subsection{ Beam Measurement}

We measured the beam profile $B(\theta, \phi)$ of our telescope
with a thermal source tethered on a balloon in the far field of
the telescope. The measurement was made in the relatively
transparent atmosphere of Antarctica. A small tethered balloon
lifted a microwave absorber (made out of eccosorb) into the
telescope beam, at a distance of about 1.5 km from the payload. We
used two sources to measure the beam, a sphere (with 45 cm
diameter) to measure the main beam, and a larger cylinder (with 76
cm diameter and 91 cm height) to map the sidelobes. The telescope
was steered by a combination of a slow azimuth scan with a very
slow elevation drift to map the beam profile of all the \bk
detectors. The position of the source relative to the
telescope was continuously measured with the tracking star camera.

\begin{figure}[p]
\begin{center}
\includegraphics[angle=90,width=14cm]{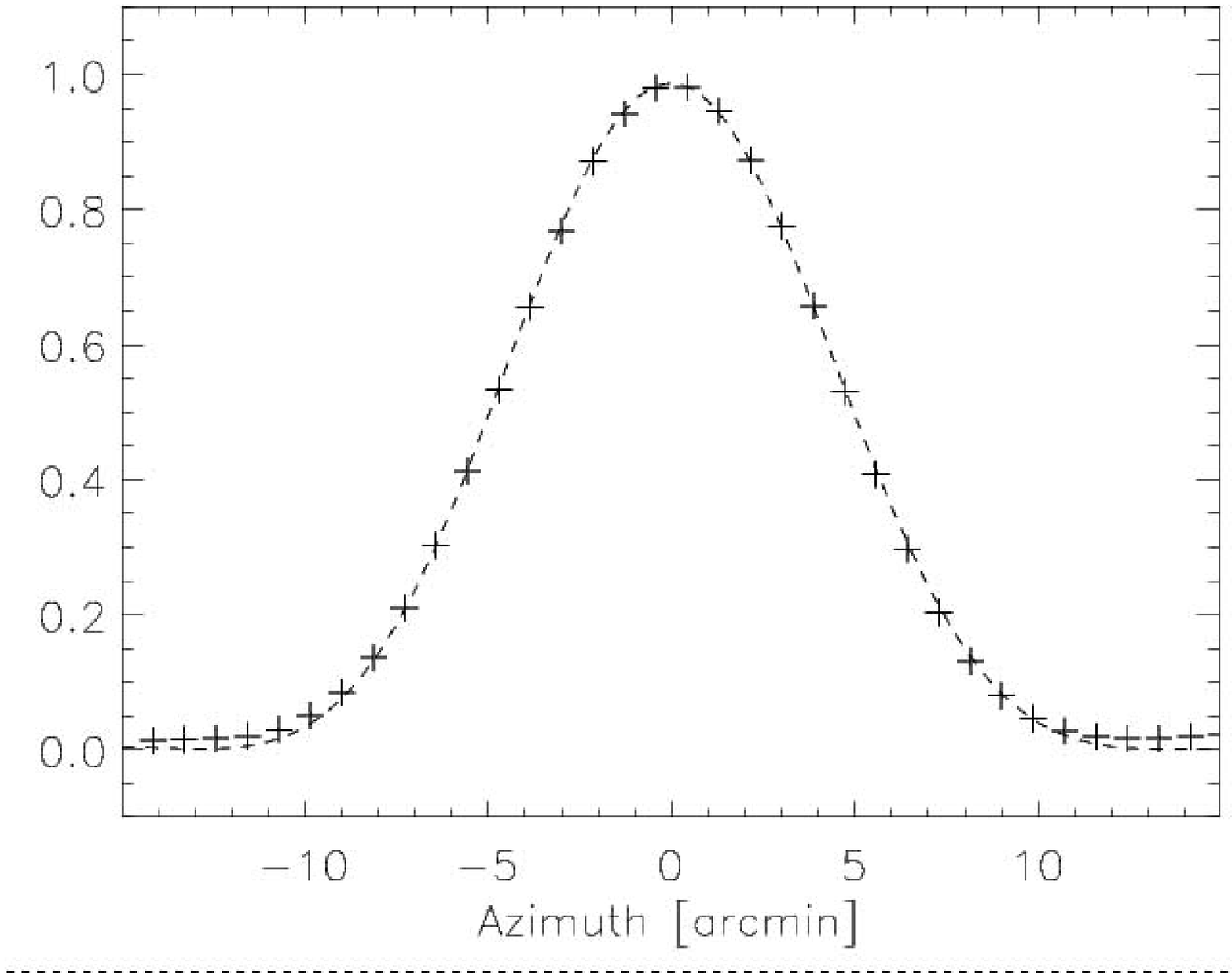}
\caption{Measurement (crosses) of the azimuth beam profile of
channel 145W1, obtained from repeated scans of a thermal source in
the far field of the telescope. The measurements fit very well the
prediction of the physical optics code BMAX (dashed line). }
\label{fig:bball}
\end{center}
\end{figure}

As an example, we plot in Fig. \ref{fig:bball} one of the 145 GHz
beam profiles $B(\theta)$, and compare it to the beam computed
from the BMAX physical optics model of our system. The agreement
is very good. This measurement, together with scans over point
sources during the flight, gives us confidence that the beam
computed with BMAX (plotted in Fig. \ref{fig:beams}) is accurate
for all our purposes.

\begin{figure}[p]
\begin{center}
\includegraphics[angle=90,width=14cm]{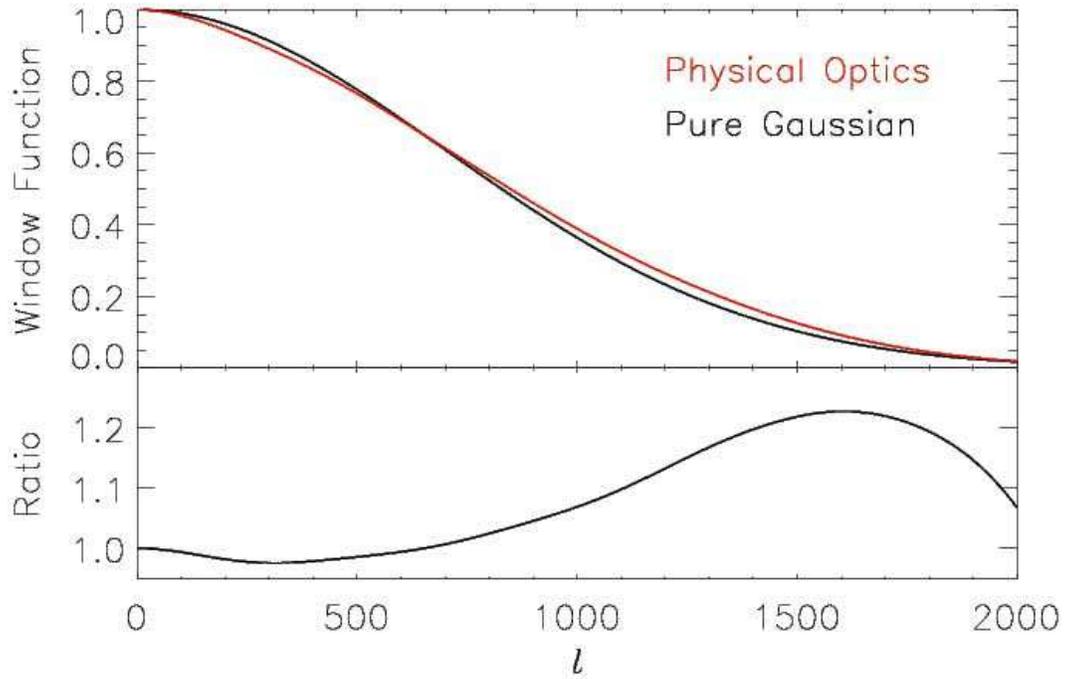}
\caption{Upper panel: Instrument response at 145 GHz in multipole
space, computed from the BMAX physical optics model (red line).
The beam is similar to a 9.8 arcmin FWHM gaussian (black line).
The ratio between the window function for the actual beam and that
for a gaussian beam is plotted in the lower panel.}
\label{fig:beam_ell}
\end{center}
\end{figure}

From the BMAX beam we compute the transfer function $B_\ell^2$ of
the instrument in multipole space. This is plotted in Fig.
\ref{fig:beam_ell}, where it is also compared to a gaussian beam
transfer function. This $B_\ell^2$ has been convolved with the
smearing resulting from pointing jitter (see \S
\ref{subs:attrec}), and used to deconvolve all our power spectrum
measurements, as reported in
~\cite[]{jones05,piacentini05,montroy05}.

\section{Observations}

The instrument was launched by NASA-NSBF on Jan. 6, 2003, from
Williams Field, near the McMurdo Station, in Antarctica (167$^o$
5.760' E ; 77$^o$ 51.760' S). The flight lasted until Jan. 21,
with a total of 311 hours.

\begin{figure}[p]
\begin{center}
\includegraphics[angle=90,width=10cm]{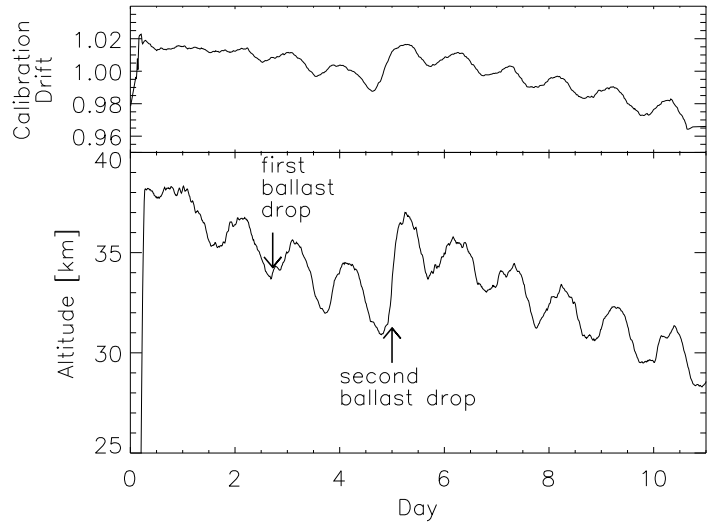}
\caption{{\bf Bottom}: The altitude of the payload during the
balloon-flight. The sinusoidal variation is due to the daily
change of elevation of the sun, while the long term trend is due
to a small leak in the balloon. The arrows indicate ballast
releases used to counter the downward drift. {\bf Top}: Drift of
detector responsivity during the flight as measured with the
on-board calibration lamp: it is contained within a few \% for the
whole length of the flight. The correlation with the altitude is
due to to the variation of atmospheric pressure with altitude,
which changes the temperature of the superfluid $^4He$ bath in the
cryostat, with a consequent change in bolometer temperature. }
\label{fig:altitude}
\end{center}
\end{figure}

The altitude of the payload during the flight is reported in Fig.
\ref{fig:altitude}. The periodic variation is due to the daily
change of elevation of the sun, while the long term trend is due
to a small leak in the balloon. We released ballast, in a moderate
amount on day 3, and then in a full drop on day 5. After day 11,
the altitude dropped below 28 km, telescope pointing become
difficult, and we had to stop observations.

We devoted a total of 119 hours to scans on the deep survey
region, a total of 79 hours to scans on the shallow survey region,
and a total of 30 hours to scans over the Galactic plane. The
remaining hours have not been used for the data analysis, due to
spurious signals after events like ballast drop or elevation
changes, and due to testing, cryogenic operations, non optimal
performance of the attitude control system at the lowest
altitudes.

\begin{figure}[p]
\begin{center}
\includegraphics[angle=90,width=14cm]{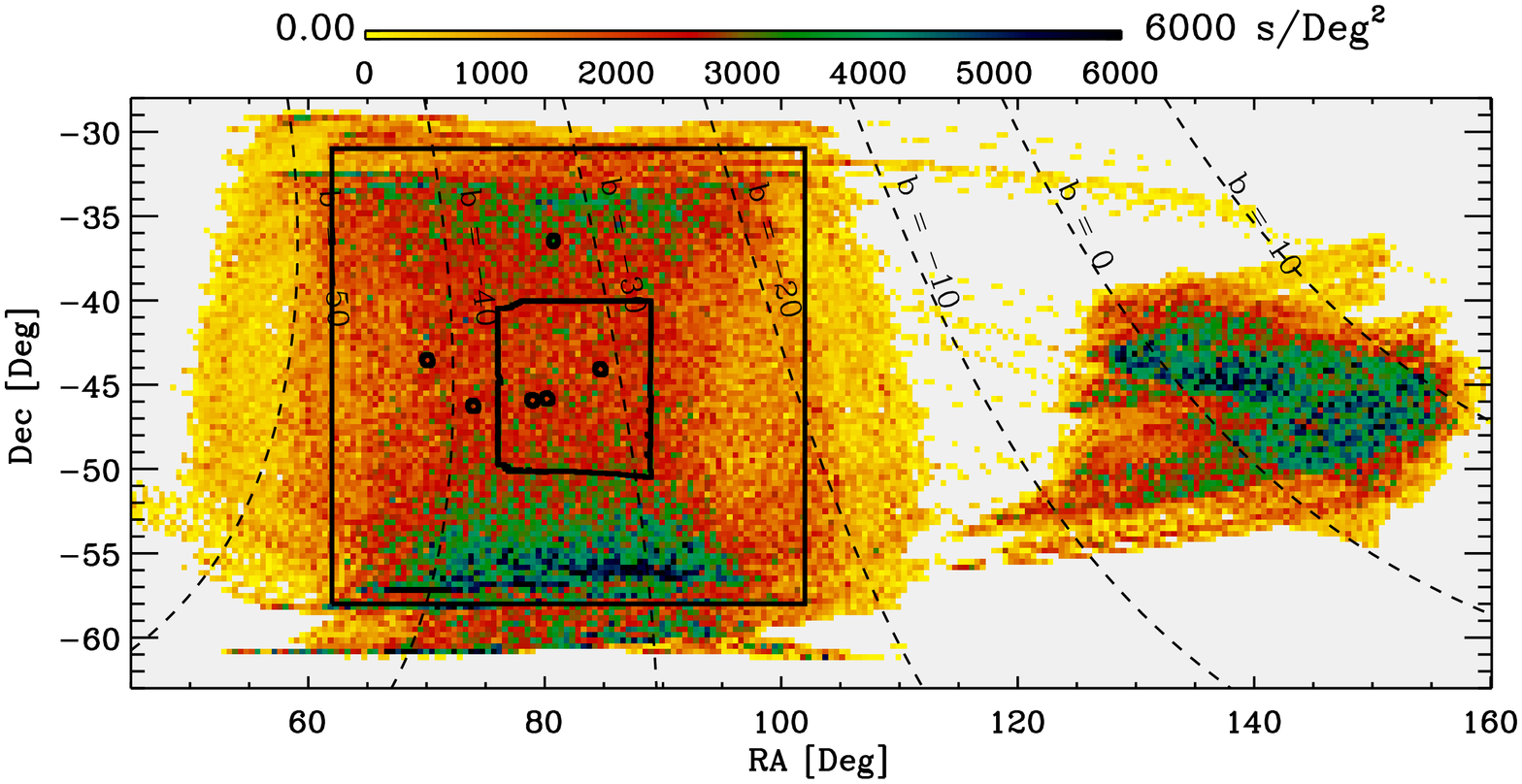}
\caption{Map of the observation time (integral over all the 145
GHz bolometers) in the "Shallow" and "Galaxy" surveys of \bk . The
larger box includes pixels of the shallow survey actually used for
the power spectrum analysis; the smaller box refers to the "Deep"
survey (see Fig. \ref{fig:deep_cov} ). } \label{fig:shal_gal_cov}
\end{center}
\end{figure}

\begin{figure}[p]
\begin{center}
\includegraphics[angle=90,width=14cm]{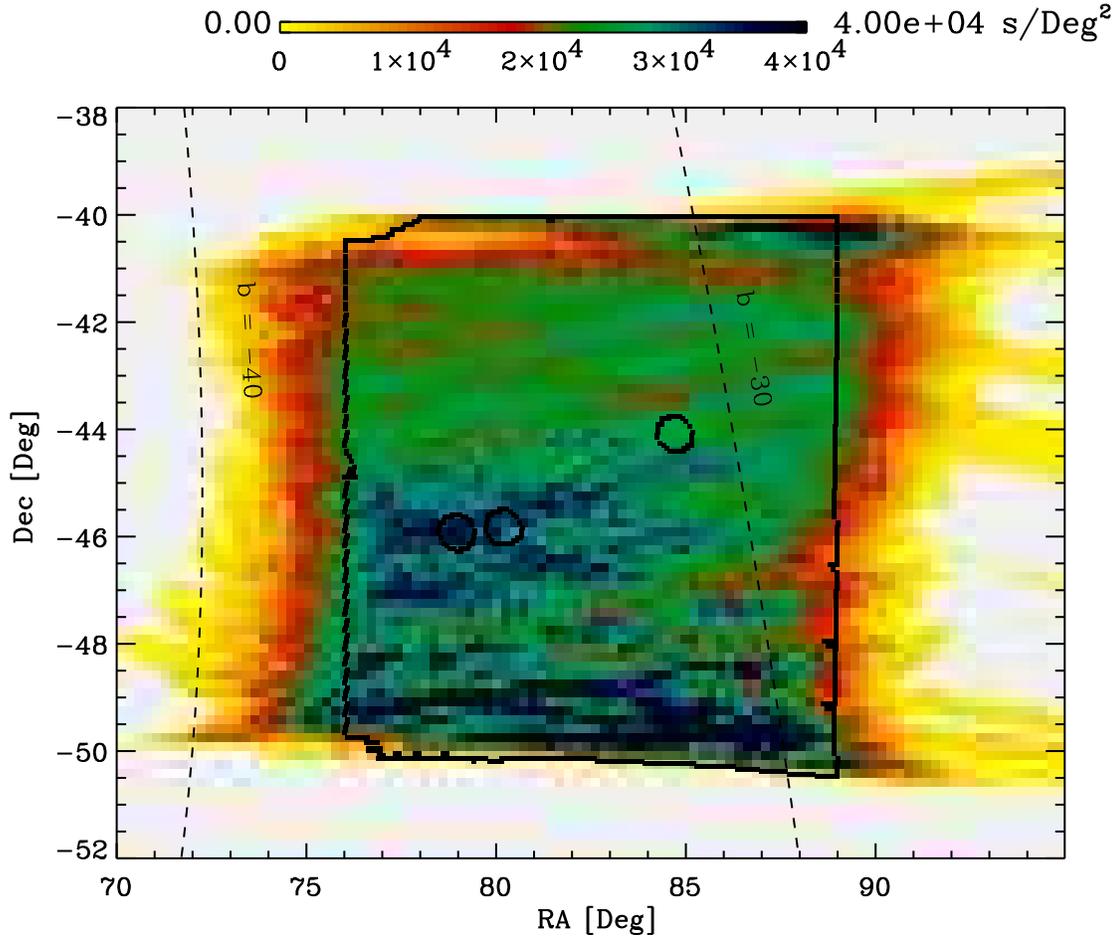}
\caption{Map of the observation time (integral over all the 145
GHz bolometers) in the "Deep" survey of \bk . This region overlaps
completely with the central part of the shallow survey region. The
box includes pixels of the deep survey actually used for the power
spectrum analysis; however, the three circular regions marked
around strong AGNs have been excised.} \label{fig:deep_cov}
\end{center}
\end{figure}

\begin{figure}[p]
\begin{center}
\includegraphics[angle=0,width=14cm]{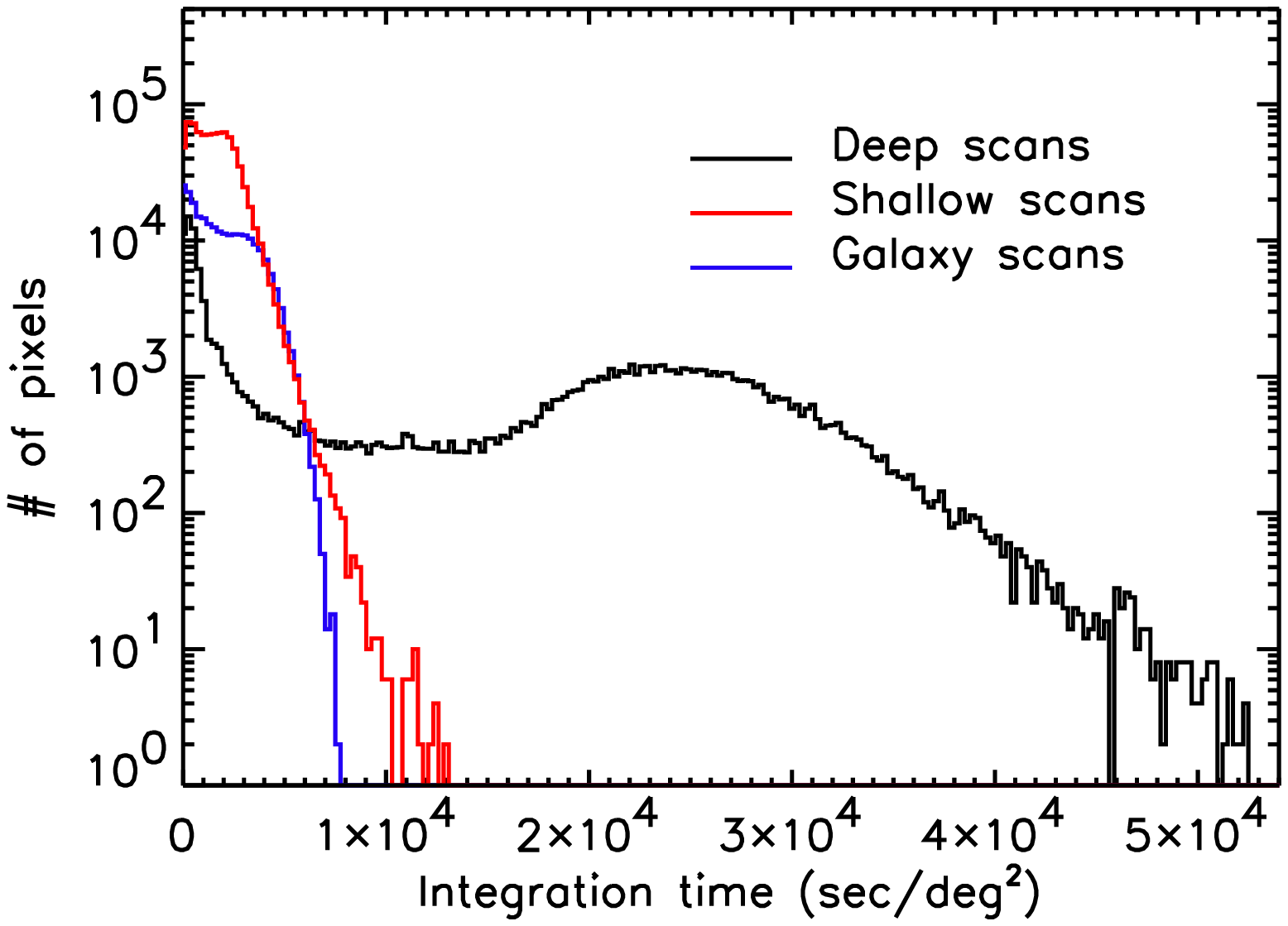}
\caption{Histogram of observation time per pixel (3.5$^\prime$
pixels) in the three surveys performed during the \bk flight. }
\label{fig:histo_coverage}
\end{center}
\end{figure}

Maps of the sky coverage for the three surveys are reported in
Fig. \ref{fig:shal_gal_cov} and in Fig. \ref{fig:deep_cov}. The
histogram of the integration time per pixel for the 3 different
regions is plotted in Fig. \ref{fig:histo_coverage}.

\subsection{Attitude Control System performance in flight}

Motion control was flawless in both the azimuth scanning and the
drive of the inner frame elevation when the payload was at nominal
altitude ($\simgt 25 km$). The GPS and FSS provided the required
coarse, $\sim$10 arcminute in-flight azimuth pointing and the
gyroscopes functioned almost continuously, aside from short
periods of dropouts or computer reboots.  Near the end of the
flight, once the payload altitude dropped below $\sim$ 23 km, high
winds made motion control impossible and the ACS was shut down.

Performance of the tracking sensors, the SC and the PSS, was
marred by communication problems in the ACS flight computer. With
both sensors operating at the same time the computer was unable to
parse the large amounts of incoming and outgoing data. To resolve
this issue the PSS was turned off early in the flight. The SC
provided the fine pointing data for the first and last thirds of
the flight. The cold temperature at low (pre-ballast drop)
altitude on day 4 caused the SC to freeze up, at which point the
PSS was turned back on, providing fine pointing for the middle
third of the flight.

\section{Raw data processing \label{sec:raw}}

The polarization signal we are looking for is very small.  In the
observations of the deep region at 145 GHz the contribution from
CMB polarization is of the order of a thousandth of the noise for
a single signal sample, and of the order of a tenth of the error
for the final measurement of a single 3.5$^\prime$ pixel. For this
reason we must properly exclude instrumental artifacts and avoid
the introduction of bias. We developed two completely independent
analysis pipelines, which will be called {\bf IT} (Italy) and {\bf
NA} (North America) in the following discussion. In this section
and in the following we outline the data analysis and specify
where the two pipelines make different choices. Despite the many
differences, the final spectra and maps are fully consistent. This
is the best test of the robustness of the dataset against
alternative data selection and analysis procedures.

The raw data acquired by the \bk telescope is convolved with a
transfer function consisting of the thermal response of the
detector and the filtering of the readout electronics. It also
contains transients caused by cosmic rays and our calibration
lamp, and longer time-scale events caused by thermal instabilities
resulting after each elevation change. At each step of the data
cleaning, we flag (mark as bad) some samples in the datastream and
fill these flagged samples with a constrained realization of the
noise. This noise filling uses linear prediction to replace
flagged samples based on the unflagged data on either side. After
removing very large cosmic rays, we flag a minimal number of
samples around each elevation change and fit out a decaying
exponential to the signal recovery.

Next, the datastreams are deconvolved with the combined thermal
and electrical transfer functions. For the {\bf NA} pipeline, we
used transfer functions determined by pre-flight lab measurements
(described in \S \ref{subs:labtransfer}). For the {\bf IT}
pipeline we used the in-flight determination of the transfer
function described in \S \ref{subs:flighttransfer}.

The samples containing signals from our calibration lamp are
flagged.  Medium amplitude cosmic rays are then flagged by passing
a simple template through the timestream and looking for spikes
larger than a certain size.  The data is then heavily band-pass
filtered to make small cosmic rays more visible.  A second
template is used to flag these small cosmic rays.  This process
might mistake bright galactic sources for small cosmic rays.  To
avoid this, we make a list of all bright source crossings in our
datastreams by using the coincidence of sources in both detectors
of a pixel pair. These samples are explicitly {\it unflagged}.
After building up a list of flagged samples, these samples are
filled with noise in the deconvolved data.

After completing this primary data cleaning, we found that each of
our time-streams contained an extraneous component that was well
correlated with the accelerations of the telescope's pitch and
roll gyroscopes.  We fit these accelerations to the data in hour
long chunks and subtracted this contribution from the
time-streams.

\subsection{Post-flight Attitude Determination \label{subs:attrec}}

Pointing reconstruction consists of determining where each
detector was pointing in the sky at each data sample. We had a
redundant number of attitude sensors that can be combined
in different ways to obtain the final pointing solution. The
two pipelines are significantly different in this respect.

\subsection{IT Pointing reconstruction}

Pointing is described by the rotation matrix $S$ that transforms
from the Celestial reference frame to the telescope reference
frame. In addition, the constant offsets of the pointing
directions of the horns in the telescope frame must be determined.
The problem is separated into two parts $S=A \cdot B$, where the
rotation matrix $A$ converts from an Earth local reference frame
to the telescope frame, and $B$ is the astronomical conversion.
$A$ is called the attitude matrix and is defined by three Euler
angles. It can also conveniently be described by a quaternion.

To derive the attitude matrix we use several attitude sensors,
measuring either absolute angles or angular velocities. In our
case the absolute sensors are a differential GPS, a Stellar Sensor
and two Solar Sensors. The differential GPS can by itself measure
the attitude matrix $A$, but suffers for low frequency drifts, of
the order of 7$^\prime / \sqrt{h}$. The solar and the stellar
sensors measure the sensor-Sun or sensor-Star angles in the
payload frame, each providing two equations for the three unknown
Euler angles. Their combination thus provides a solvable system.
The angular velocities are obtained from the laser gyroscopes,
which have a bias due to their orientation in the Earth's magnetic
field.

We combine the signals of the different sensors taking into
account their noise/drift properties, by means of an optimized
Kalman filter \cite[]{kalman60} and by properly flagging the
sensors signals. The Kalman filter is based on the propagation
from the status at time $t$ to the status at time $t + \Delta t$
as predicted by the dynamic measurements (from gyroscopes), in
combination with the direct measurements of the status $t + \Delta
t$ (update). The weighting of the update is given by the noise
properties of the gyroscopes and of the sensors.

The direction of lines of sight of the different detectors in the
telescope frame are derived from observation on known sources,
i.e. AGNs and Galactic HII regions.

A final correction to the IT pointing solution is obtained by
dividing the timestream in hour-long chunks. For each chunk we
make a sky map, and find the Azimuth and elevation offsets which
best fit a single template map derived from the \boom data from
the 1998 flight. An IRAS template was used for the scans on the
Galactic plane. The pointing solution is then corrected using
these best fit offsets, which are of the order of $\simlt$ 1
arcmin.

The accuracy of the IT pointing solution is tested by means of
the signals detected in the direction of bright AGNs (see \S
\ref{subs:effbeam}), and is of the order of $(2.4 \pm 0.3)
^\prime$ rms.

\subsection{The NA Pointing Solution Pipeline}

Spikes in the raw pointing data are removed, and where possible
(over a few samples) the data can be linearly interpolated.
Post-flight re-calibrations are applied to the SC, PSS and FSS
data.  For the SC/PSS azimuth and elevation this is a simple
matter of rotating the coordinates through a small angle until
correlated signal is minimized.  The FSS is re-calibrated using
azimuth data from the GPS and the SC.  A look-up table of sun
elevation versus raw FSS azimuth is constructed. Each element of
the table contains a GPS/SC derived sun azimuth relative to the
gondola, averaged over the whole flight.  Raw FSS data is replaced
by the corresponding element in the table.  Another useful derived
quantity is obtained from the GPS up, north and east velocity
data.  The relatively stable LDB environment allows one to model
the gondola as a pendulum.  With this model first estimates of
gondola pitch and roll, $pitch_{GV}$ and $roll_{GV}$ may be
calculated from the GPS velocity data.

Clean, calibrated pointing fields are combined to determine the
gondola attitude.  The sensors used in the final analysis include
the FSS azimuth, the SC elevation, $pitch_{GV}$, $roll_{GV}$ and
the integrated azimuth, pitch and roll gyro data.  At frequencies
below $\sim$50 mHz the pointing solution is based on the best fit
azimuth, pitch and roll to the sun and star positions as
determined by the star camera, the FSS, $pitch_{GV}$ and
$roll_{GV}$.  Gaps in the pointing data less than $\sim$ 40
seconds long are filled with integrated gyro data and above $\sim$
50 mHz the pointing solution is strictly gyro signal.  Gaps longer
than $\sim$ 40 seconds are flagged.

The elevation encoder signal is added to the gondola pitch,
thereby translating gondola (outer frame) attitude into telescope
(inner frame) attitude.  The beam offsets for each detector are
obtained from fits to the five brightest QSOs in the CMB field.
Galactic and CMB source centroid offsets reveal a 0.1$^\circ$
shift in gondola pitch after the mid-flight ballast drop.  To
account for this approximately 6 hours of data during and after
the ballast drop are flagged and a pitch shift is applied to all
pointing data preceding the drop.  The reconstructed elevation and
azimuth of each beam on the sky, along with the measured
polarization angles for each detector, and  the GPS latitude,
longitude and time are combined to determine the right ascension,
declination and psi angle for each beam.

The CMB field pointing error, based on comparisons of analytical
beams with observed beams, is $\sim$2.5' rms in azimuth and
$\sim$1.5' rms in elevation (see \S \ref{subs:effbeam}).

\section{In-flight Calibration}

\subsection{Transfer function \label{subs:flighttransfer}}

The transfer function of each detector, including the readout, has
been measured in flight using a procedure similar to the one
described in \cite{crill03}. Even though the detectors are
designed to minimize cosmic rays cross section, each detector
produces detectable cosmic ray hits at the rate of about one every
two minutes, leaving on the data-stream a typical signature which
is the response of the system to an impulsive input. This
signature is the transfer function of the system in real space.

\begin{figure}[p]
\begin{center}
\includegraphics[angle=0,width=10cm]{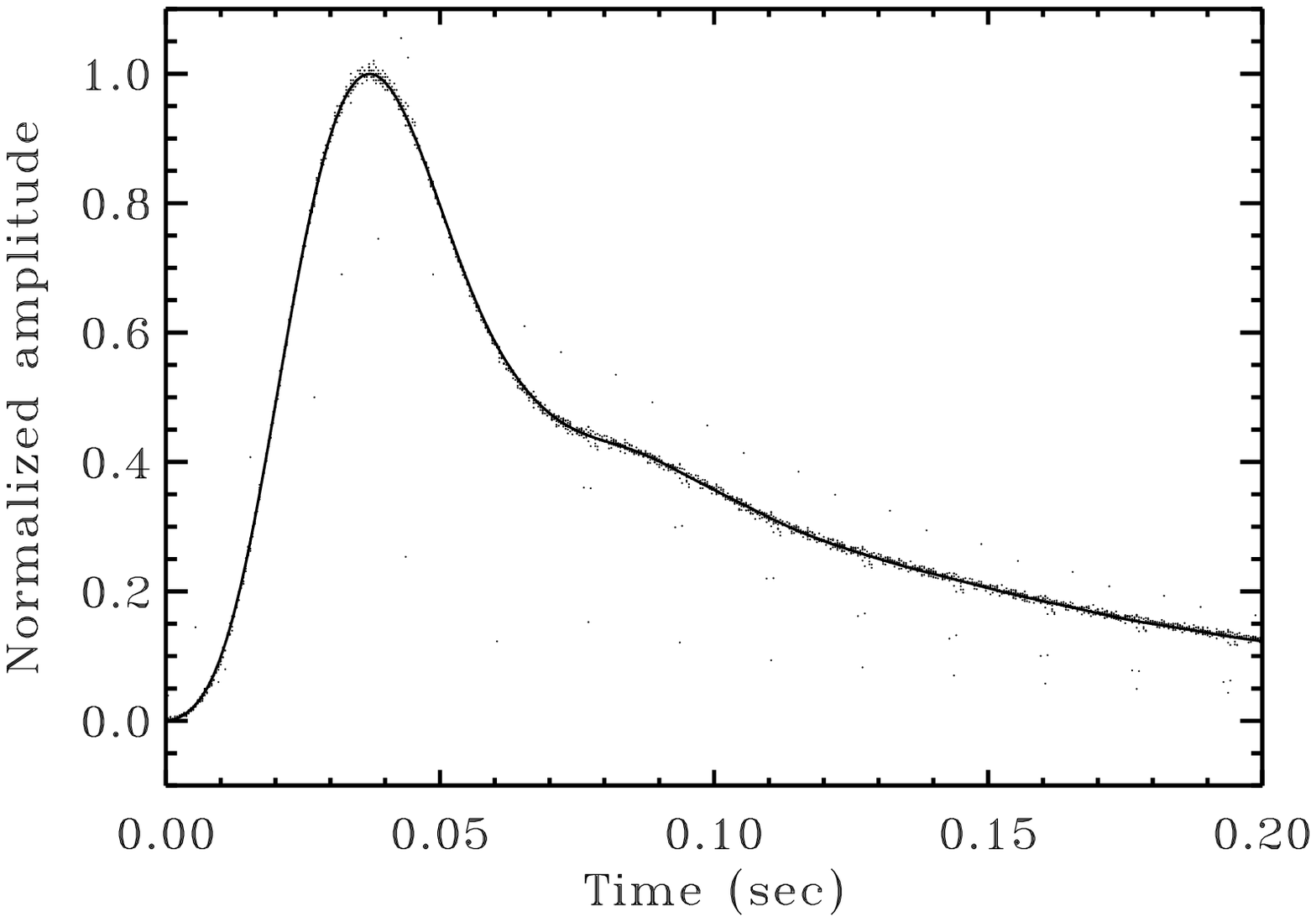}
\caption{In-flight response of the 145W1 channel to an impulsive
event. The frequency response of the system is the Fourier
Transform of this response. The points are accumulated from
several cosmic-rays events shifted and normalized to fit the same
template.} \label{fig:cr_impulse}
\end{center}
\end{figure}

Of all the cosmic rays events in a given channel, only the subset
producing a spike in the (0.5-4) Volts range is selected for this
analysis, in order to neglect the effects of noise and to avoid
saturated events. Each event in the database is shifted in time
and normalized to minimize the chi square to a first order
approximation of the impulse response. The shift is performed on a
grid much finer than the 60~Hz sampling rate of the Data
Acquisition System. The combination of all the hits provides a
template of the impulsive response of each detector. As an
example, we plot in Figure~\ref{fig:cr_impulse} the data for
detector 145W1. The result is insensitive to the choice of the
first order approximation. The Fourier transforms of those
templates are the transfer functions of the detectors. The method
is sensitive in a frequency range between $\sim$ 0.1 and $\sim$
250~Hz. The resulting transfer function is similar (but not
exactly equal) to the one described in \S \ref{subs:labtransfer}.
The small differences are probably due to the internal time
constant of the detector absorber, which affects in different ways
the response to mm-wave photons (depositing their energy on the
entire absorber) and the response to particles (depositing their
energy on a small localized part of the absorber). The IT pipeline
uses this transfer function, complementing it with lab calibration
data (\S \ref{subs:labtransfer}) at $f < 0.1 Hz$. The NA pipeline
uses the transfer function derived from pre-flight measurements.

\subsection{Effective Beam Calibration \label{subs:effbeam}}

The effective \bk beam profiles are estimated combining physical
optics modelling and observations of bright extragalactic point
sources.

The physical beam is computed with a physical optics simulation of
the telescope (see \S \ref{subs:telescope} and Fig.
~\ref{fig:bball}); its shape is similar to a Gaussian function
with a FWHM of 9.8 arcmin for the 145~GHz channels and of
5.6$^\prime$ for the 245 and 345~GHz channels, with residuals of
the order of 2\% (see Fig. \ref{fig:beam_ell}).

Pointing errors produce a jitter on the map that we assumed to be
gaussian and uniform in all the observed sky regions. Under this
assumption the jitter is defined by a single parameter,
$\sigma_{jitter}$.

The observed profile of a point source is the convolution of the
physical beam with the gaussian describing the pointing jitter.
For a gaussian and isotropic physical beam we would have
\begin{equation}
B(\theta, \sigma_{jitter}) = \exp \left[ - {{ \theta^2}\over {2
(\sigma_{beam}^2+ \sigma_{jitter}^2) } } \right]
\label{eq:convolution}
\end{equation}
Using the actual physical beam gives a result very similar to
equation \ref{eq:convolution}. We measured $\sigma_{jitter}$ by
comparing the expected beam to the profiles of five sources
derived from the 8 PSBs maps at 145~GHz. We did not use the higher
frequency channels to estimate the jitter because of their higher
noise. The jitter, estimated as the weighted average of the 5 best
fit values, is $\sigma_{jitter} = (2.5 \pm 0.3) $~arcmin. The FWHM
of the effective beam used in the final analysis is $(11.2 \pm
0.3)$ arcmin for the 145 GHz channels, $(7.8 \pm 0.3)$ arcmin for
the 245 and 345 GHz channels.

\subsection{Gain drift from Calibration Lamp \label{subs:callamp}}

The responsivity of the bolometers depends on the operating
temperature and on the radiative loading. Both these quantities
can change during the flight.

For this reason, we include a calibration lamp in the cold
optics of \boom , that flashes every $\sim 15$ min. during the flight.
Details on the lamp, which is a composite bolometer structure
suspended in the center of a 1 cm hole in the tertiary mirror, are
in \cite{crill03}. Its intrinsic stability depends on the fact
that the temperature achieved during the flashes is significantly
higher than the 2K base temperature. In this limit, the emission
of the lamp depends only on the heater current, which is stable to
better than $0.1 \%$.

The amplitude of the calibration signal changes only by a few
percent during the flight, as shown in the top panel of Fig.
\ref{fig:altitude}. To demonstrate that these fluctuations are not
intrinsic to the source, we have analyzed the DC signal across the
bolometers. This is proportional to the resistance of the
bolometers, which in turn is proportional to the responsivity. For
all the detectors we found a very good correlation between the
bolometer responsivity estimated from the calibration lamp and the
DC level: a clear indication that the signal of the calibration
lamp is constant through the flight.

\subsection{Bolometer Performance \label{subs:flightnoise}}

We compute in-flight bolometer parameters with a bolometer model
based on \cite{holmes98}, \cite{jones53}, and \cite{mather82} (see
\cite{jones05a} for details).   The receiver model computes the total optical load on the
detectors using the resistance and thermal
conductivity as functions of temperature measured on the ground,
and the in-flight DC bolometer voltage, bias current, and noise.

We further model the optical load on the detectors
due to the atmosphere, telescope, and Cosmic Microwave Background,
and assume that leftover optical load is due to internal parasitic
loading.  The optical load and responsivity change by several percent during
the flight (see \S \ref{subs:callamp}) due to changes in emission
from the atmosphere and thermal radiation from cryostat
components.

We model the contributions to the noise from photon shot noise,
Johnson noise, phonon shot noise in the thermal link between the
bolometer and the cold stage, and amplifier noise and compute the
NEP (Noise Equivalent Power) contributed by each. The results of
these calculations, for typical values of the in-flight optical
load, are reported in table \ref{tab:boloinflight}.

\begin{table}[p]
\begin{center}
\begin{tabular}{lcrrr}
\hline
& units &  145 & 245 & 345 \\
\hline
Optical power\dotfill &pW &     0.496  &   0.846  &   6.158 \\
Equivalent R-J temperature\dotfill&K&4&4&16\\
\hspace{5mm}CMB/total\dotfill &\%&    23.0  &    6.2  &   0.9 \\
\hspace{5mm}Atmos./total\dotfill &\%&     4.1  &  32.6 &   64.7 \\
\hspace{5mm}Refl./total\dotfill &\%&    21.4  &  17.2  &   10.0 \\
\hspace{5mm}Internal/total\dotfill &\%&   51.5  &  44.0  &  24.4\\
\hline
NEP$_{\mathrm{total}}$\dotfill &$10^{-17}$ W/$\sqrt{\mathrm{Hz}}$ & 2.49 & 3.89 & 8.95 \\
\hspace{5mm}NEP$_{\mathrm{photon}}$\dotfill &$10^{-17}$ W/$\sqrt{\mathrm{Hz}}$  & 0.99 & 1.64  &   5.31 \\
\hspace{10mm}Bose term\dotfill &\% &  14.8  & 7.2 &  16.4 \\
\hspace{5mm}NEP$_{\mathrm{phonon}}$\dotfill &$10^{-17}$ W/$\sqrt{\mathrm{Hz}}$ &   1.003  &   1.636  &   2.891 \\
\hspace{5mm}NEP$_{\mathrm{johnson}}$\dotfill &$10^{-17}$ W/$\sqrt{\mathrm{Hz}}$&    1.307 &    1.839 &    3.669 \\
\hspace{5mm}NEP$_{\mathrm{amplifier}}$\dotfill &$10^{-17}$ W/$\sqrt{\mathrm{Hz}}$&  1.578 &   2.518 &   5.494 \\
NEP$_{\mathrm{bolo}}$ / NEP$_{\mathrm{background}}$ \dotfill &   &  1.67  &   1.50  &   0.88 \\
\hline
NEFD\dotfill &mJy/$\sqrt{\mathrm{Hz}}$ &     37   &  43 &   177 \\
NET$_{\mathrm{cmb}}$\dotfill & $\mu\mathrm{K}_{\mathrm{CMB}}\sqrt{\mathrm{s}}$  &    151 &   341 &   551 \\
NET$_{\mathrm{RJ}}$\dotfill  &  $\mu\mathrm{K}_{\mathrm{RJ}}\sqrt{\mathrm{s}}$ &     89  &   92  &   51 \\
NET$^{\mathrm{inst}}_{\mathrm{cmb}}$\dotfill  & $\mu\mathrm{K}_{\mathrm{CMB}}\sqrt{\mathrm{s}}$ & 53  &  171  &  275 \\
\hline
Voltage noise\dotfill &nV/$\sqrt{\mathrm{Hz}}$ & 18.0  & 17.7  &   18.5 \\
\hline
\end{tabular}
\end{center}
\caption[In-flight receiver performance]{\small The receiver
  performance for a typical channel from each of the {\sc Boomerang} bands,
  computed from a receiver model using the observed average in-flight loading conditions.
  The performance measured in flight is consistent with these estimates.}
\label{tab:boloinflight}
\end{table}

A source of impulsive noise is cosmic rays. All the \bk detectors
have been designed to minimize their cross section to cosmic rays.
As a result, even in the polar stratosphere, the average rate of
cosmic-rays hits in PSBs is one every two minutes. Over all the
identified events, $83\%$ produce transients with amplitude at the
ADC $0.05< A < 1 V$; 10$\%$ have amplitude $1V<A< 2V$, $5\%$ have
$2V<A< 4V$ and $3\%$ have $A > 4V$. All these events are flagged
and removed from the analysis as explained in \S \ref{sec:raw}.

\subsection{Noise Estimation \label{subs:noisest}}

\begin{figure}[p]
\begin{center}
\includegraphics[angle=0,width=12cm]{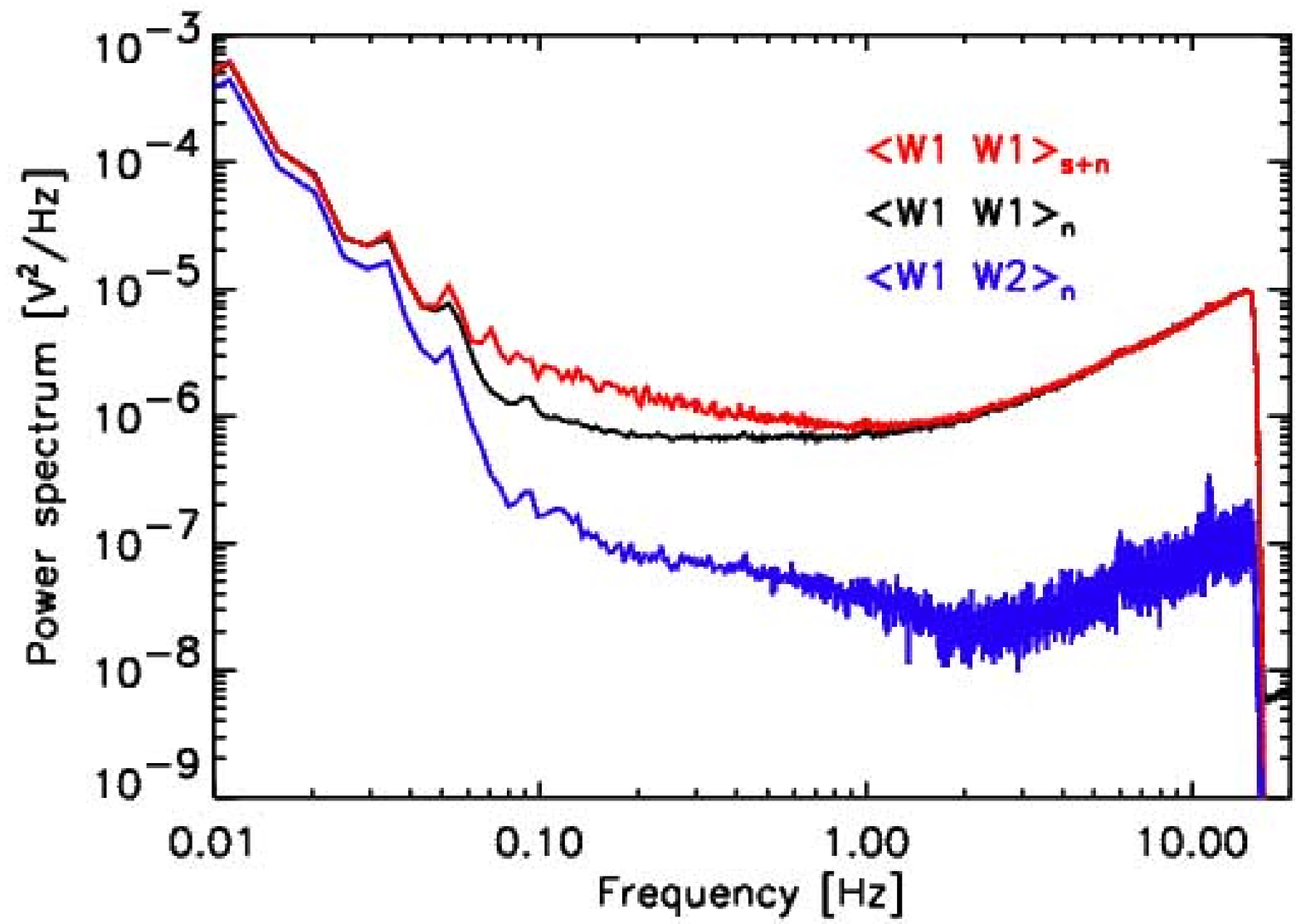}
\caption{Flight-average of the auto power spectrum of the data
time-stream, from channel 145W1 (top line, red). The data are
recorded at the ADC input. The total gain of the readout chain
between the bolometer and the ADC input is 50000. The middle line
(black) is the auto power spectrum of the time stream after
subtraction of the signal estimated from the best fit map: this is
an estimate of the instrumental noise. Comparing the two, it is
evident that most of the CMB signal is encoded in the 0.05-1~Hz
range. The lowest line (blue) is the cross correlation of the
noise (time-stream - best fit map) signal from 145W1 and 145W2. In
the frequency range of interest for the CMB signal, the
cross-correlated noise signal is at least one order of magnitude
smaller than the auto-correlated signal. The correlated noise
rises at very low frequencies, where atmospheric effects, thermal
drifts, pendulations and scan-related microphonics can be present.
At frequencies higher then 2~Hz the noise correlation remains well
below the noise of the detector. \label{fig:nep}}
\end{center}
\end{figure}

The measurement of the actual in-flight noise is obtained via a
Fourier transform of the raw datastream. A sample power spectrum
of the raw data (deconvolved from system frequency response, as
estimated in \S \ref{subs:labtransfer} and \S
\ref{subs:flighttransfer}) is shown in Fig. \ref{fig:nep}. Signal
contributions are present above the noise, at the scan frequency
and at its harmonics, and in the $\sim 0.1$ Hz range. The
Generalized Least Squares (GLS) map-making method relies on
knowledge of the noise correlation function, which is related to
the noise power spectral density by a Fourier transform. This
quantity must be estimated from the data themselves, which are a
combination of noise and signal.

The standard way to characterise the underlying noise is by
subtracting an estimate for the signal (see~\cite{ferreira00}).
Since the signal can be estimated by making a map, the two
problems are entangled, and a possible approach is iterative
(see~\cite{prunet01} and \cite{dore01}; see however
\cite{natoli02} for an alternative approach).

In the IT pipeline, a rough estimate of the signal is obtained by
na\"ively coadding a band-pass filtered version of the timeline;
the resulting map is used as a baseline to obtain an estimate of
the noise power spectral density. This is in turn used to make a
new, GLS, map and the process can be iterated as desired. We have
found empirically that, for \bk, no significant improvement is
obtained by iterating more than five times. Furthermore, tests on
simulated data suggest that, under these conditions, the
underlying noise properties are recovered without any substantial
bias (see~\cite{degasperis05}).

The following scheme is employed to optimally use the information
produced by all detectors. The above procedure is repeated for
each uncalibrated bolometer timeline (e.g. eight times, one for
each 145~GHz bolometer). Then, a first estimate of the relative
calibration factors between channels at a given frequency are
computed as explained in \S \ref{subs:pixpixcal}. A multi-channel,
high signal to noise ratio, relatively calibrated map is produced
and used as an estimate for the underlying signal to evaluate a
more precise noise power spectral density for each bolometer. As a
further step, relative calibration factors are recomputed with
these noise estimates, and a new multichannel map is made.  We
have verified that iterating further is useless since if this last
map is used to produce further noise estimates, the latter do not
change significantly.

A sample noise power spectrum is shown in Fig. \ref{fig:nep}. With
this method we estimate the noise of all the \bk channels in
$V/\sqrt{Hz}$. This is converted into NEP (see table
\ref{tab:nep}) using the measured in-flight responsivity (see \S
\ref{subs:pixpixcal} and \S \ref{subs:speccal}).

\begin{table}[p]
\begin{center}
\begin{tabular}{|c|c|}
\hline
Channel  & NET ($\mu K_{CMB} \sqrt{s}$)\\
\hline
145W1 & 140 \\
145W2 & 137 \\
145X1 & 156 \\
145X2 & 150 \\
145Y1 & 151 \\
145Y2 & 159 \\
145Z1 & 182 \\
145Z2 & 281 \\
\hline
245W & 281 \\
245X & 358 \\
245Y & 316 \\
245Z & 331 \\
\hline
345W & 459 \\
345X & 424 \\
345Y & 620 \\
345Z & 309 \\
\hline
\end{tabular}
\end{center}
\caption{ In flight detectors performance: \small We report the
noise equivalent temperature at 1 Hz, derived from the noise
equivalent voltage measured in flight as in \S
\ref{subs:flightnoise}, from the responsivity estimated as in \S
\ref{subs:pixpixcal} and from the spectral calibrations of \S
\ref{subs:fts}. In our survey, this frequency corresponds to
multipoles $400 \simlt \ell \simlt 1200$. The noise increases at
lower and at higher frequencies, as shown in Fig. \ref{fig:nep}.}
\label{tab:nep}
\end{table}

There is a component of noise correlated between different
detectors. This is especially visible at very low frequencies. In
Fig.\ref{fig:nep} we include an example. This noise has a similar
level for all detector couples (roughly independent of whether the
detectors are members of the same PSB pair). It may originate in
scan synchronous microphonics and in fluctuations of the residual
atmosphere. The levels of correlated noise in the interesting
frequency band (0.1-2 Hz), for all possible couples of 145 GHz
bolometers, are compared to the levels of noise of each bolometer
in table \ref{tab:noise_x}. The presence of this correlated noise
is neglected in the map-making, but is taken into account for our
estimates of the power spectra (see \cite{jones05},
\cite{piacentini05}, \cite{montroy05}).

\begin{deluxetable}{l|cccccccc}
\tablecolumns{4} \tablewidth{0pc}
\tablehead{\colhead{~}&\colhead{W1} &\colhead{W2} &\colhead{X1} &
  \colhead{X2}&\colhead{Y1}&\colhead{Y2} &\colhead{Z1} &\colhead{Z2}}
\startdata
W1 &  1.0  \\
W2 &  6.0E-02 &  0.85\\
X1 &  3.4E-02 &  3.1E-02 &  1.1\\
X2 &  3.3E-02 &  3.1E-02 &  5.9E-02 &  1.0\\
Y1 &  4.8E-02 &  4.0E-02 &  3.5E-02 &  3.5E-02 &  1.2\\
Y2 &  4.3E-02 &  3.6E-02 &  3.5E-02 &  3.5E-02 &  1.6E-01 &  1.2\\
Z1 &  3.7E-02 &  3.3E-02 &  3.8E-02 &  3.7E-02 &  4.1E-02 &
4.1E-02 &
1.5\\
Z2 &  6.0E-02 &  5.4E-02 &  6.3E-02 &  6.1E-02 &  6.6E-02 &
6.6E-02 &
9.6E-02 &  4.3    \\
\enddata
\tablecomments{Absolute value of the noise cross-power-spectrum
for the 145 GHz detectors, averaged in the 0.1-2~Hz range. The
numerical values have been normalized to the noise
auto-power-spectrum of detector 145W1, which is $1.2 \times
10^{-6} V^2/Hz$ at the ADC input. } \label{tab:noise_x}
\end{deluxetable}

\subsection{CMB dipole}

\begin{figure}[p]
\begin{center}
\includegraphics[angle=90,width=10cm]{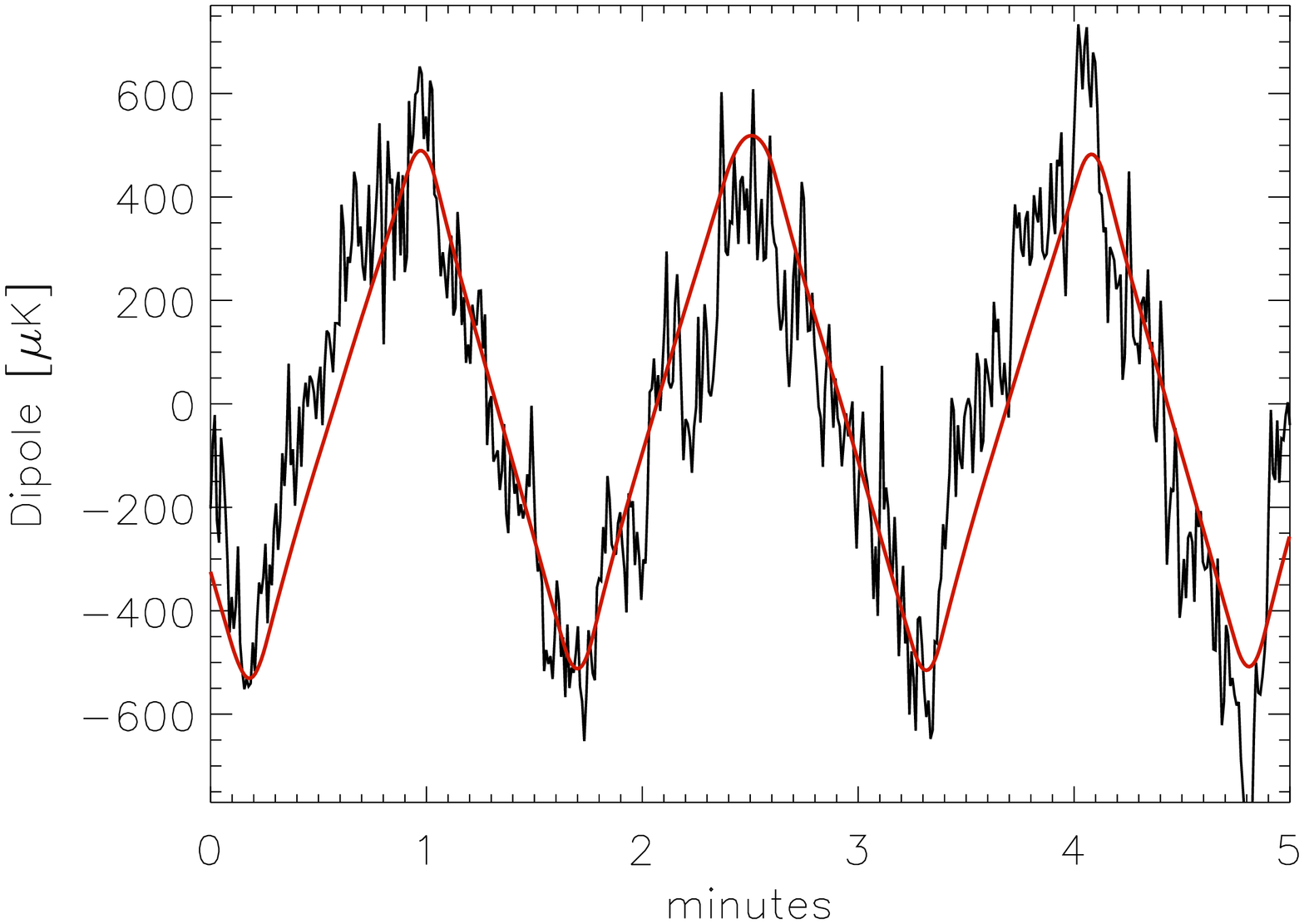}
\caption{Raw-data time-stream from channel 145W1. It is evident
the scan-synchronous signal from the CMB dipole. The (red) line is
the CMB dipole, measaured by WMAP and COBE, along this scan. There
is clear agreement between the predicted CMB dipole and the
observed scan synchronous signal. \label{fig:dipole}}
\end{center}
\end{figure}

The dipole of the CMB is visible as an approximately linear drift
along our short scans. It produces a scan-synchronous triangle
wave (see Fig. \ref{fig:dipole}) that we have removed from the
time-streams before proceeding with the analysis. Since it is
scan-synchronous, as some possible systematics, we cannot use it
for a precise calibration of the instrument: the calibration
accuracy achievable with the dipole is $\simgt 15 \%$, not
sufficient for our purposes. However, its amplitude is consistent
with the amplitude of the CMB Dipole measured by COBE and \wmap.

\section{Stokes Parameters Maps }

\subsection{The Map Making Pipeline \label{subs:mapmaking}}

A Generalized Least Square (GLS) method is used to jointly
estimate the Stokes parameter sky maps $I$, $Q$ and $U$ from time
ordered data. The best maps are obtained by combining data from
all detectors available at a given frequency. This approach is
implemented in the ROMA map making code (\cite{degasperis05}) for
the IT pipeline, and in the DIQU code \cite[]{jones05b} for the NA
pipeline. Following eq. \ref{eqn:vcross}, we assume the following
data model:
\begin{equation}
  \mathcal D_{t}^i = \frac{{\cal S}^{i}}{2}
  A_{tp}^i \left[I_p + \frac{1-\epsilon_i}{1+\epsilon_i}
\left( Q_p \cos 2\alpha_{t}^i + U_p \sin 2\alpha_{t}^i \right)
\right] + n_{t}^i. \label{eq:TOD_xpol}
\end{equation}
Here $i$, $t$ and $p$ label channel, time and map pixel
respectively; $D_{t}^i$ are the time ordered data for a given
channel, related to the sky maps $I_p$, $Q_p$ and $U_p$ by the
pointing operator $A_{tp}^i$, which we assume to be as simple as
possible: its elements are equal to one when a pixel is observed,
and are zero otherwise. Eq.~(\ref{eq:TOD_xpol}) can be recast in a
more compact formalism by defining a generalized pointing matrix
$\mathbf A_{tp}$ which includes the trigonometric functions and a
map triplet $ \mathbf{S_p} \equiv (I_p,Q_p,U_p)$:
\begin{equation}
\mathbf {\mathcal D}_t = \mathbf A_{tp} \mathbf S_p + \mathbf n .
\label{eq:TODgen}
\end{equation}
The GLS estimator for $\mathbf{S_p}$ is then
(see~\cite{degasperis05})
\begin{equation}\label{eq:MMsolution}
  \mathbf {\widetilde S_p} = \left( \mathbf A^t
\mathbf{N}^{-1} \mathbf A\right)^{-1}
  \mathbf A^t \mathbf{N}^{-1} \mathbf {\mathcal D},
\end{equation}
where $\mathbf{N}$ is the noise covariance matrix, whose
estimation is discussed in Sect.~\ref{subs:noisest} above:
\begin{equation}
  \mathbf N \equiv \left\langle \mathbf n_t \mathbf n_{t^\prime}\right\rangle=
  \left(
    \begin{array}{ccc}
      \left\langle  n_t^1  n_{t^\prime}^1 \right\rangle &
      \cdots &
      \left\langle  n_t^1  n_{t^\prime}^n \right\rangle \\
      \vdots & \ddots & \vdots  \\
      \left\langle  n_t^n  n_{t^\prime}^1 \right\rangle &
      \cdots &
      \left\langle  n_t^n  n_{t^\prime}^n \right\rangle
    \end{array}
  \right).
\label{eq:gen_noise}
\end{equation}
This matrix becomes block diagonal assuming that there is no noise
correlation between different detectors:
\begin{equation}
  \left\langle n_t^i n_{t^\prime}^j
  \right\rangle\ = \left\langle n_t^j n_{t^\prime}^i \right\rangle = 0 \;\;\;
\left( i\neq j \right).
\end{equation}
Given the size of the problem, in the IT pipeline the map making
normal equations~(\ref{eq:MMsolution}) are solved by implementing
a Fourier-based preconditioned conjugate gradient iterative
solver, which only needs to perform matrix to vector products.
This approach critically depends on the assumption that the noise
is stationary (\cite{natoli01}). In the NA pipeline a Jacobi
solver is used.

In practice, there are many further details. Since the solver
works in Fourier space, the timeline must be continuous;
 bad or missing data chunks (``gaps'') are
replaced by a constrained realization of Gaussian noise
(\cite{hoffman91}), designed to mimic the correct noise behavior
at the gap's boundaries. The samples belonging to these chunks are
then flagged and the solver is instructed not to take them into
account, as they do not contain any useful data. The noise
correlation functions (also called noise filters) are band pass filtered
in the range $70$~mHz -- $14.5$ Hz before being given to
the map making code.

Each detector has its own noise filter, which is kept constant
over the full flight (i.e.\ data are reduced as if the noise were
stationary within a single detector) for the IT pipeline, while the NA
pipeline estimates noise separately for each ($\sim$ hour long) chunk of
data. The NA approach treats the non stationarity of the
data, at the cost of a reduced accuracy in the estimate of the
noise filter. The differences in the resulting maps are
negligible.

For the polarization maps, detectors forming PSB pairs are treated
as independent detectors in the IT pipeline, while the difference
of their signals is used by the NA pipeline (see \cite{jones05b}
for details).

The ROMA code employs about 20 minutes (and $\sim$ 200 iterations)
to produce a set of $I$, $Q$, $U$ maps from an eight bolometer
(full flight) timeline, when running a 128 processor job on an IBM
SP machine featuring 450 MHz Power~3 processors.

%

\subsection{Responsivity Calibration}

The first step in obtaining our final product maps of I, Q, U, is
the relative calibration of all the detectors. As explained in \S
\ref{subs:general} , we need a relative calibration accuracy of
the order of $\simlt 2 \%$. We start by producing single detector
maps of I using one of the map-making procedures. We then proceed
in two ways: in pixel space and in multipole space.

\subsubsection{Relative calibration from pixel-pixel correlation of
different PSBs \label{subs:pixpixcal}}

We carry out this analysis for the PSBs in the deep region. We
use the 145 GHz $W_1$ channel as the reference channel. The
characteristics of these channels (beam, frequency response,
noise, polarization efficiency) are so similar that their data can
be directly compared, pixel to pixel. For channel $j$ we scatter
plot the $I_{i,j}$ values (where $i$ is the pixel index) versus
the $I_{i,W_1}$ values, and we fit the data to a straight line. We
estimate the error on each $I_{i,j}$ as $\sigma_j/\sqrt{N_{j,i}}$
where $\sigma_j$ is obtained by integrating the TOD noise power
spectrum, and $N_{j,i}$ is the number of observations of pixel
$i$. The slope of the best fit line is the relative calibration
factor $\mathcal{R}_{j,W_1}$. Simulations show that this procedure
results in an accurate (within 1\%) estimate of the relative
calibration. The results for all channels are reported in table
\ref{table:pixpixcal}.

\begin{table}[p]
\begin{center}
\begin{tabular}[h]{|c|c|c|}
\hline
channel & $\mathcal{R}_{*W_1}^{pixel}$ &  $\mathcal{R}_{*W_1}^{spectrum}$ \\
\hline
$145W2$   & 0.956$\pm $0.028 & $0.950\pm 0.014$ \\
\hline
$145X1$   & 0.985$\pm $0.029 & $0.980\pm 0.016$\\
\hline
$145X2$   & 0.749$\pm $0.022 & $0.731\pm 0.013$\\
\hline
$145Y1$   & 0.920$\pm $0.030 & $0.932\pm 0.020$\\
\hline
$145Y2$   & 0.912$\pm $0.029 & $0.906\pm 0.019$\\
\hline
$145Z1$   & 0.674$\pm $0.021 & $0.665\pm 0.013$\\
\hline
$145Z2$   & 0.407$\pm $0.023 & $0.422\pm 0.011$\\
\hline
\end{tabular}
\end{center}
\caption{Relative calibration $\mathcal{R}$ of the PSB channels
obtained from the pixel-pixel scatter plots with NSIDE=256 (second
column, see \S \ref{subs:pixpixcal}) and from the cross-spectrum
(third column, see \S \ref{subs:speccal}), using $W_1$ as the
reference channel. For $\mathcal{R}_{*W_1}^{spectrum}$, 1-$\sigma$
errors are used. For $\mathcal{R}_{*W_1}^{pixel}$, we also include
in the error the possible bias due to a conservative $\pm 20\%$
error in the estimate of the noise. \label{table:pixpixcal}}
\end{table}

We also study the effect of a mis-estimate of the noise in the
responsivity estimate. This depends strongly on the signal to
noise per pixel of the data. For 7$^\prime$ pixels, the signal to
noise per pixel for CMB anisotropies is $\sim 3$. In these
conditions, simple simulations show that in order to bias the
relative calibration by more than $2\%$ the estimate of the noise
must be wrong by more than $20\%$. We conclude that this method
provides a robust estimate of the relative calibration. The errors
on the relative calibration results reported in table
\ref{table:pixpixcal} include the effect of a conservative 20\%
misestimate of the errors per pixel.

If, instead, the S/N per pixel is $\simlt 1$, then a $20\%$
misestimate of the noise induces $\simgt 20\%$ errors in the
calibration. This happens when an absolute calibration is
attempted by fitting $I_{i,j}$ versus the $I_{i,WMAP}$. Since the
angular resolution of \wmap is worse than the one of \bk, and the
noise per pixel is higher, the slope of the best fit line is
significantly biased. Also, working in pixel space it is not
trivial to take into account the difference in beam size and shape
between \bk and \wmap.

\subsubsection{Absolute and Relative calibration from ratios of
cross-spectra \label{subs:speccal}}

We can compare the signals detected by different detectors
observing the same sky in multipole space rather than in pixel
space. A cross-power spectrum is very useful in this case, due to
its unbiased nature: the noise properties of the two signals do
not affect the results. Moreover, the use of the cross-power
spectrum allows us to properly take into account in a simple way
the effect of the different angular resolution, and map-making
transfer functions. All these effects can be included in a
function $F_\ell^k$, which can be different for each channel $k$.
Since we do see CMB fluctuations in all the \bk maps, this method
can be used to calibrate the \bk detectors at 145, 245 and 345
GHz, and to find the absolute calibration factor by comparing \bk
maps to \wmap maps.

We expand the uncalibrated maps of all channels in spherical
harmonics:

\begin{equation}
 V_k =  \sum_{\ell,m} a_{\ell,m}^{k} Y^\ell_m F_\ell^k
\end{equation}

We compute the cross power spectrum $\langle a_{\ell,m}^{k}
F_\ell^k \times a_{\ell,m}^j F_\ell^j \rangle$ and compare it to
the cross power spectrum $\langle a_{\ell,m}^{j} F_\ell^{j} \times
a_{\ell,m}^{W_1} F_\ell^{W_1} \rangle$: the ratio is the relative
calibration of channel $k$ against the reference channel $W_1$:

\begin{equation}
{\mathcal R}_{k,W_1} = \frac{\langle a_{\ell,m}^{k} F_\ell^k
\times a_{\ell,m}^j F_\ell^{W_1} \rangle}{\langle a_{\ell,m}^j
F_\ell^j \times a_{\ell,m}^{W_1} F_\ell^{W_1} \rangle}
\label{eq:crossps}
\end{equation}

This is iterated for all $j \ne k$ for consistency checks; the
uncertainty on the result is estimated with a bootstrap method
(see \cite{polenta04} for details). The relative calibrations
obtained in this way are reported in table \ref{table:pixpixcal}.
These numbers are consistent with the ones obtained in \S
\ref{subs:pixpixcal}. We have also verified that using either of
the two does not change the combined map.

Having measured the relative calibrations for all channels of the
same band, we can combine them into an uncalibrated optimal map
of the sky, using the optimal map-making described in \S
\ref{subs:mapmaking}.

We then estimate the absolute calibration of this map by comparing
it to the \wmap map in the same region. The \wmap all-sky maps
are calibrated with remarkable accuracy (0.5\%, see
\cite{bennett03}). So we can compare the angular power spectra
measured by \bk and \wmap in the same sky region in order to
measure the absolute responsivity $\mathcal{S}$ of \bk.

Assuming that the uncalibrated signal in \bk (in Volt) is related
to the CMB temperature measured by \wmap as $T_{WMAP} = V_{B}/
{\mathcal S}$, we can estimate $\mathcal S$ from different
combination of \bk and \wmap power spectra. We use the following
one:

\begin{equation} {1 \over {\mathcal S}} =
\frac {\langle a_{\ell,m}^{B} F_\ell^{B} \times a_{\ell,m}^{WMAP}
F_\ell^{WMAP} \rangle} {\langle a_{\ell,m}^{B,WX}F_\ell^{WX}\times
a_{\ell,m}^{B,YZ} F_\ell^{YZ}\rangle} \label{eq:crossps2}
\end{equation}

For the 145 GHz PSBs, we compute the cross-power spectrum of
$\langle a_{\ell,m}^{B} F_\ell^{B} \times a_{\ell,m}^{WMAP}
F_\ell^{WMAP} \rangle$ using optimal combined maps of all channels
for both the experiments, and the cross-power spectrum of $\langle
a_{\ell,m}^{B,WX} F_\ell^{WX}\times a_{\ell,m}^{B,YZ}
F_\ell^{YZ}\rangle$ using optimal maps obtained combining the $W$
and $X$ PSBs for the first map, and the $Y$ and $Z$ PSBs for the
second one. Using this denominator in place of $\langle
a_{\ell,m}^{B} F_\ell^{B} \times a_{\ell,m}^{B} F_\ell^{B} \rangle
$ we avoid the need to remove a possible bias due to correlated
noise. This choice, based on half of the detectors, does not
degrade the precision of the result, which is already limited by
cosmic variance.

\begin{figure}[p]
            \begin{center}
            \includegraphics[width=14cm]{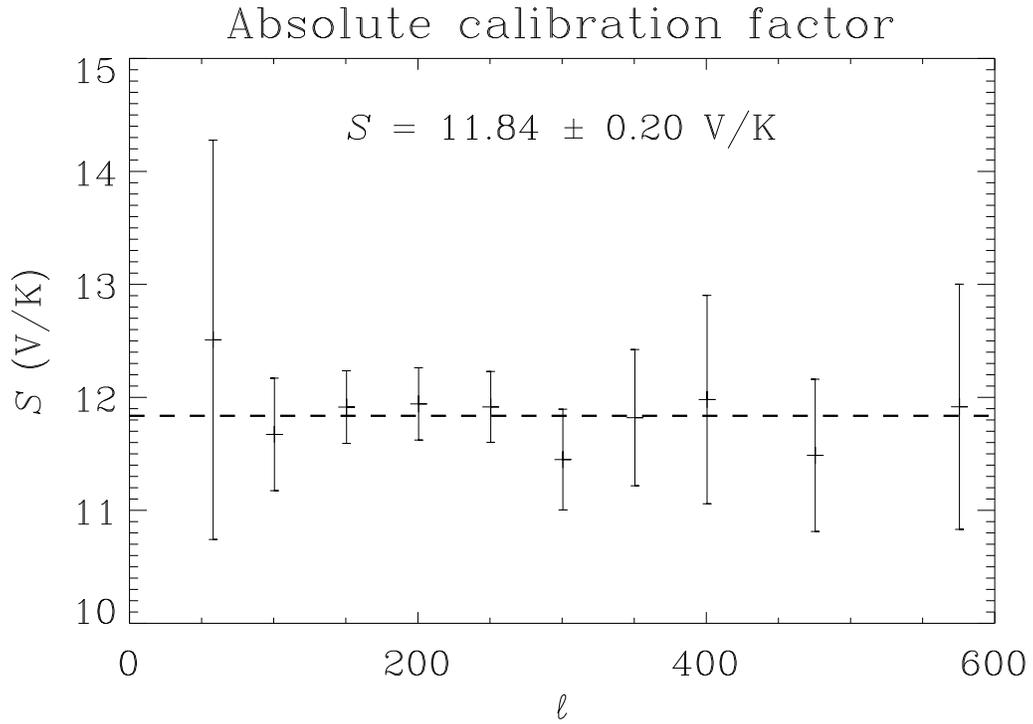}
            \caption{Absolute calibration factors ${\mathcal S}_{\ell}$ based on \wmap measurement.
            The cross-power spectra are computed using the shallow
            integration region with uniform weight. The final value for ${\mathcal S}$ is the weighted average of the ${\mathcal
S}_{\ell}$.}
            \label{fig:abscal}
            \end{center}
\end{figure}

For the 145 GHz T map, we show in Figure~\ref{fig:abscal} the
measurement of $\mathcal S$ obtained from eq. \ref{eq:crossps2},
i.e. the absolute calibration factor as a function of $\ell$. Its
flatness confirms that we are properly correcting for beam,
pixelization and noise effects.

We  confirm the robustness of the result by computing calibration
factors from the shallow and deep surveys separately and by
applying different weighting schemes. The final result for the 145
GHz map is \begin{equation} {\mathcal S}_{145} = (11.84 \pm 0.20)
V/K \end{equation} where the 1-$\sigma$ error already incorporates
the effect of the calibration uncertainty of \wmap. The absolute
responsivity of a single channel $k$ can be obtained from the
values listed in table \ref{table:pixpixcal} as ${\mathcal
S}_k={\mathcal S}{\mathcal R}_{k,W_1}$.

The calibrations of 245 and 345 GHz maps are found in a similar
way. We obtain \begin{equation} {\mathcal S}_{245} = (4.27 \pm
0.20) V/K \end{equation}  and \begin{equation} {\mathcal S}_{345}
= (3.04 \pm 0.27) V/K \end{equation}

\subsection{Differences in the two pipelines \label{subs:differences}}

Before discussing the results of the analysis, we summarize in
table \ref{tab:differences} the most important differences between the
two pipelines. Despite the many different choices, the final
results of the analysis are very consistent for both the maps (see
\S \ref{subs:galaxy} and following) and the power spectra
\cite[]{jones05,piacentini05,montroy05}.

\begin{table}[p]
\begin{center}
\begin{tabular}{|c|c|c|}
\hline
item & IT pipeline & NA pipeline \\
\hline
{\bf cross-polar response:} & full beam (table \ref{tab:crosspol}, col.4) & axial (table \ref{tab:crosspol}, col.5)\\
{\bf time response:} & in-flight, from cosmic rays (\S \ref{subs:flighttransfer}) & from lab measurements (\S \ref{subs:labtransfer})\\
{\bf noise:} & estimated on the full flight &  estimated in $\sim$ hour long chunks \\
{\bf Q,U map-making:} & all detectors treated individually & PSB pairs differences \\
{\bf map-making solver:} & Conjugate Gradients & Jacobi \\
\hline
\end{tabular}
\end{center}
\begin{center}
\caption{\small Summary of the different choices made for the two
data analysis pipelines. \label{tab:differences}}
\end{center}
\end{table}

\subsection{Observations of the Galactic Plane \label{subs:galaxy}}

\subsubsection{Brightness Maps of the Galactic Plane}

\begin{figure}[p]
\begin{center}
\includegraphics[angle=0,width=5cm]{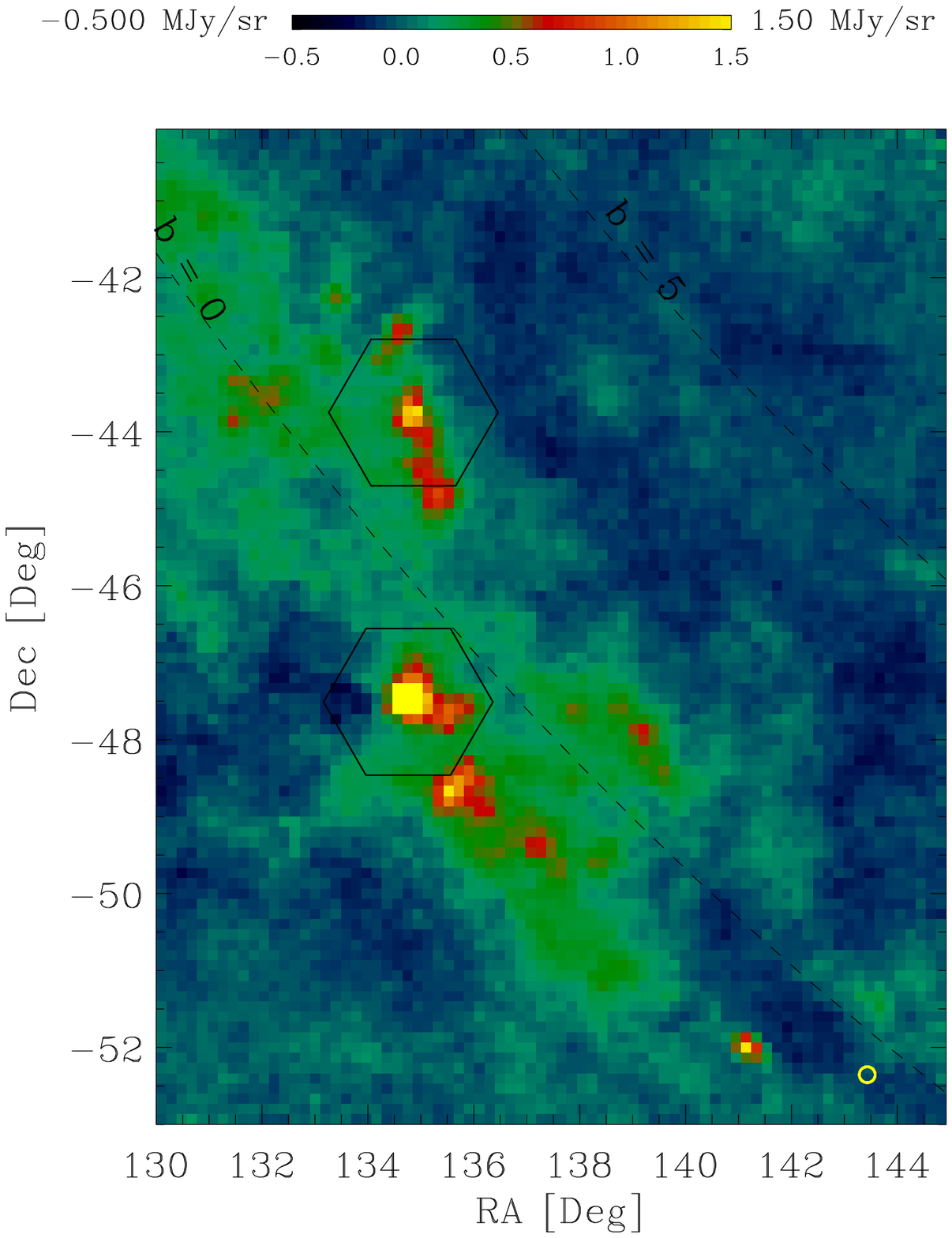}
\includegraphics[angle=0,width=5cm]{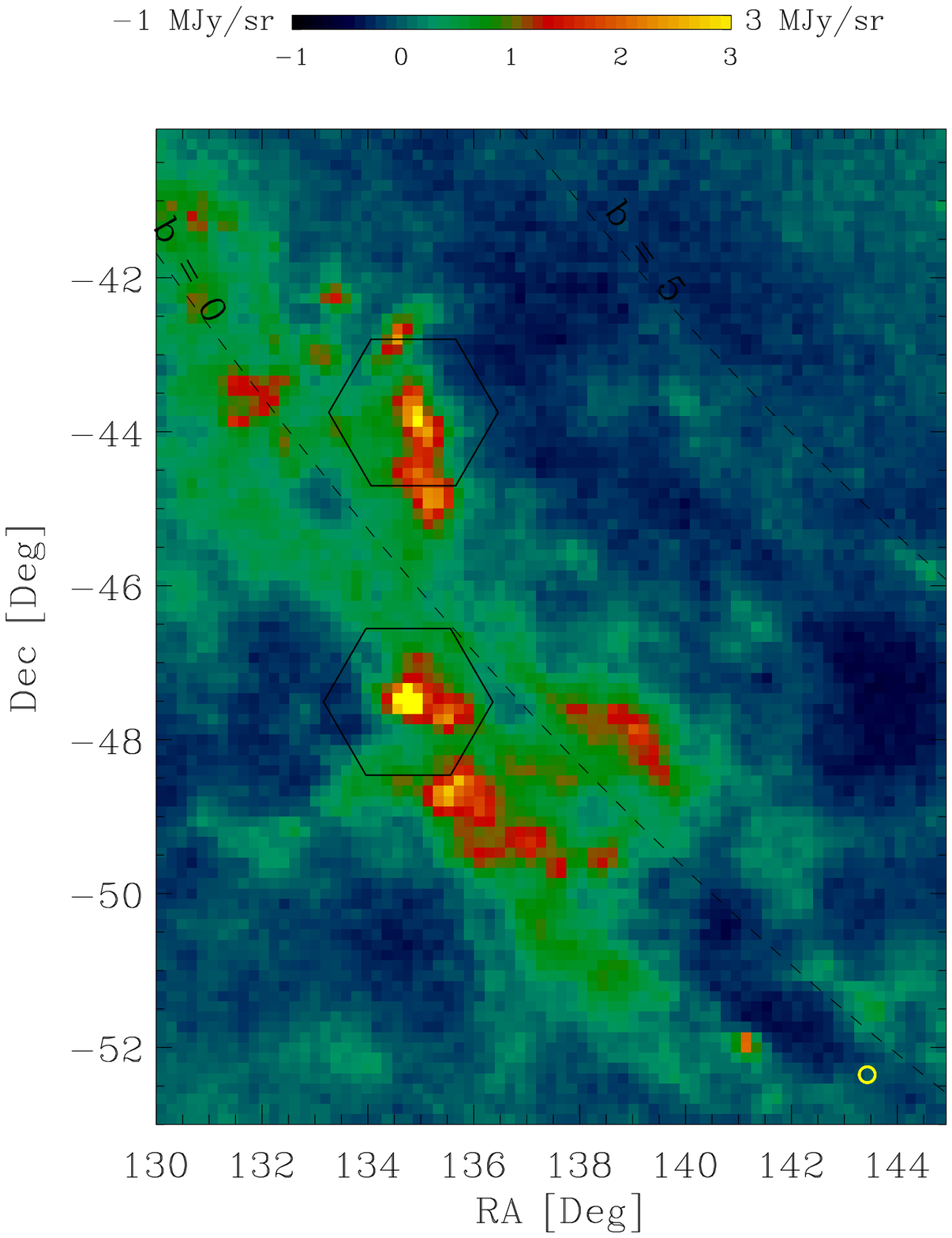}
\includegraphics[angle=0,width=5cm]{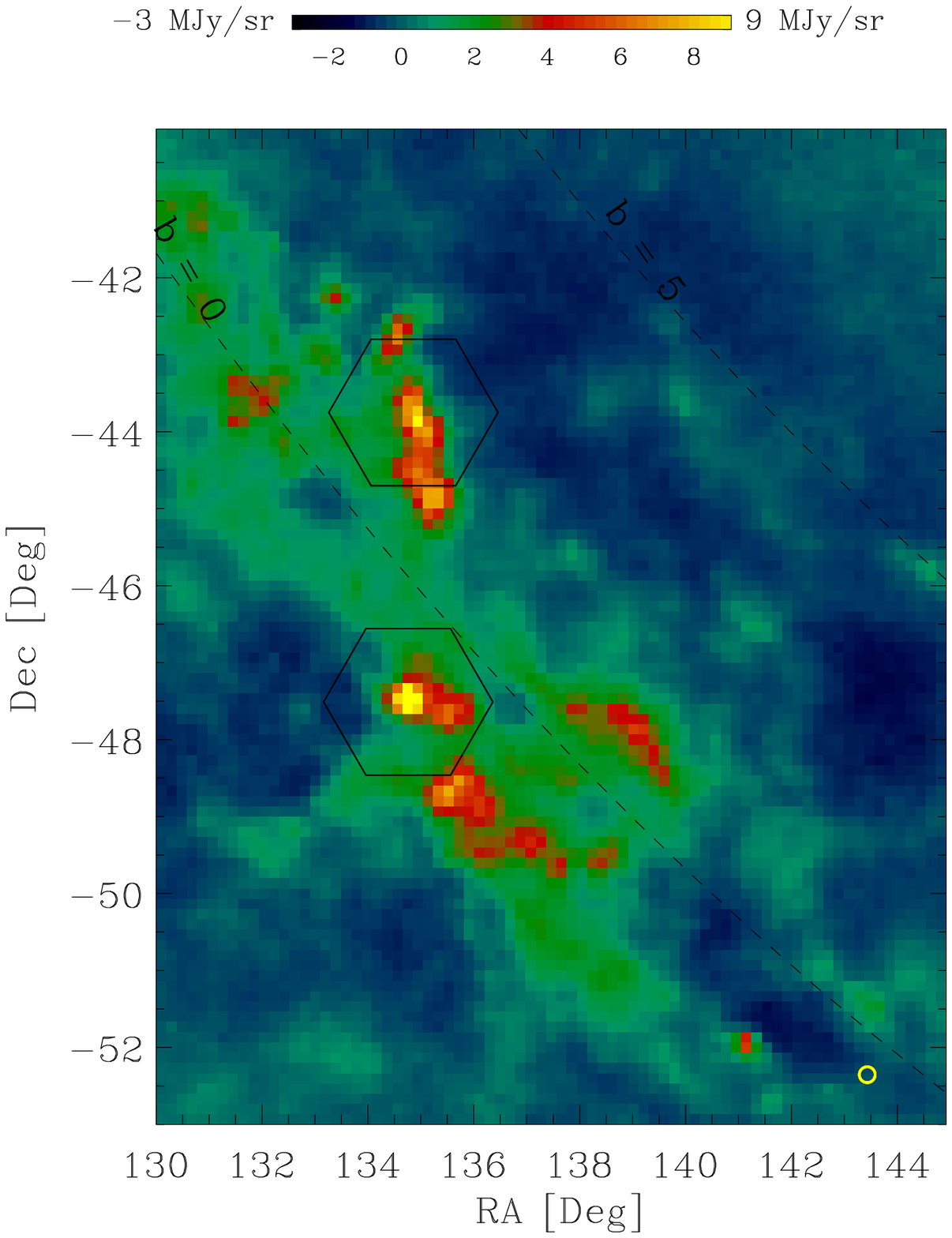}
\includegraphics[angle=0,width=5cm]{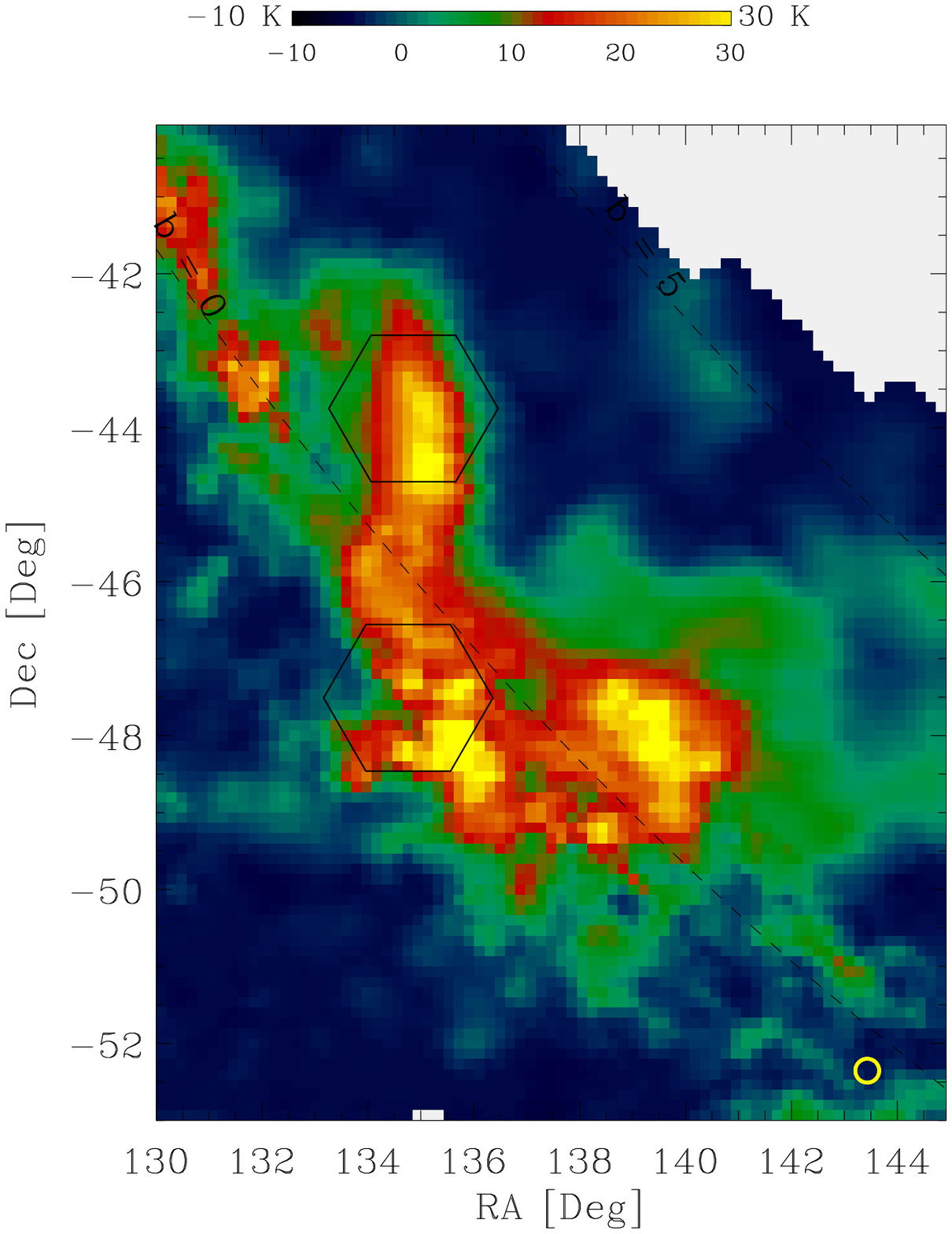}
\includegraphics[angle=0,width=5cm]{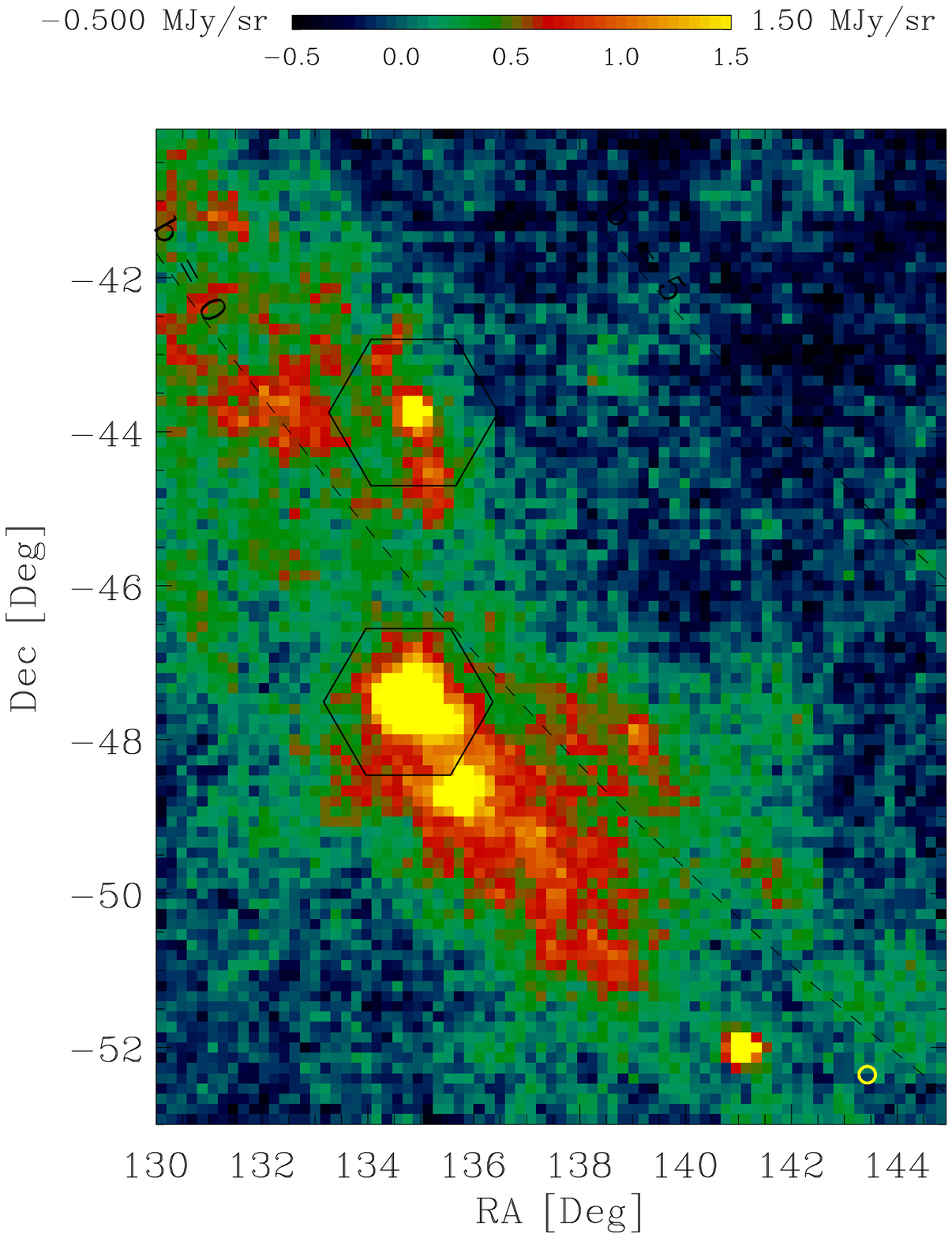}
\includegraphics[angle=0,width=5cm]{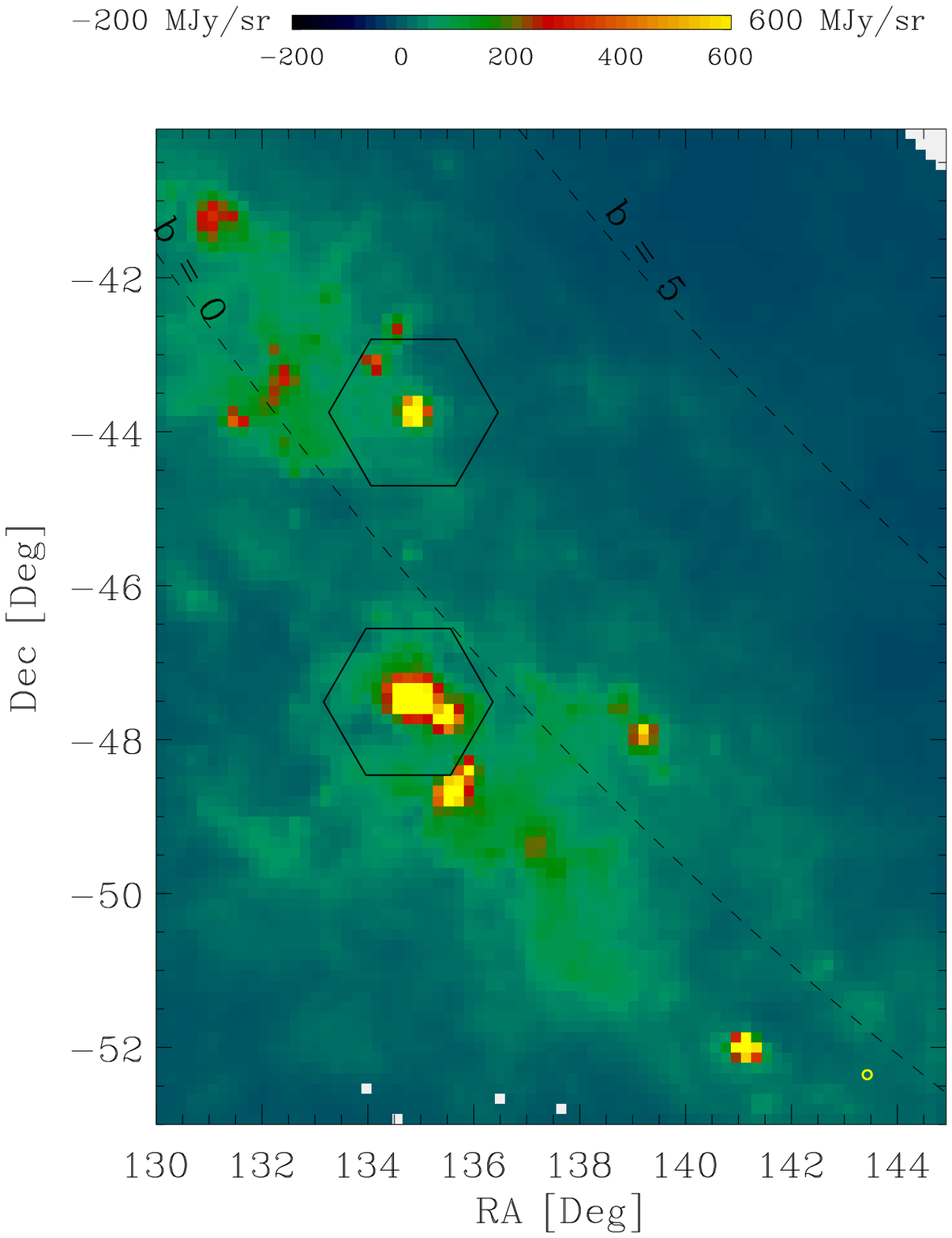}
\caption{\bk survey of the Galactic plane. In the top row we
display (left to right): \bk maps at 145, 245 and 345 GHz. The
brightness units are MJy/sr. The conversion factors from MJy/sr to
$mK_{CMB}$ are given in table \ref{tab:optical_efficiencies}. In
the bottom row we report, for comparison, maps of the same region
from (left to right): the CO survey of \cite{dame01}, the WMAP 94
GHz channel, and the IRAS 3000 GHz survey. All the maps have been
convolved with a 6$^\prime$ FWHM gaussian filter. The equivalent
FWHM resolution of the maps is indicated by the yellow circle on
the bottom right of each panel. The hexagons locate RCW38 ($dec
\sim -47.5^o$) and IRAS08576 ($dec \sim -43.8^o$).}
\label{fig:galaxy}
\end{center}
\end{figure}

Brightness maps of a section of the Galactic plane at 145, 245 and
345 GHz are shown in Fig.\ref{fig:galaxy}, in comparison with
other relevant maps of emission from dust and associated molecular
gas. Several known HII regions are evident in the maps.
Particulary luminous are RCW38 (at RA=134.76$^o$, dec=-47.47$^o$)
and IRAS08576 (at RA=134.92$^o$, dec=-43.75$^o$). Diffuse emission
is also evident. It is instructive to compare the \bk maps to
monitors of different components of the ISM, like the IRAS/DIRBE
maps for the dust continuum and the 115 GHz CO map of
\cite{dame01} for the molecular gas (see bottom row of
Fig.\ref{fig:galaxy}). There is a good correlation between the \bk
maps and the IRAS/DIRBE 3000 GHz map for most of the diffuse
structure observed. A striking exception is the cloud evident in
all \bk channels at RA=135.24$^o$ and dec=-44.80$^o$: this is very
dim in the IRAS map, while it is bright in the CO map and is also
visible in the WMAP 94 GHz map: the signature of a cold dust cloud
associated with molecular gas.

\begin{figure}[p]
\begin{center}
\includegraphics[angle=0,width=10cm]{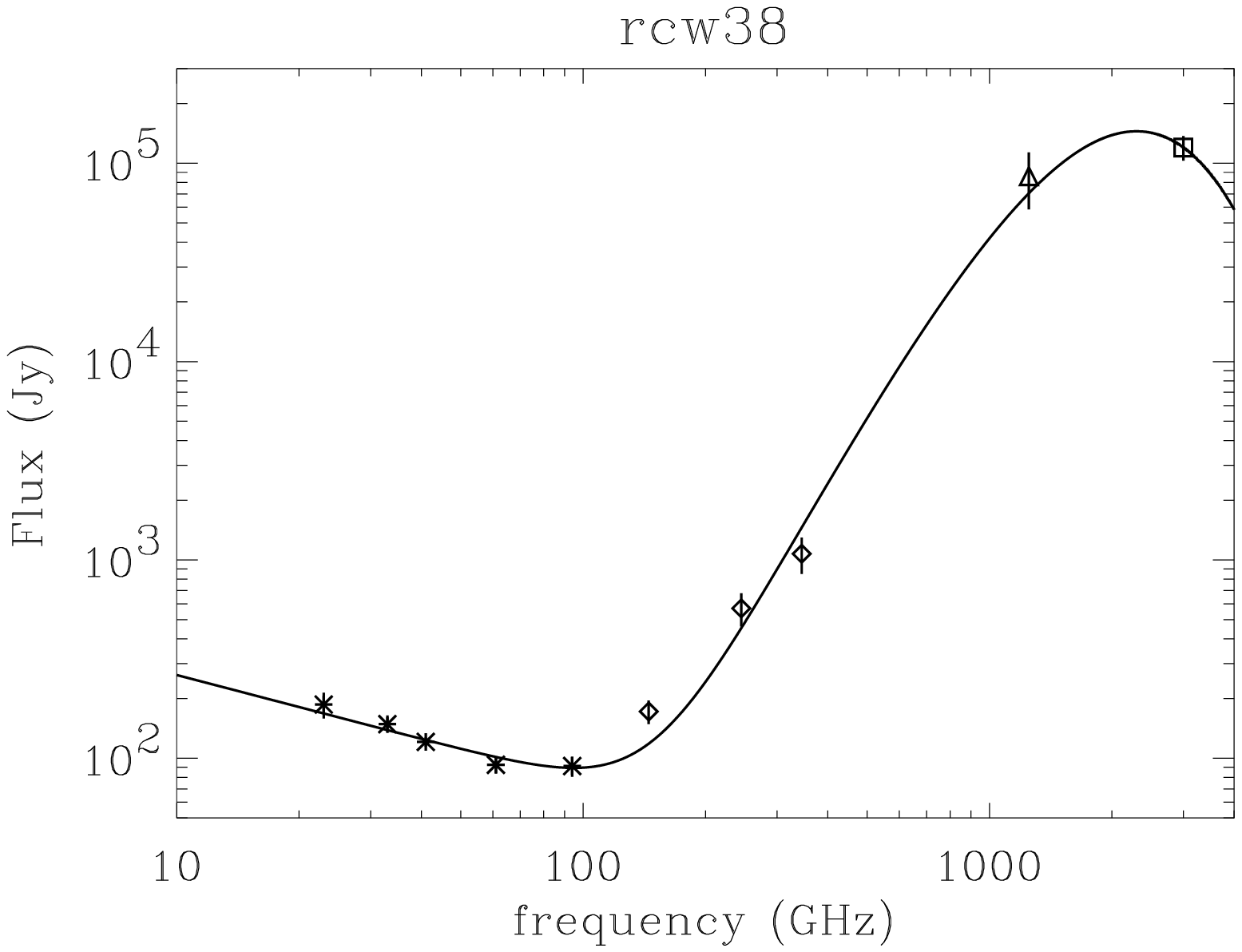}
\includegraphics[angle=0,width=10cm]{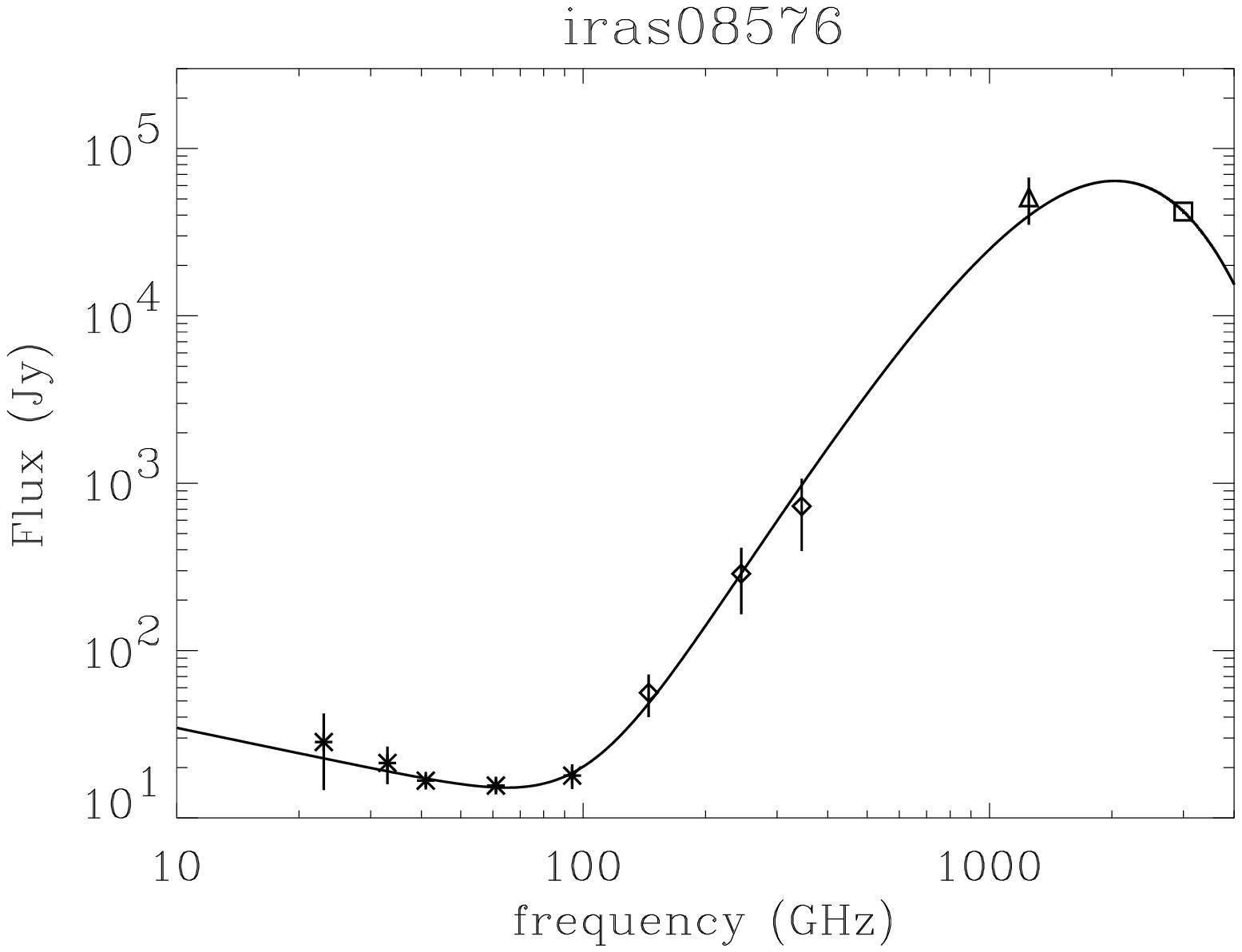}
\includegraphics[angle=0,width=10cm]{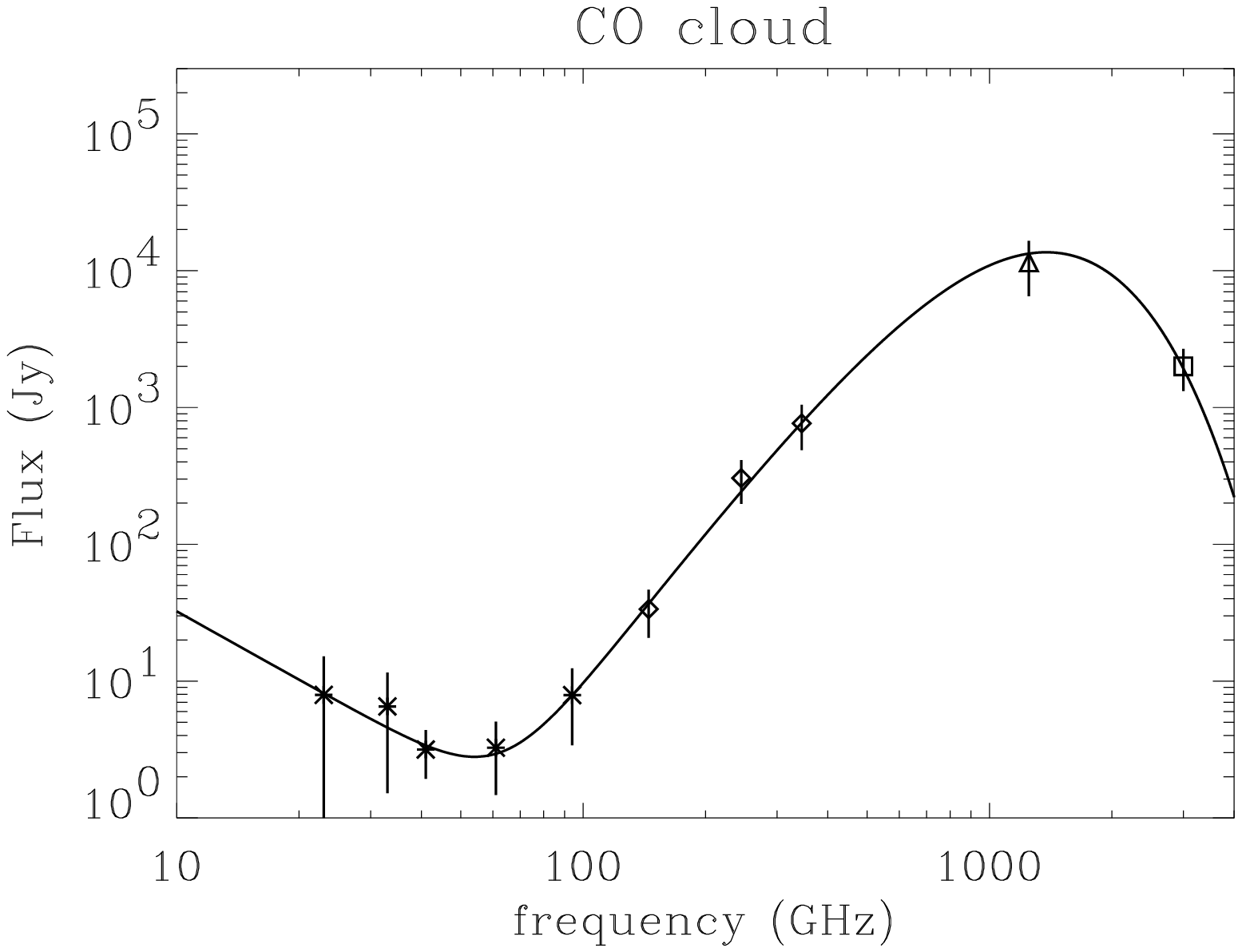}
\caption{Measured integrated fluxes for selected Galactic sources
observed by \bk (diamonds). The other data points are from WMAP
(stars), from DIRBE (triangle) and from the IRAS/DIRBE map
(square). The continuous line is the best fit SFD obtained as the
sum of a power law at low frequencies ($\sim \nu^a$) plus a
thermal dust spectrum dominating at high frequencies ($\sim \nu^2
B(\nu, T_d)$). \label{fig:galsources} }
\end{center}
\end{figure}

The measured spectral flux density (SFD) of the main Galactic
sources measured by \bk is reported in Fig. \ref{fig:galsources}.
The fluxes have been obtained integrating the brightness maps on
disks centered on the sources. The diameter of the disks is chosen
as the maximum between the apparent size of the source and twice
the FWHM of the experiment beam. The error in the determination of
the flux is dominated by the presence of diffuse emission, which
must be subtracted to estimate the net flux of the source: this is
less critical for RCW38, which is a relatively isolated source.
This results in error bars larger than those from calibration and
pointing jitter.

In order to gain insight on the physical processes of emission
operating in the sources, we combined the \bk FSD data with data
from WMAP, DIRBE and IRAS (see Fig. \ref{fig:galsources}). A
combination of a power law dominating at low frequencies ($\sim
\nu^a$), and a dust-spectrum at high frequencies ($\sim \nu^2
B(\nu, T_d)$) produces a good fit to the data.

We find dust temperatures of $(22.4 \pm 0.9)K$, $(19.6 \pm 0.7)K$,
and $(13.3 \pm 0.6)K$ for RCW38, IRAS08576, and the CO cloud
respectively. The spectral index of the power law is $(-0.5 \pm
0.1)$, $(-0.5 \pm 0.3)$, and $(-1.7 \pm 1.3)$ for RCW38,
IRAS08576, and the CO cloud respectively.

A word of caution is necessary for the power-lax indices derived
here. We are using data from experiments with different angular
resolution. In particular the low frequency data have poorer
angular resolution, so the measured flux can be contaminated by
nearby sources entering these wider beams. For this reason the
spectra shown in Fig. \ref{fig:galsources} could be increasingly
contaminated (biased high) at low frequencies. This issue will be
analyzed in a future publication.

\subsubsection{Polarization of RCW38}

RCW38 is the brightest source we observed. At 145 GHz the
polarization of RCW38 is very low. We have carried out an analysis
on the 145 GHz W, X, and Y PSBs, analyzing 53 scans over the
source, and assuming that there is a constant polarization over
the size of the source. For each scan and for each of the 6
bolometers we fit a maximum amplitude of the detected signal. We
then solve for Q and U of each scan using the detector signal
differences W1-W2, X1-X2, Y1-Y2, taking into account the relative
calibration of the bolometers, the color corrections, and the
orientation angles. The resulting average values for $Q$ and $U$
are $Q_{RCW38}=(-0.046 \pm 0.076) MJy/sr$ and $U_{RCW38}=(0.054
\pm 0.082) MJy/sr$. Since the average brightness for RCW38 on the
same scans is 3.7 MJy/sr, the upper limit for the polarization
degree is $p_{RCW38} < 6 \%$ (2-$\sigma$ U.L.).

\subsection{Brightness Maps at High Galactic  Latitude}

\subsubsection{Maps of CMB Temperature Anisotropy at High Galactic
Latitude \label{subs:maps}}

\begin{figure}[p]
\begin{center}
\includegraphics[angle=0,width=15cm]{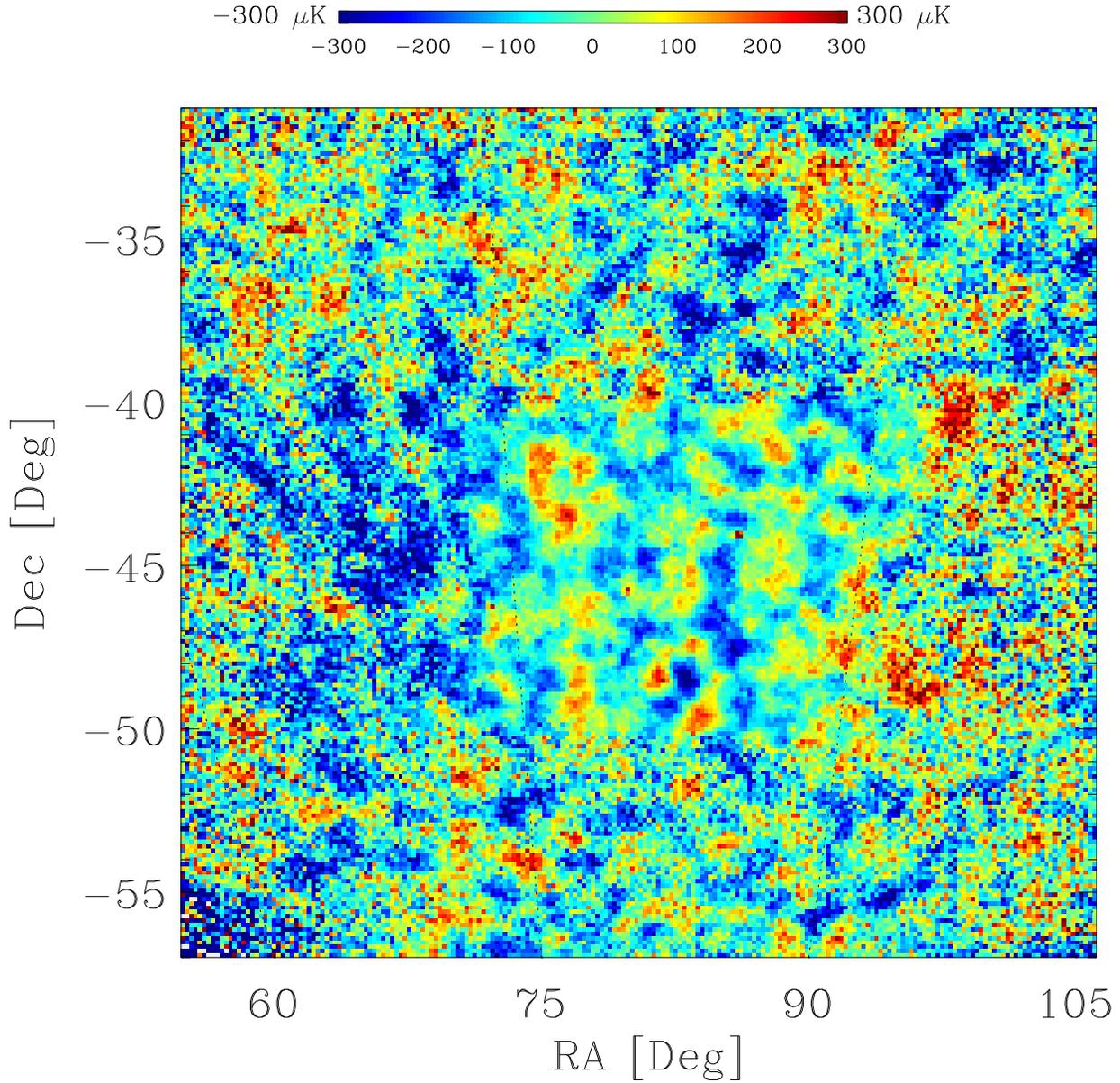}
 \caption{145 GHz I map from all the PSB bolometers of \bk.
I is encoded in the false color scale in thermodynamic temperature
units for a 2.725K blackbody. The pixel size is 3.4 arcmin. The
data of this map will be made publicly available (together with a
set of realistic simulations needed for quantitative analysis) at
the \bk web servers: http://oberon.roma1.infn.it/boomerang/b2k and
http://cmb.phys.case.edu/boomerang .
 \label{fig:deep_145_i}}
\end{center}
\end{figure}

The main product of this experiment is the 145 GHz map, which is
shown as a large image in Fig. \ref{fig:deep_145_i}. The structure
visible in the map with high S/N is CMB anisotropy. ISD is
negligible in comparison to the CMB, as we show below.

\begin{figure}[p]
\begin{center}
\includegraphics[angle=0,width=5cm]{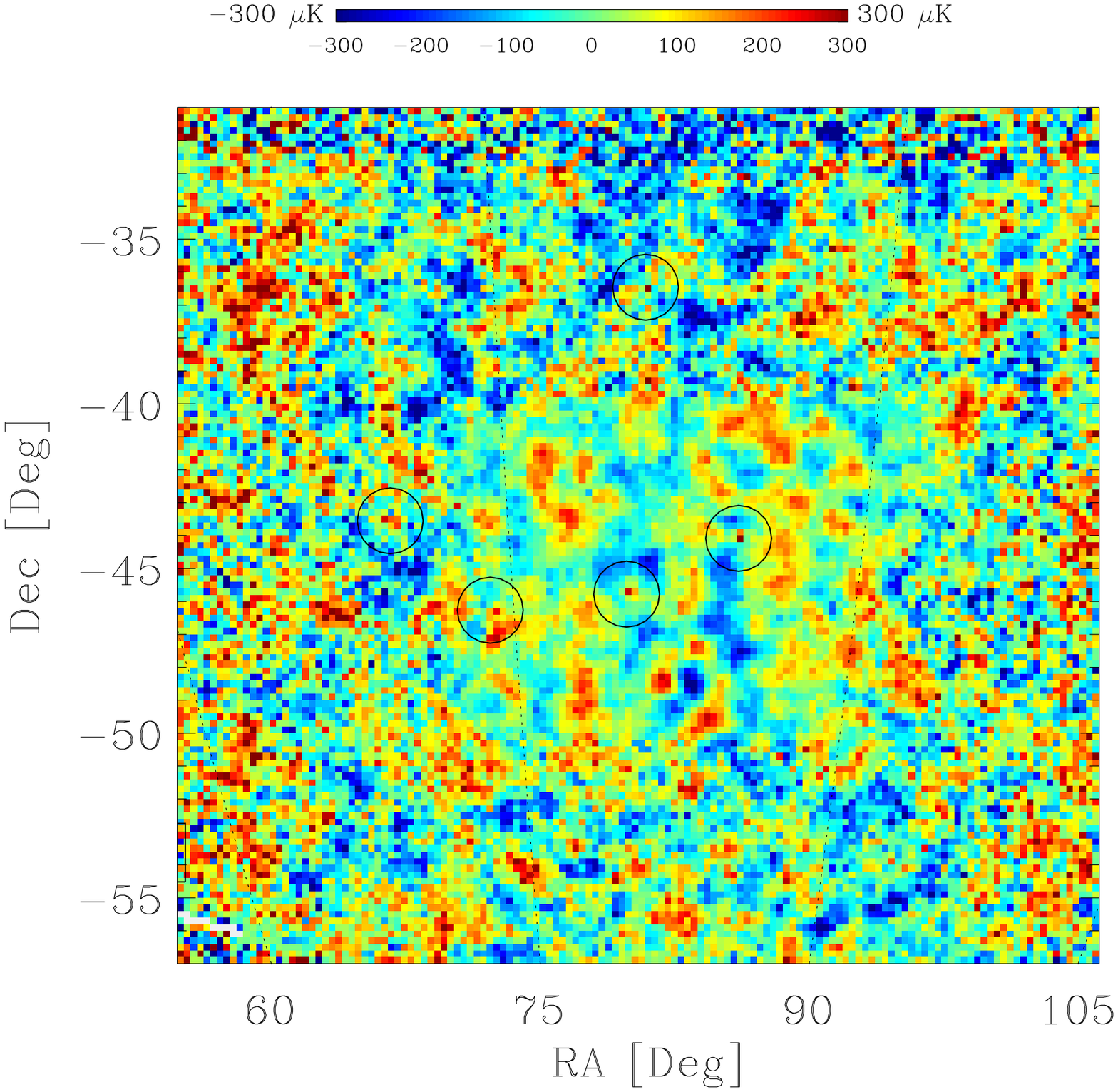}
\includegraphics[angle=0,width=5cm]{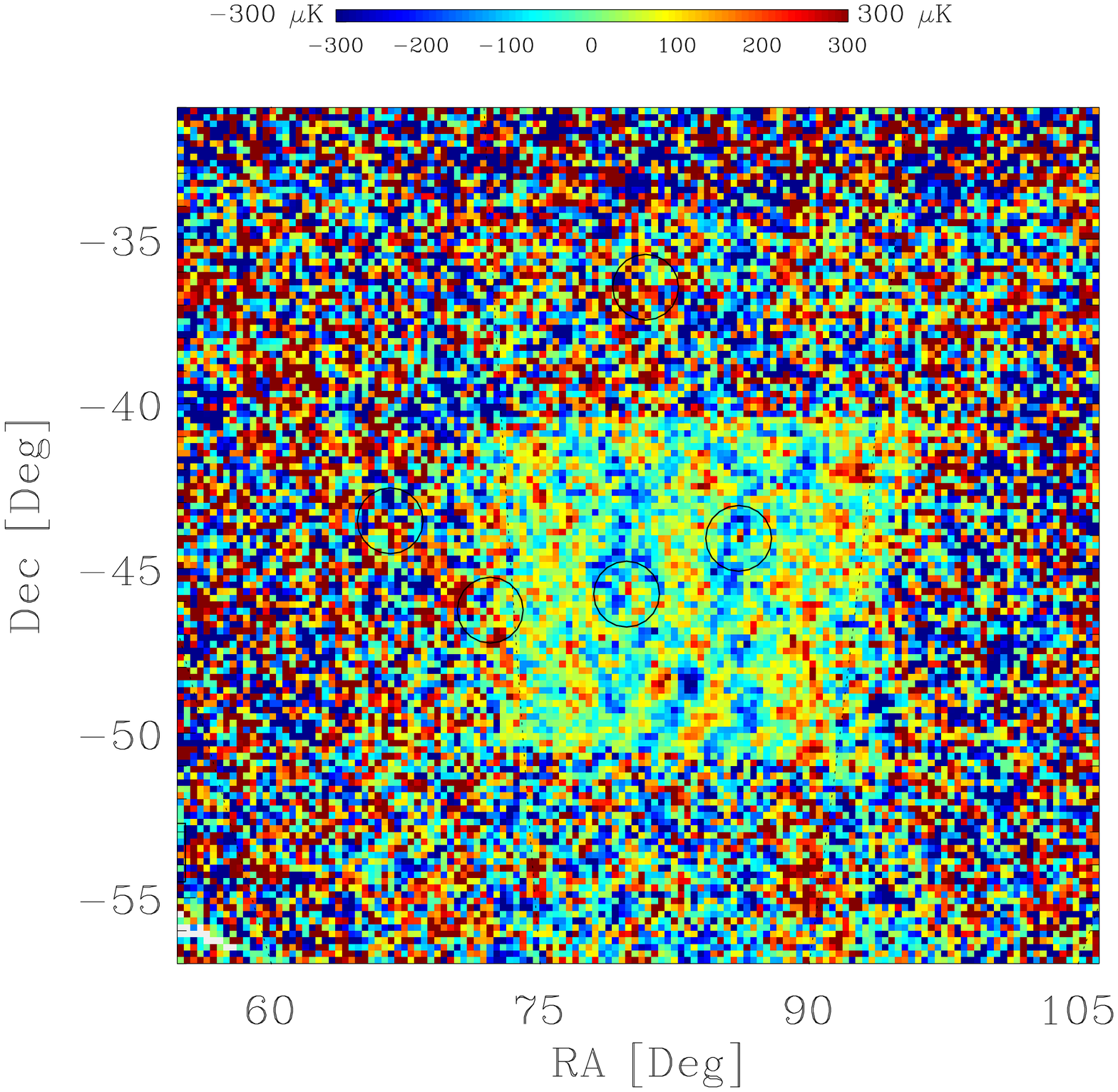}
\includegraphics[angle=0,width=5cm]{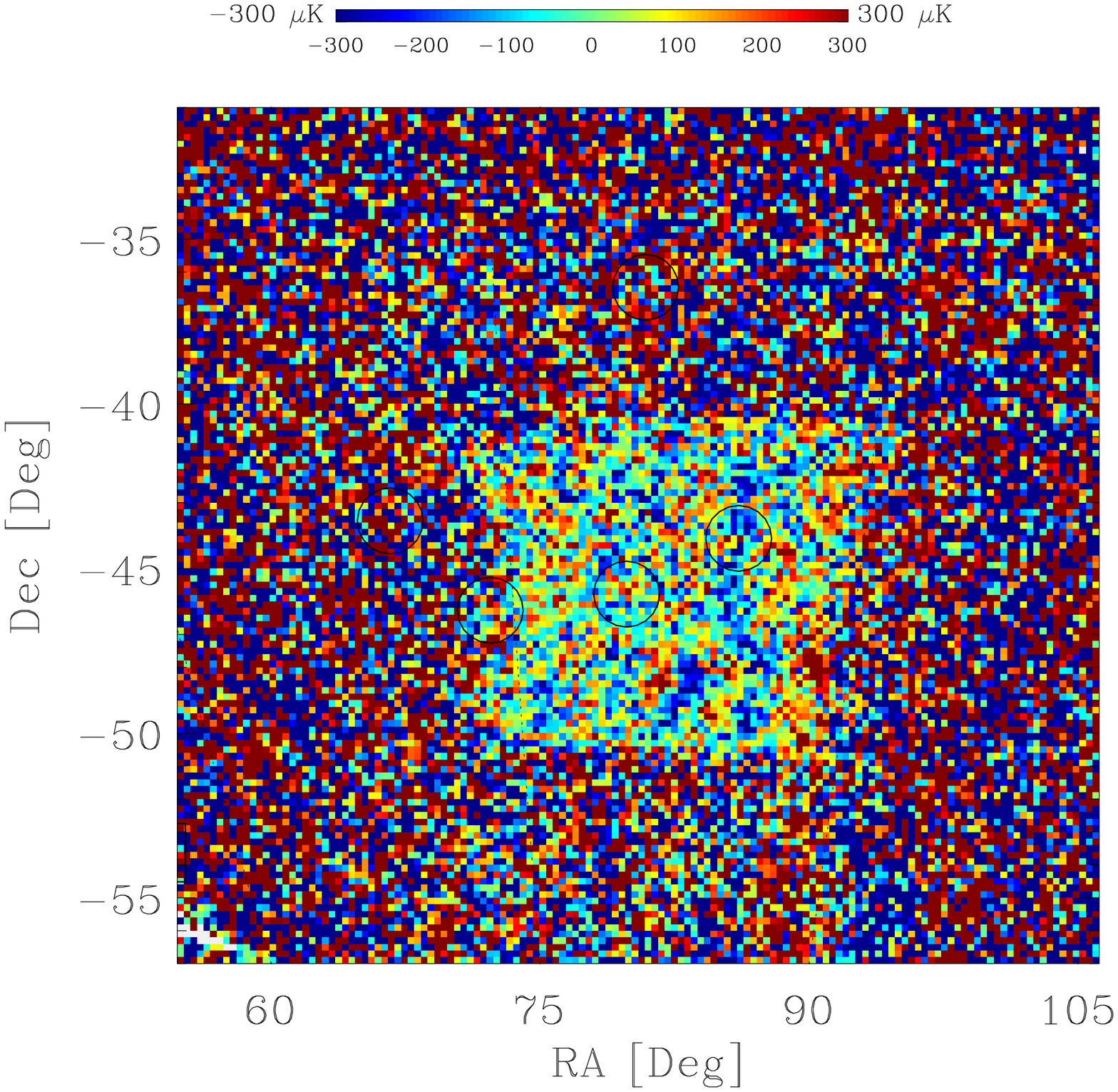}
\includegraphics[angle=0,width=5cm]{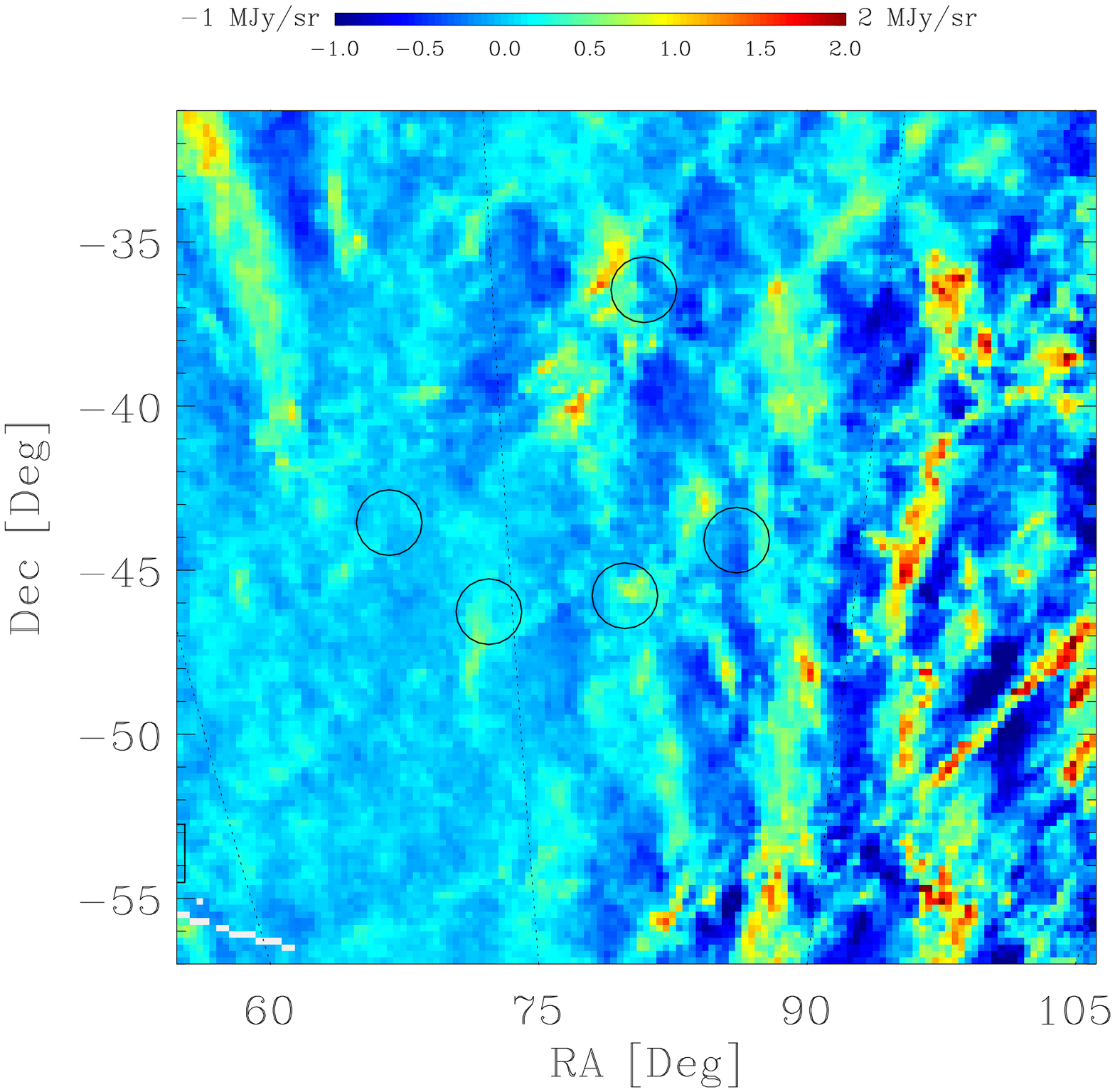}
\includegraphics[angle=0,width=5cm]{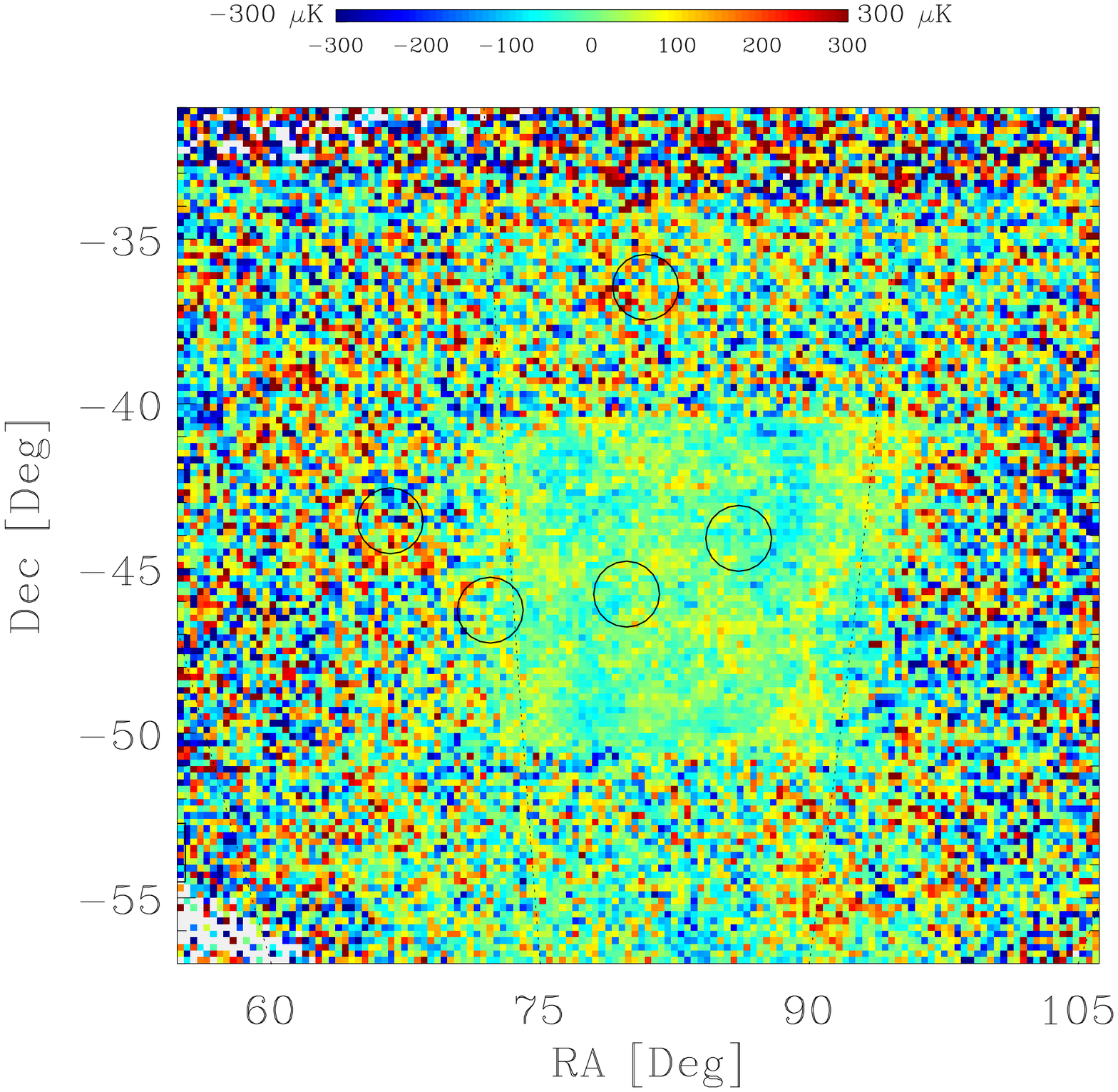}
\includegraphics[angle=0,width=5cm]{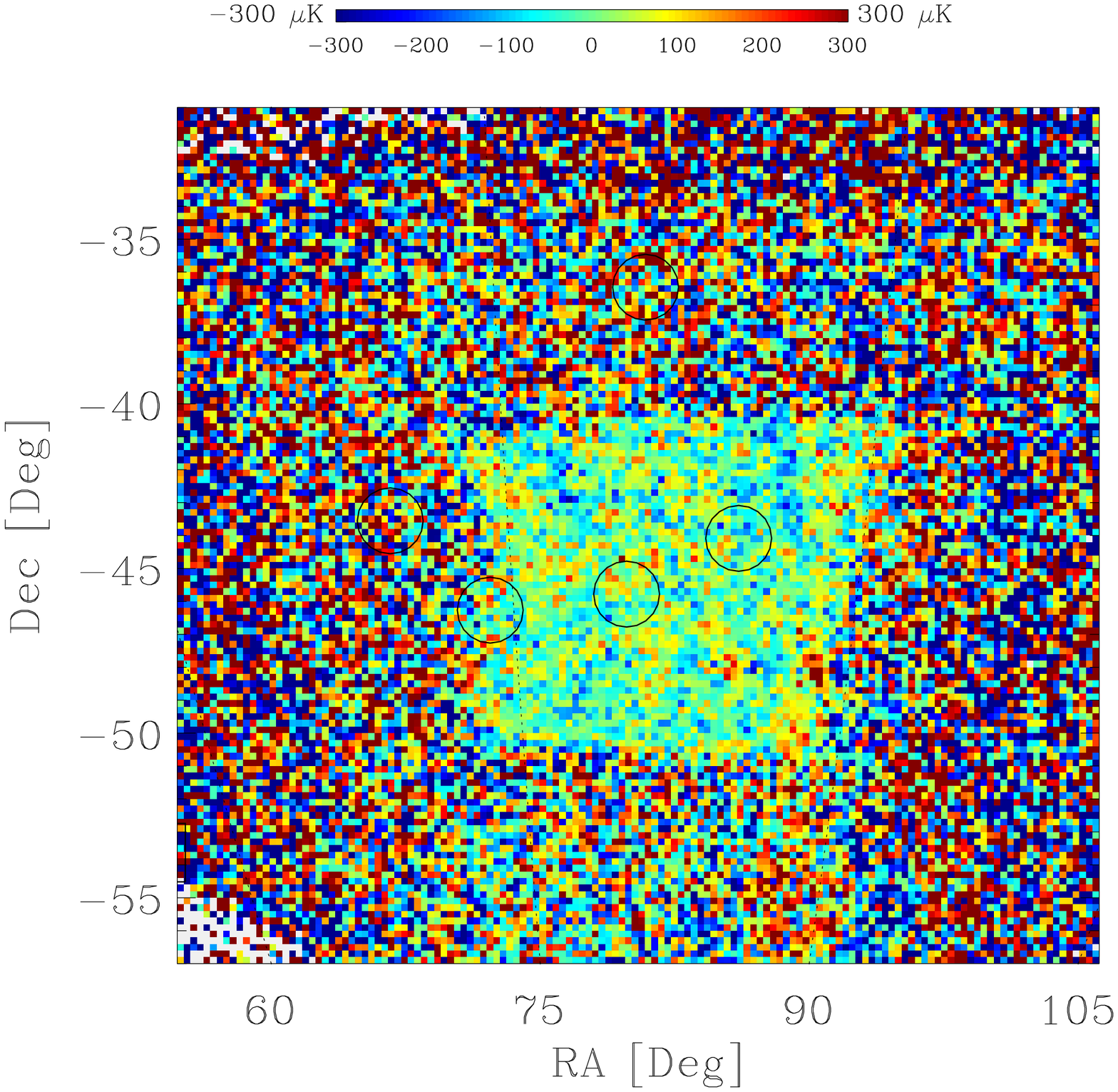}
 \caption{The shallow and deep surveys of \bk.
Top row: I maps from \bk. From left to right: 145 GHz; 245 GHz;
345 GHz. I is encoded in the false color scale in thermodynamic
temperature units for a 2.725K blackbody. The pixel size is 3.4
arcmin. Bottom row, from left to right: IRAS/DIRBE image of the
same region at 3000 GHz; \bk difference map obtained by
subtracting the 145 GHz map from the 245 GHz map, in order to
remove CMB anisotropy from the resulting map; \bk difference map
obtained as 345 GHz map minus 145 GHz map. The central region,
where the noise is evidently lower, is the deep survey. In Fig.
\ref{fig:deep_i} we zoom on such region. The circles label the
positions of known AGNs. To remove large-scale gradients, a
high-pass filter at 7.5 times the scan frequency was applied on
the time-ordered data: this filter is needed to remove artifacts
due to scan synchronous noise in the 245 and 345 GHz maps. A
gaussian filter with 7$^\prime$ FWHM was used as anti-aliasing.
 \label{fig:shallow_i}}
\end{center}
\end{figure}

\begin{figure}[p]
\begin{center}
\includegraphics[angle=0,width=5cm]{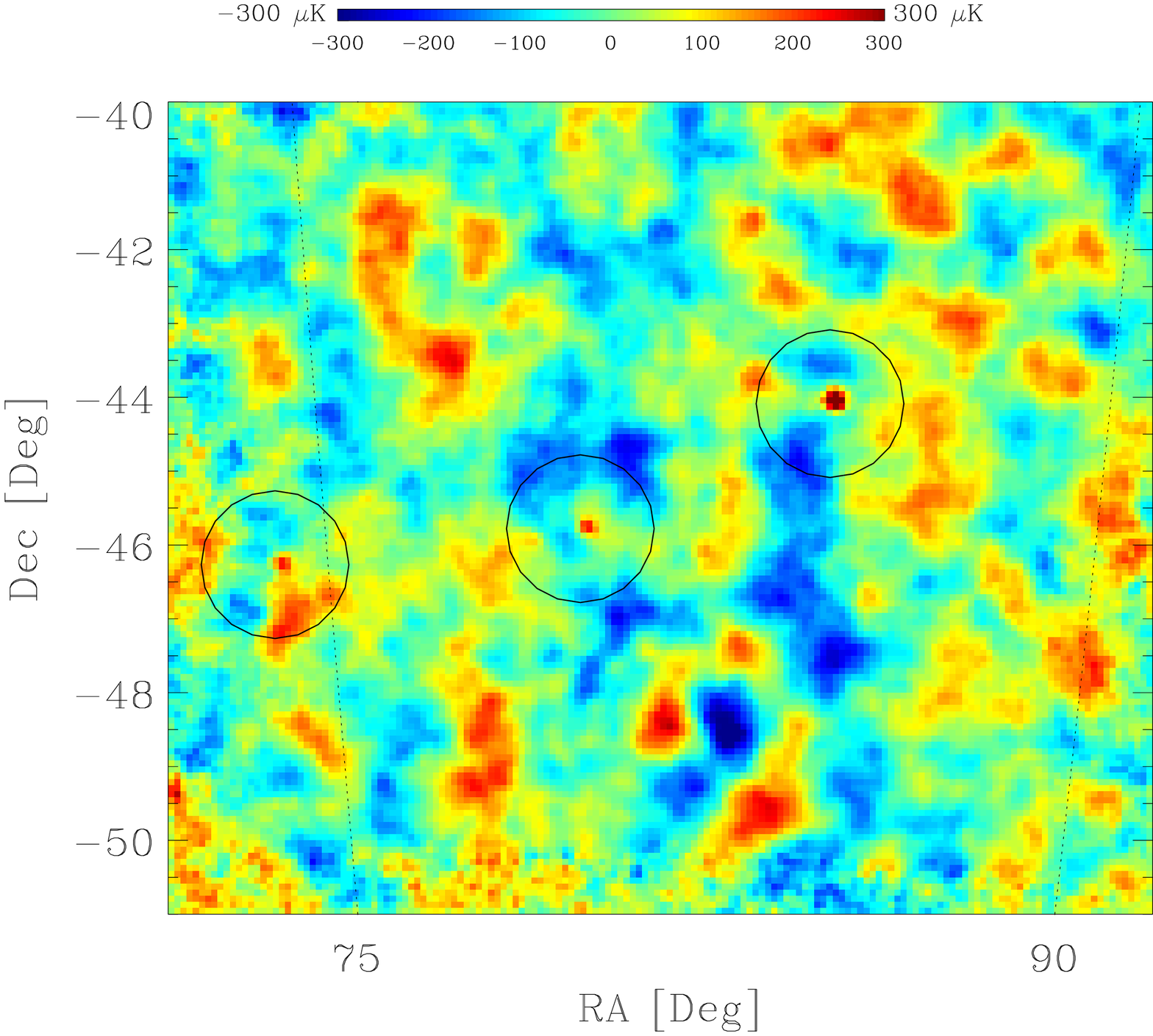}
\includegraphics[angle=0,width=5cm]{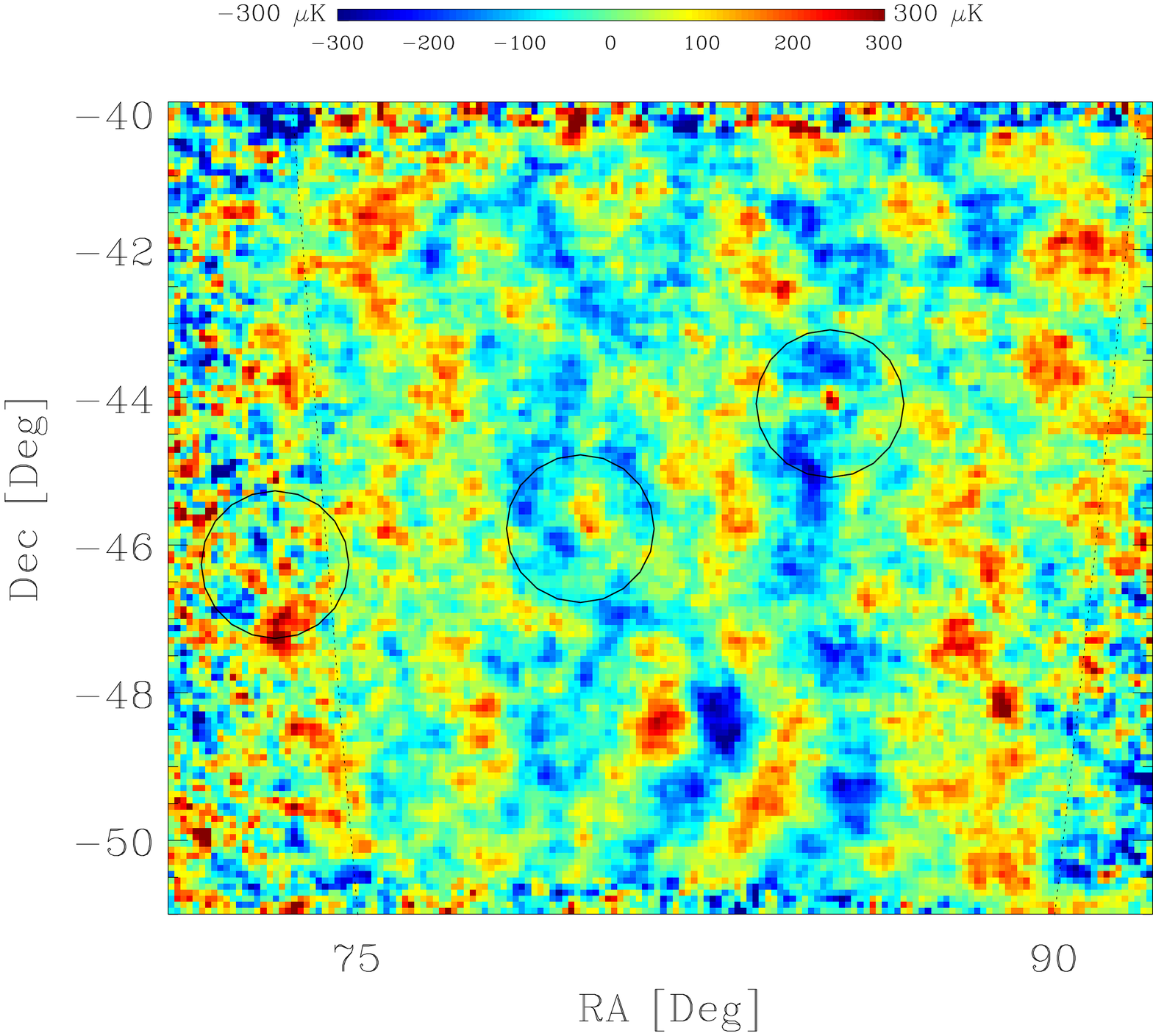}
\includegraphics[angle=0,width=5cm]{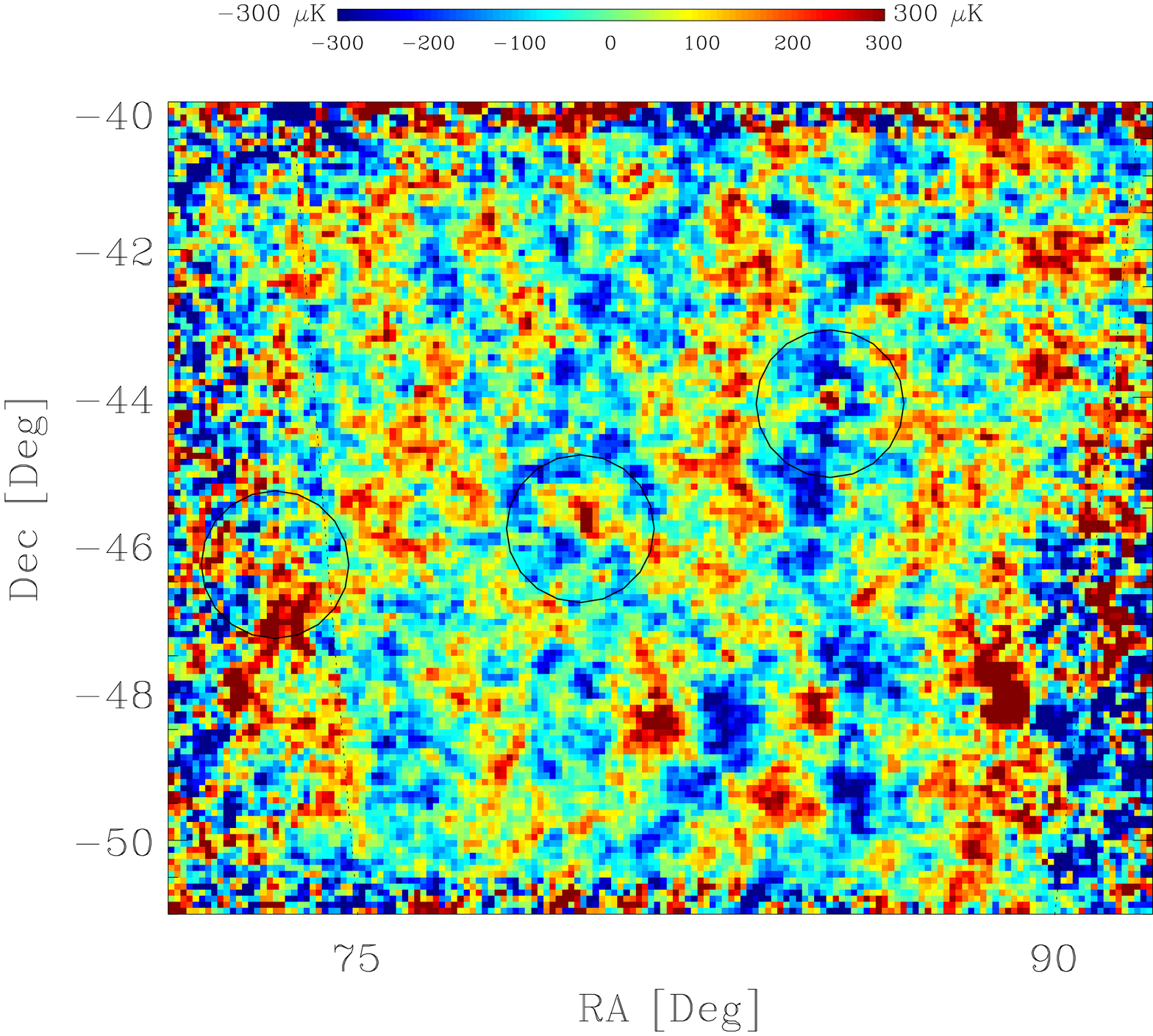}
\includegraphics[angle=0,width=5cm]{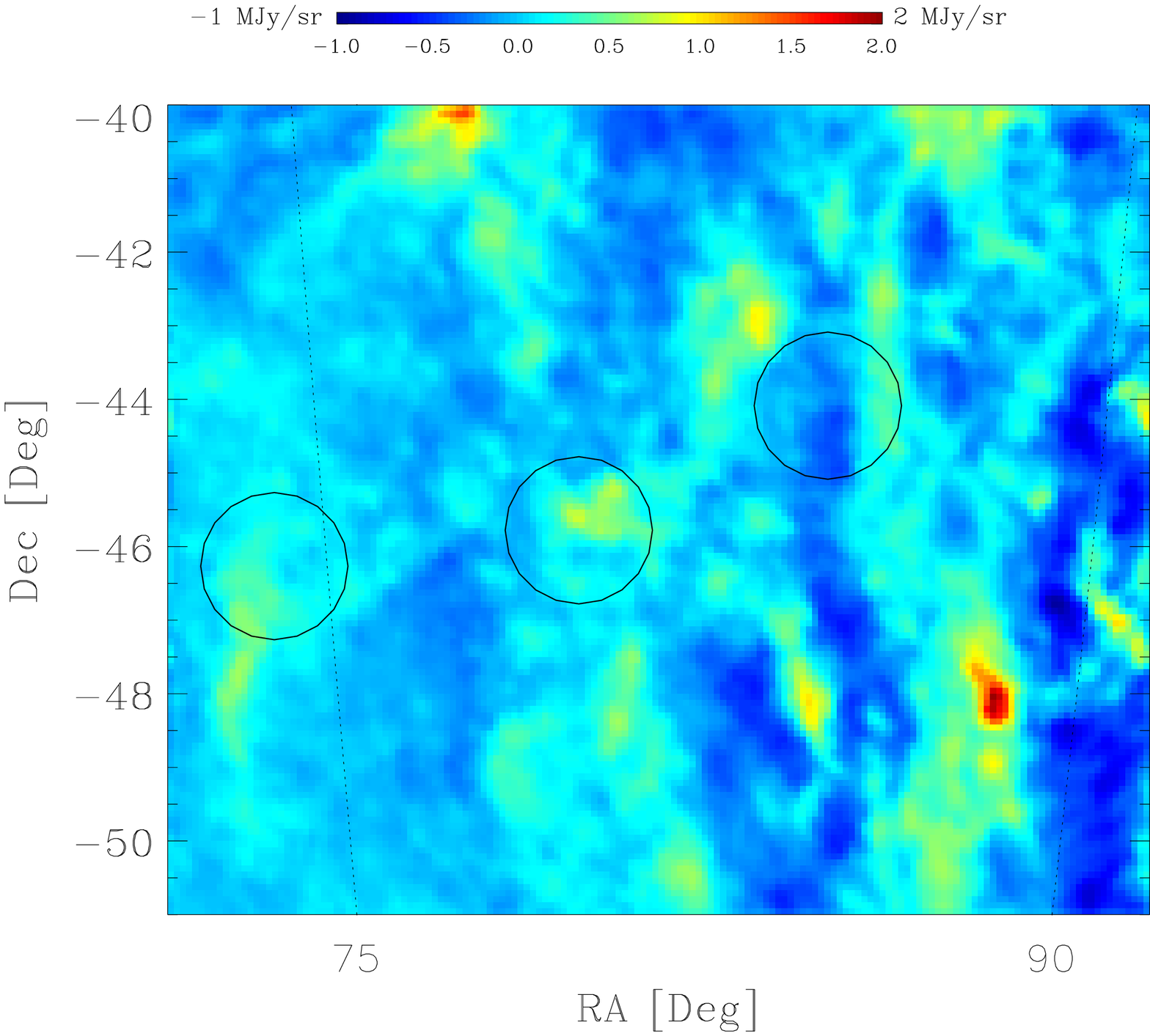}
\includegraphics[angle=0,width=5cm]{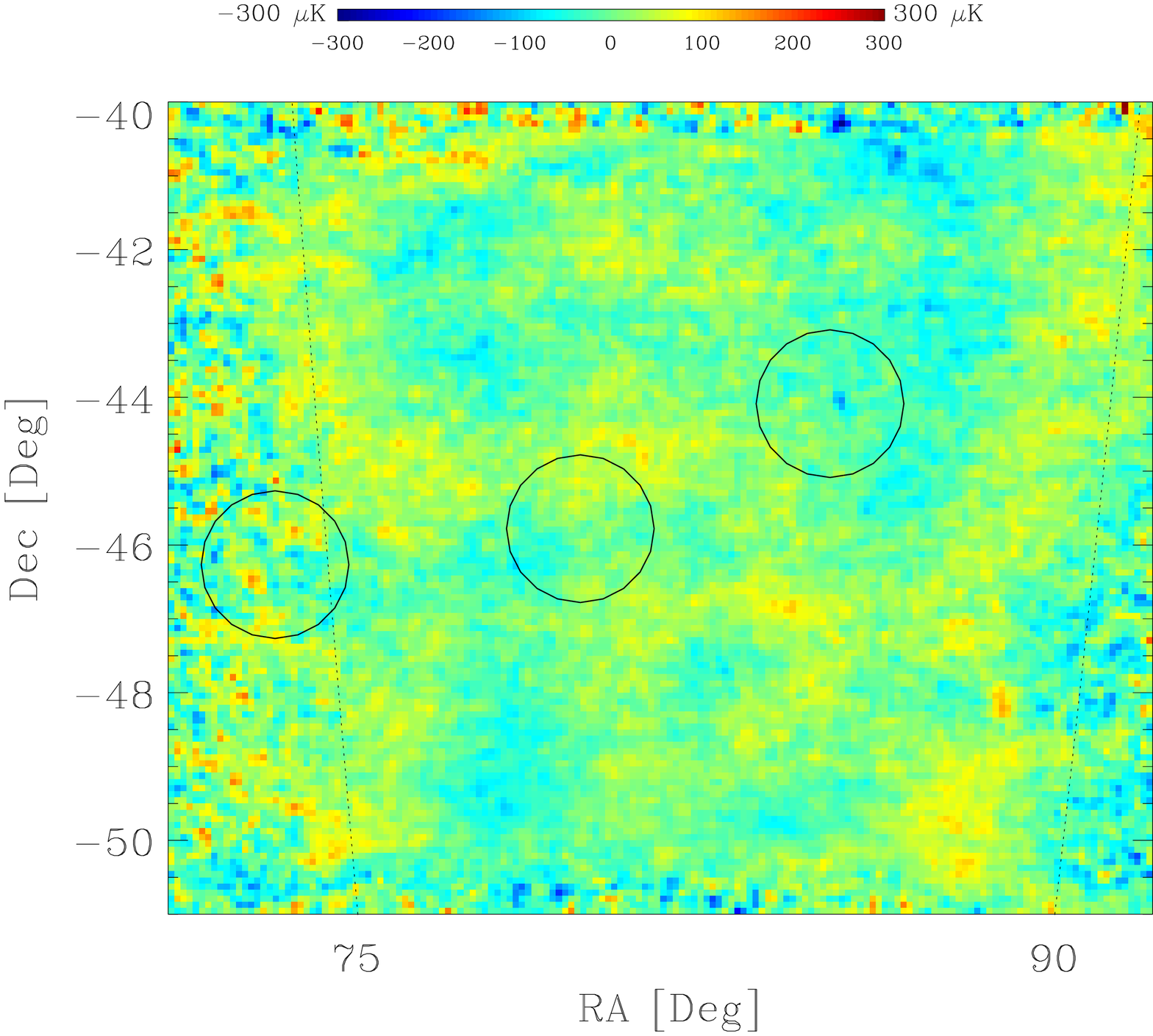}
\includegraphics[angle=0,width=5cm]{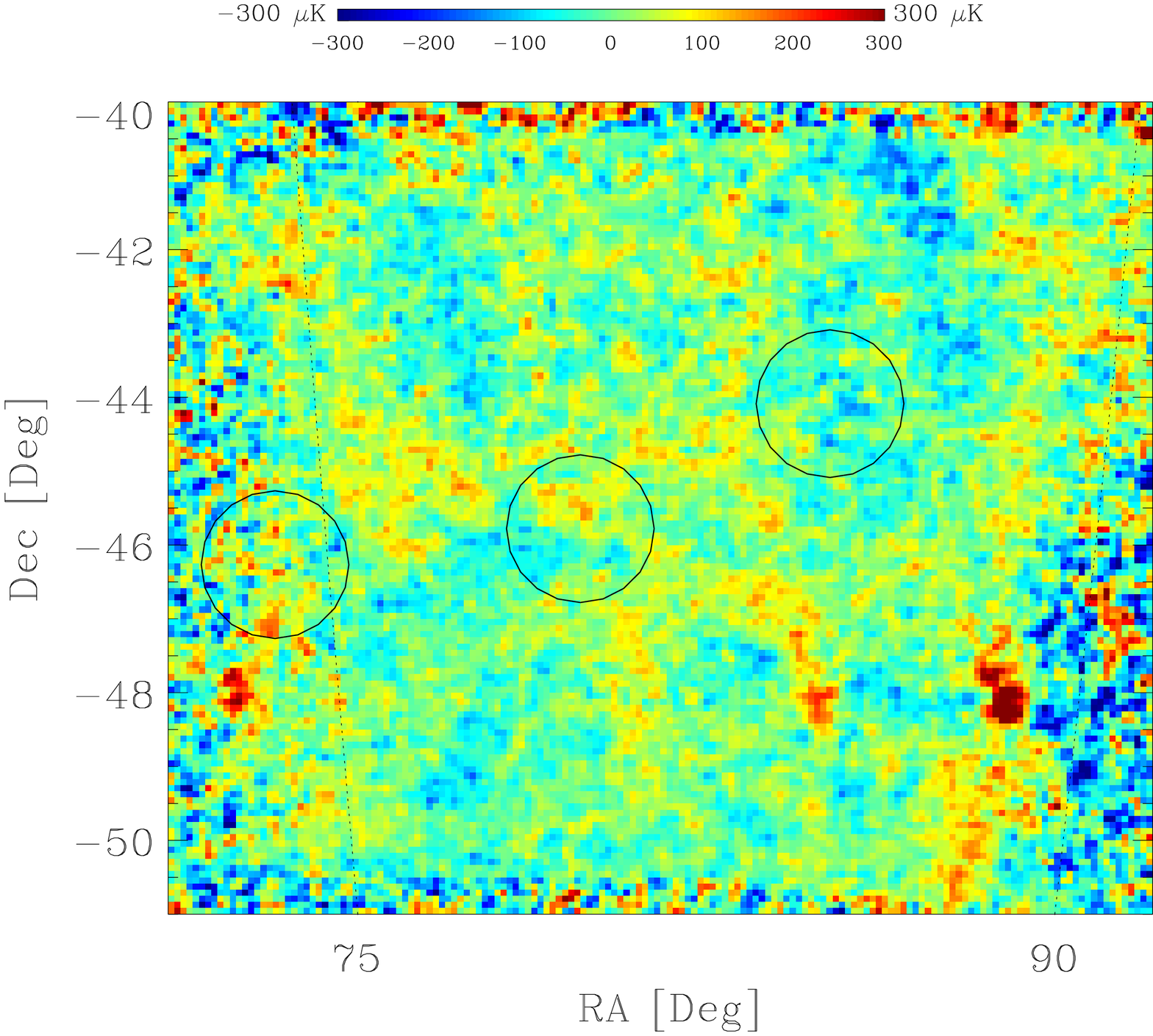}

 \caption{The deep survey of \bk.
Top row: I maps from \bk. From left to right: 145 GHz; 245 GHz;
345 GHz. I is encoded in the false color scale in thermodynamic
temperature units for a 2.725K blackbody. The pixel size is 3.4
arcmin. Bottom row, from left to right: IRAS/DIRBE image of the
same region at 3000 GHz; \bk difference map obtained by
subtracting the 145 GHz map from the 245 GHz map, in order to
remove CMB anisotropy from the resulting map; \bk difference map
obtained as 345 GHz map minus 145 GHz map. All the maps have been
filtered in the same way. To remove large-scale gradients, a
high-pass filter at 7.5 times the scan frequency was applied on
the time-ordered data: this filter is needed to remove artifacts
due to scan synchronous noise in the 245 and 345 GHz maps. A
gaussian filter with 7$^\prime$ FWHM was used as anti-aliasing.
\label{fig:deep_i}}

\end{center}
\end{figure}

Maps of the shallow and deep regions at 145, 245 and 345 GHz are
compared in Figs \ref{fig:shallow_i} and \ref{fig:deep_i}
respectively. For a meaningful comparison, all maps have been
filtered in the same way. Aggressive high-pass filtering of the
time ordered data (cut-on at 7.5 times the scan frequency) was
required for the 245 GHz and 345 GHz bolometers, in order to avoid
artifacts due to scan synchronous noise. The brightness at 245 and
345 is due to a mixture of CMB and emission from interstellar dust
(ISD), as evident from the comparison of the CMB-subtracted maps
to the IRAS 3000 GHz map in the botton rows of Fig.
\ref{fig:shallow_i}.

\begin{figure}[p]
\begin{center}
\includegraphics[angle=0,width=5cm]{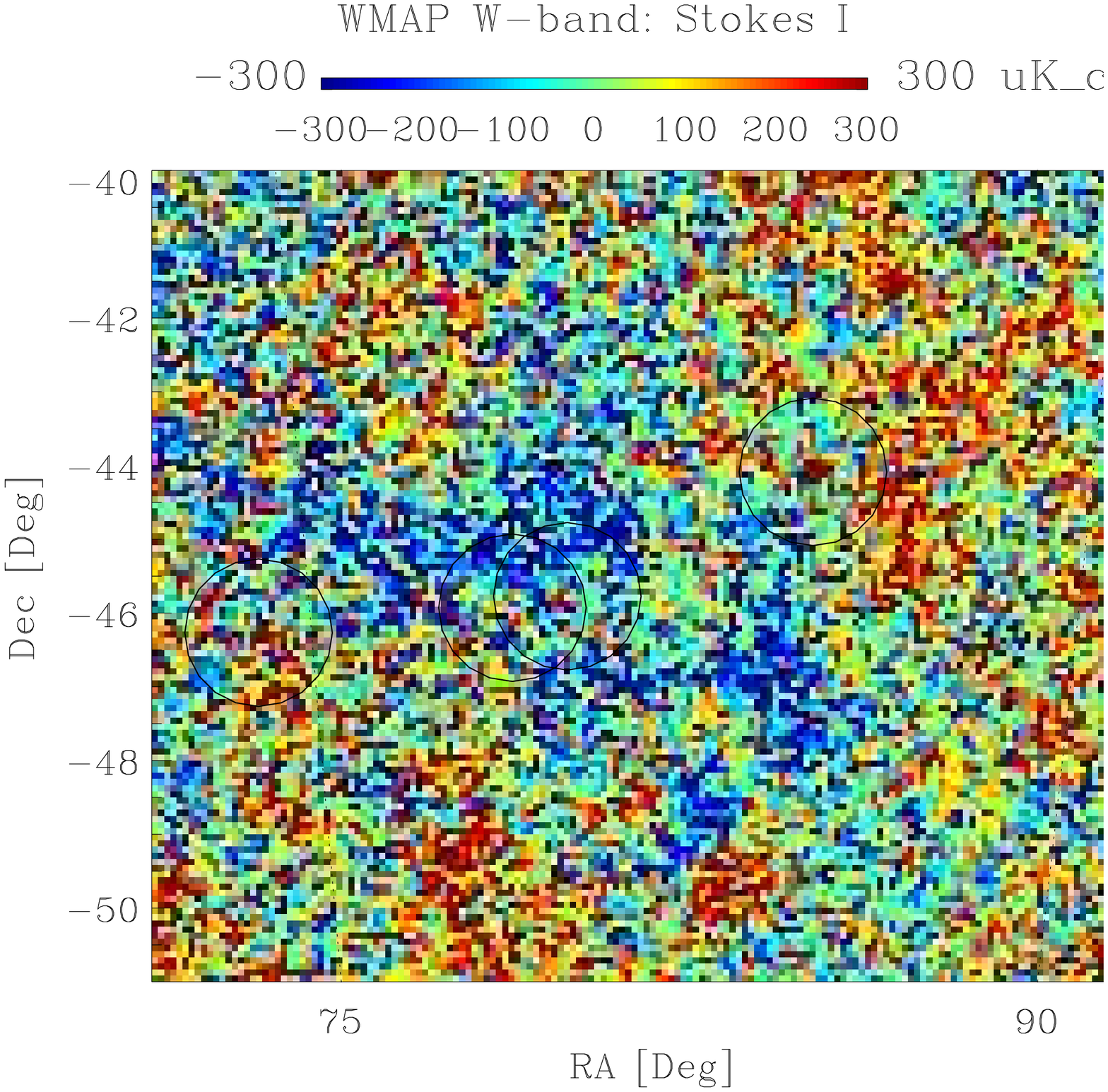}
\includegraphics[angle=0,width=5cm]{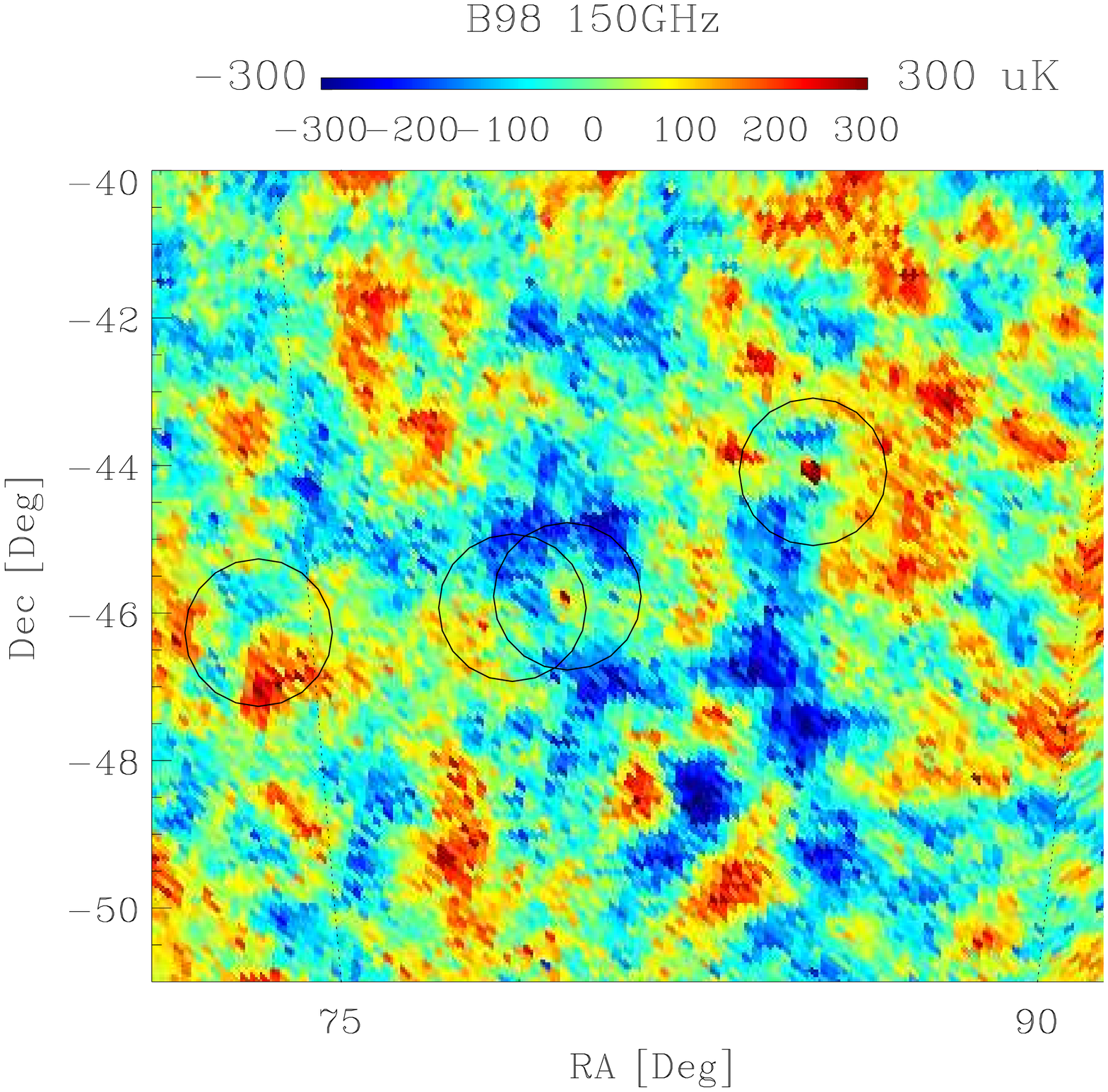}
\includegraphics[angle=0,width=5cm]{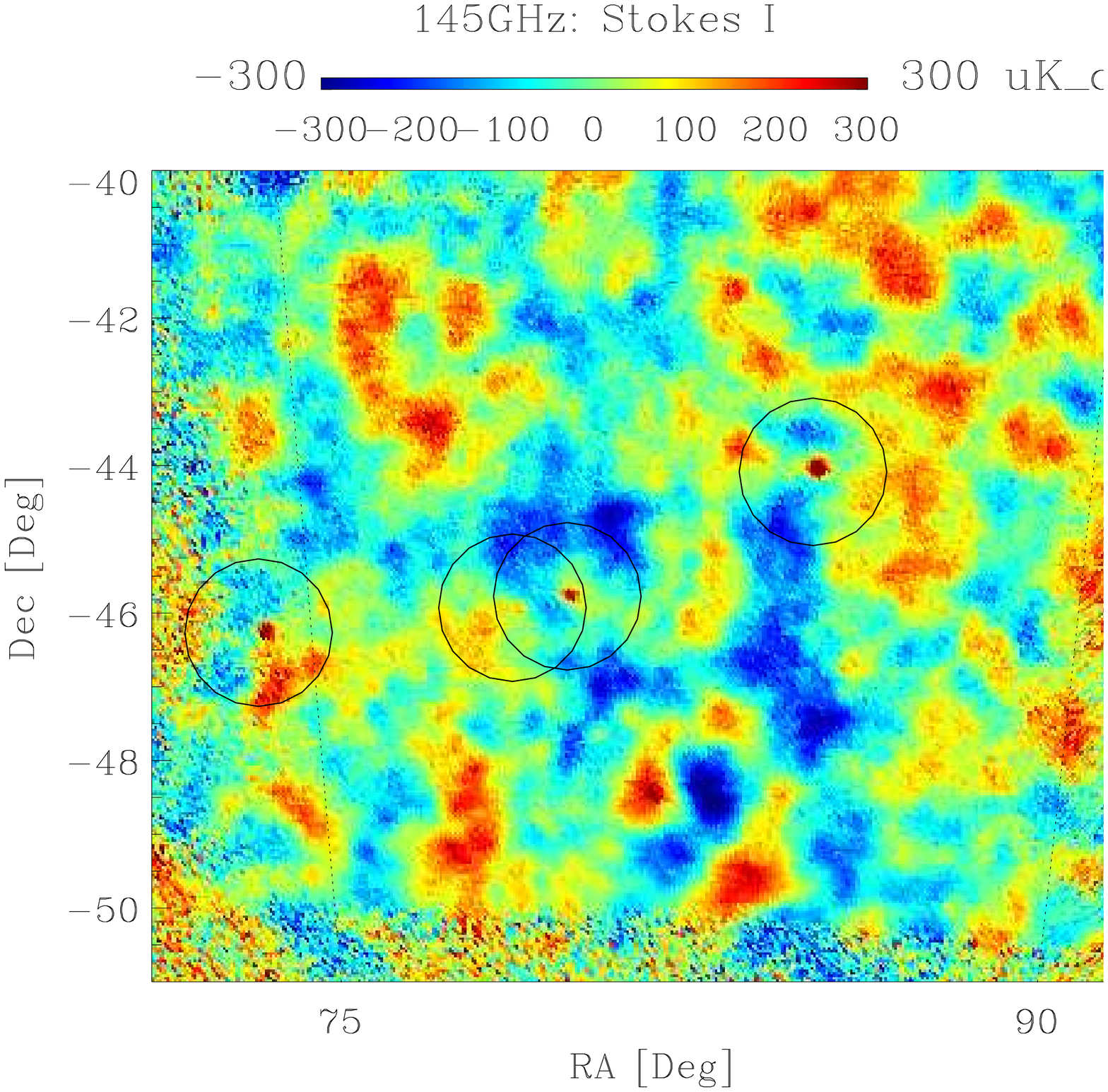}
\caption{Comparison of maps of the deep survey region from \wmap
94 GHz (left), the \boom 98 data at 145 GHz (center) and the \bk
145 GHz data (right). All maps have 7$^\prime$ pixels. The
improvement in the S/N of \bk is evident. A remarkable morphology
agreement of the structures detected in three independent
experiments is also evident. In all maps I is encoded in the false
colors scale in thermodynamic temperature units for a 2.725K
blackbody. \label{fig:compare}}
\end{center}
\end{figure}

The temperature anisotropy of the CMB is measured with high S/N in
the deep 145 GHz map. Fig. \ref{fig:compare}  compares the 145 GHz
deep map with maps of the same region obtained from \boom-98 at
150 GHz, and from \wmap at at 94 GHz.  There is excellent
morphological agreement between all three maps.

\begin{figure}[p]
\plotone{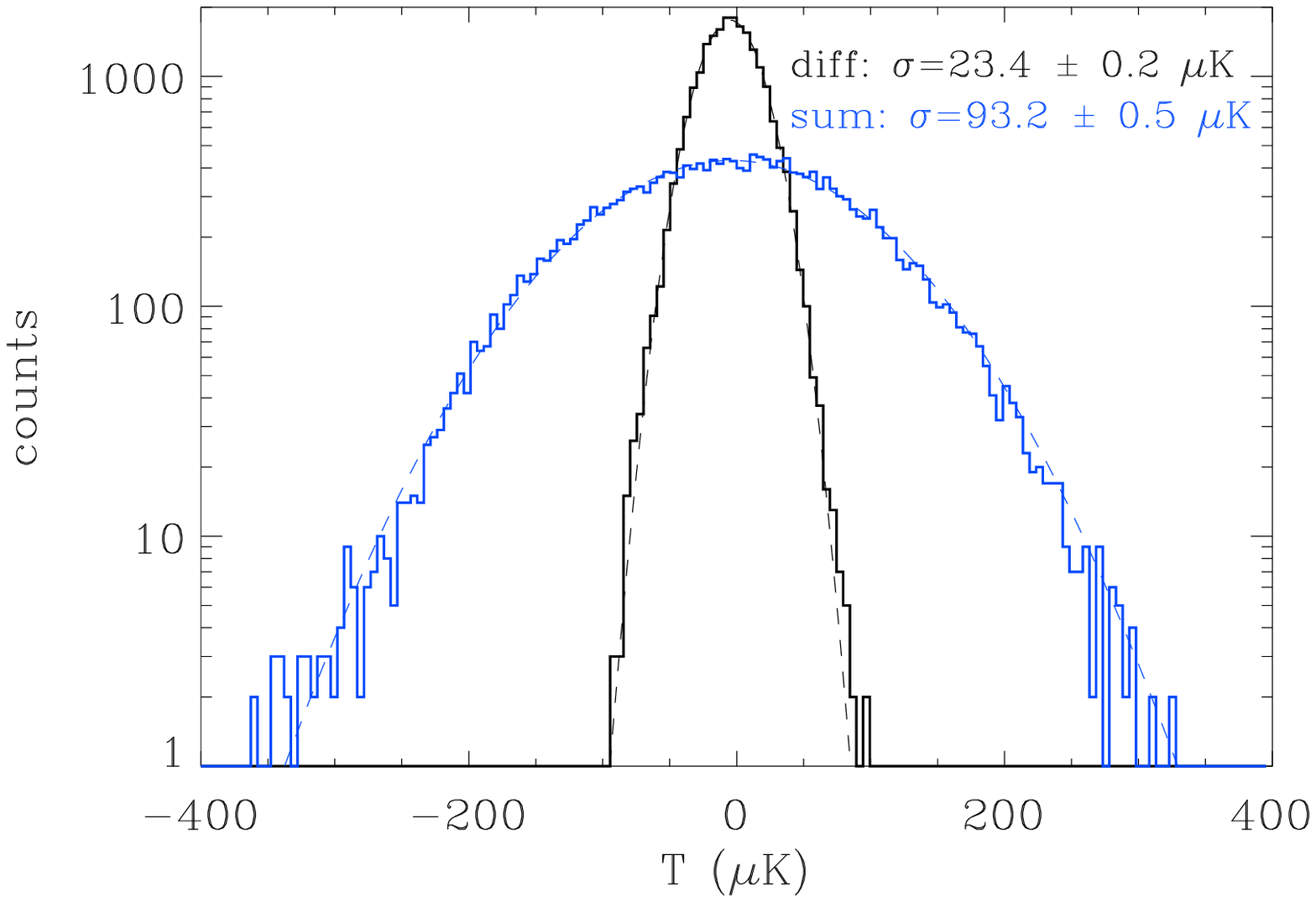} \caption{ Comparison of the 145 GHz CMB I
maps obtained from the two halves of the deep survey. The
pixelization of the map is 3.5 arcmin (HEALPIX NSIDE=1024). We
histogram pixel temperatures resulting from the difference and sum
maps. Only noise and systematic effects contribute to the former,
while both signal and noise contribute to the latter. The two
curves are labelled with their standard deviations. The standard
deviation of the signal is thus $\sqrt{93.2^2-23.4^2}=90.2 \mu K$.
\label{fig:jack_T}}
\end{figure}

To estimate more precisely the S/N of the 145 GHz deep map, we
jackknife the data to provide two independent maps: one using data
only from the first half of the deep survey (D1) and one using
data only from the second half of the deep survey (D2). We then
form the sum and difference maps, (D1+D2)/2 and (D1-D2)/2. The sum
map contains both signal and noise, the difference map only noise.
For all maps we have used HEALPIX NSIDE=1024, i.e. a pixel side of
3.5'. Histograms of both maps, shown in Fig. \ref{fig:jack_T}, are
gaussian-distributed. The rms in the sum and difference maps is
$\sigma_+ = (93.2 \pm 0.5) \mu K$ and $\sigma_- = (23.4 \pm 0.2)
\mu K$, respectively. The rms of the sky signal is thus
$\sqrt{93.2^2-23.4^2}=(90.2 \pm 2.3) \mu K$, where the error now
includes the uncertainty in absolute calibration. Realistic
simulations of our observations, including all the details of scan
speed, coverage, detectors noise, filtering, pixelization, etc.
can be used to estimate the expected rms of CMB anisotropy in
these observations, for the standard "concordance" model best
fitting WMAP. We used the same simulations used for our spectral
analysis \cite[]{jones05,piacentini05,montroy05}. The result is
$(94 \pm 13) \mu K$ (including cosmic variance). A similar result
is obtained integrating the power spectrum of the "concordance"
model over the B03 window function shown in Fig.
\ref{fig:beam_ell}.

\begin{figure}[p]
\begin{center}
\includegraphics[width=10cm]{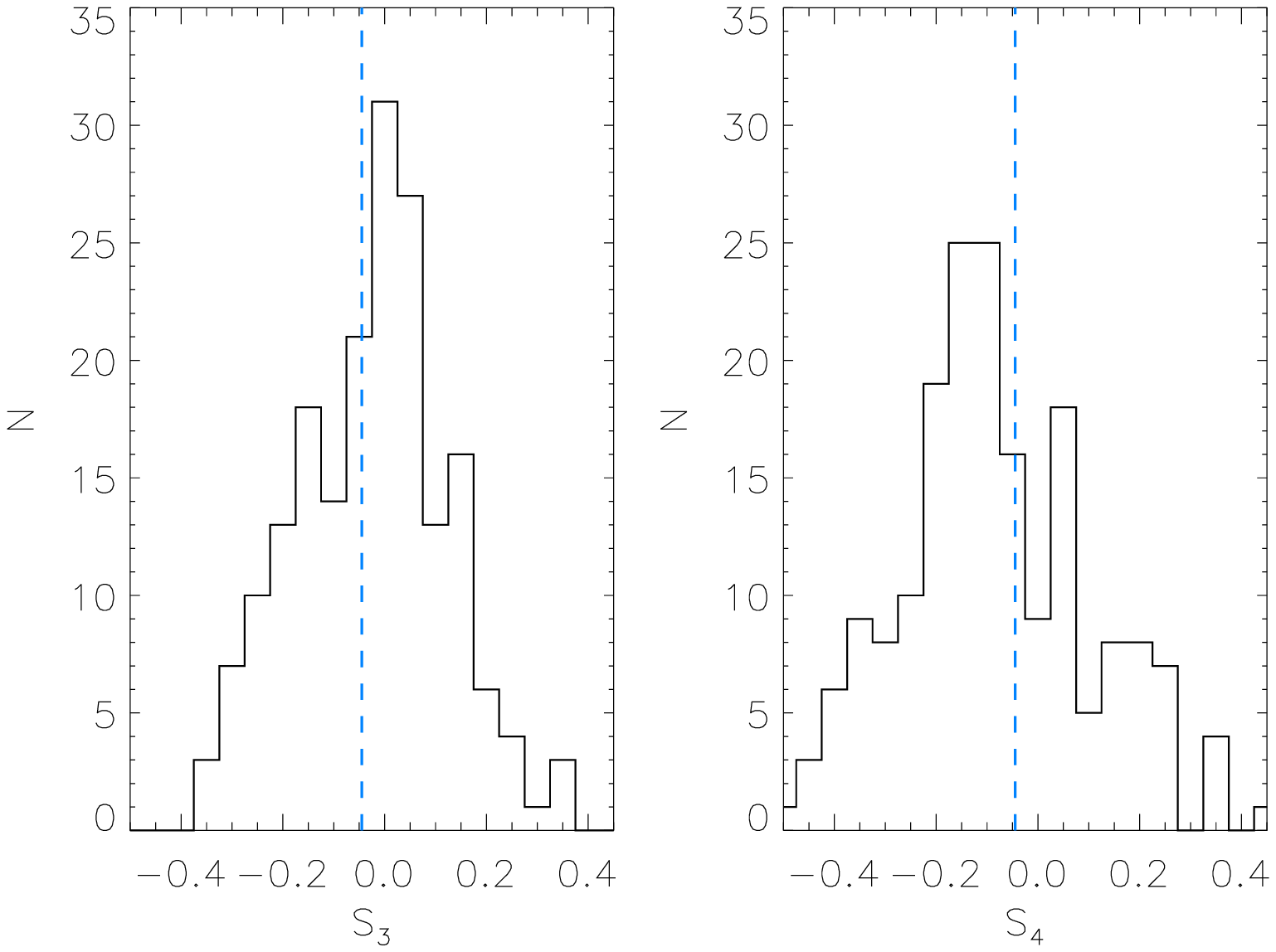}
\caption{The vertical lines represent the measured skewness (left
panel) and kurtosis (right panel) of the pixel temperature
distribution in the 145 GHz deep survey map. The histograms derive
from realistic simulations of the measurement, assuming an
underlying (gaussian) "concordance" model. \label{fig:s3s4}}
\end{center}
\end{figure}

The distribution of the measured brightness in the deep 145 GHz
map is accurately gaussian. The simplest gaussianity test is the
evaluation of the skewness $S_3$ and kurtosis $S_4$ of the pixel
temperature distribution, computed as
\begin{equation}
S_3= {1 \over N} \sum_{i=1}^N \left[ { T_i - \langle T \rangle
\over \sigma_T} \right]^3 ~~~~~ ; ~~~~~~ S_4= {1 \over N}
\sum_{i=1}^N \left[ {T_i - \langle T \rangle  \over \sigma_T}
\right]^4-3
\end{equation}
In Fig. \ref{fig:s3s4} we compare the measured skewness
($S_3=-0.062$) and kurtosis ($S_4=-0.053$) to the distribution of
$S_3$ and $S_4$ evaluated from our realistic simulations set. The
agreement is very good: no hints for any deviation from
gaussianity. This also confirms that systematic effects and
foreground contamination are negligible in this map.

The CMB structures mapped at 145 GHz are evident also in the 245
and 345 GHz maps (Fig. \ref{fig:deep_i}), even if with lower S/N,
due to the higher background on the bolometers, to the lower
number of detectors in these bands, and to the presence of
increasing contamination from interstellar dust emission.

\subsubsection{Foreground Contamination by Diffuse ISD Emission at
High Latitude}

Our multiband maps allow us to carry out basic tests on the level
of foreground contamination. We consider first the deep survey
(carried out at Galactic latitudes $|b| > 25^o$). We use the SFD
IRAS map $D_{i,j}$  at 100 $\mu m$, corrected using DIRBE data
\cite[]{schlegel99}, as a template for ISD emission. We use the
\wmap 94 GHz map $W_{i,j}$ as a template for CMB anisotropy. Both
maps have been sampled along the scans of the \bk, and then
high-pass and low-pass filtered using the transfer function of the
145 GHz \bk detectors, in order to create the corresponding
synthesized time-streams. These have been processed in the same
way as the \bk detectors (see \S \ref{subs:maps}), to obtain the
maps shown in Figs. \ref{fig:shallow_i},  \ref{fig:deep_i} and
\ref{fig:compare}. For this particular analysis we use a 14'
pixelization, in order to avoid problems with the different beam
size of the instruments, and carry out a linear fit in pixel
space:
\begin{equation}
I_{i,j}=A W_{i,j}+B D_{i,j} + C \label{eq:bestfit}
\end{equation}
where $I_{i,j}$ is one of the three \bk maps (145, 245, 345 GHz).
The results of the fits are reported in table
\ref{tab:fitwmapiras}. The errors have been estimated by dividing
the deep survey region into nine sub-regions, performing the fit
for each subregion, and computing the standard deviation of the
best fit parameters. In this way we account for variations in the
properties of the foreground over the survey region. The $A$
values are not reported since they are biased low, due to the
relatively high noise of the WMAP map. Simulations show that, with
the noise levels present here, this problem does not bias the
estimates of the $B$ values by more than a few \%. This error is
added in quadrature to the statistical error, which dominates the
total error.

\begin{table}[p]
\begin{center}
\begin{tabular}{|c|c|c|c|c|}
\hline
Frequency & $R(A)$ & $R(B)$  & $B[\mu K/(MJy/sr)]$ & $ \langle \Delta T_{dust}^2 \rangle^{1/2} (\mu K) $ \\
\hline
145 GHz & 0.634 &  0.074 & $(18 \pm 23) $ & $< 10 $        \\
245 GHz & 0.452 &  0.173 & $(49 \pm 21) $ & $12 \pm 5$   \\
345 GHz & 0.331 &  0.452 & $(202 \pm 18)$ & $50 \pm 5$   \\
\hline
145 GHz & 0.588 & -0.085 & $(-5 \pm 9)  $ & $< 5$        \\
245 GHz & 0.273 &  0.143 & $(37 \pm 11)  $ & $21 \pm 6$    \\
345 GHz & 0.123 &  0.380 & $(183 \pm 24)$ & $ 104 \pm 14 $\\
\hline
\end{tabular}
\end{center}
\begin{center}
\caption{\small Pixel-space fits of the \bk maps with a
combination of a CMB template (from \wmap) and a dust template
(from IRAS), see eq. \ref{eq:bestfit}. For the three upper rows,
the fit has been performed in the deep survey region ( $74^o < RA
< 90^o$; $-50^o < dec < -40^o$ , $\langle |b| \rangle \sim 32^o$,
for a total of 2172 15' pixels). In this region, the rms
fluctuation of the IRAS/DIRBE map is 0.25 MJy/sr. For the three
lower rows, the fit has been done in the part of the Shallow
Survey closer to the Galactic Plane ($90^o < RA < 105^o$; $-50^o <
dec < -40^o$, $\langle |b| \rangle \sim 23^o$, 2027 pixels of
15'). Here the rms fluctuation of the IRAS/DIRBE map is 0.57
MJy/sr. $R(A)$ is the correlation coefficient for the CMB
template, while $R(B)$ is the correlation coefficient for the dust
template. The last column gives the estimated brightness
fluctuation due to the ISD component correlated to the IRAS/DIRBE
map, in CMB temperature units. \label{tab:fitwmapiras} }
\end{center}
\end{table}

The fact that the correlation coefficient for the WMAP template is
always significant means that we have detected CMB anisotropy in
all of our channels. The correlation coefficient for the IRAS
template, instead, is statistically significant for the 245 and
for the 345 GHz survey. This means that the best fit values for
$B$ at 145 GHz provide only an upper limit for ISD contamination.
As an example of a region where ISD fluctuations are larger, we
can consider the part of the Shallow Survey closer to the Galactic
Plane: $90^o < RA < 105^o$; $-50^o < dec < -40^o$. Proceeding as
before, in this region we obtain the results reported in the lower
part of table \ref{tab:fitwmapiras}, which confirm the results
found for the deep survey.

We have carried out an analysis similar to the one just described,
using the difference maps $M_{345}=I_{345}-I_{145}$ and
$M_{245}=I_{245}-I_{145}$ as "CMB-subtracted" maps. Since our 145
GHz map is dominated by CMB anisotropy and features very low
noise, $M_{345}$ and $M_{245}$ monitor all non-CMB brightness, and
have lower noise than what we can obtain subtracting the 94 GHz
WMAP map as a CMB template. The fit of equation
\begin{equation}
M_{i,j}=B D_{i,j} + C \label{eq:bestfit2}
\end{equation}
gives results very similar to the fits to eq. \ref{eq:bestfit}
reported in table \ref{tab:fitwmapiras}.

The rms fluctuation of the dust template in the deep survey region
is about 0.25 $MJy/sr$, so we get $\Delta T_{dust,rms}=(12 \pm 5)
\mu K_{CMB}$ at 245 GHz and $\Delta T_{dust,rms}=(50 \pm 5) \mu
K_{CMB}$ at 345 GHz. The upper limit at 145 GHz is $\Delta
T_{dust,rms} < 10 \mu K_{CMB}$ (1-$\sigma$ u.l.).

\begin{figure}[p]
\plotone{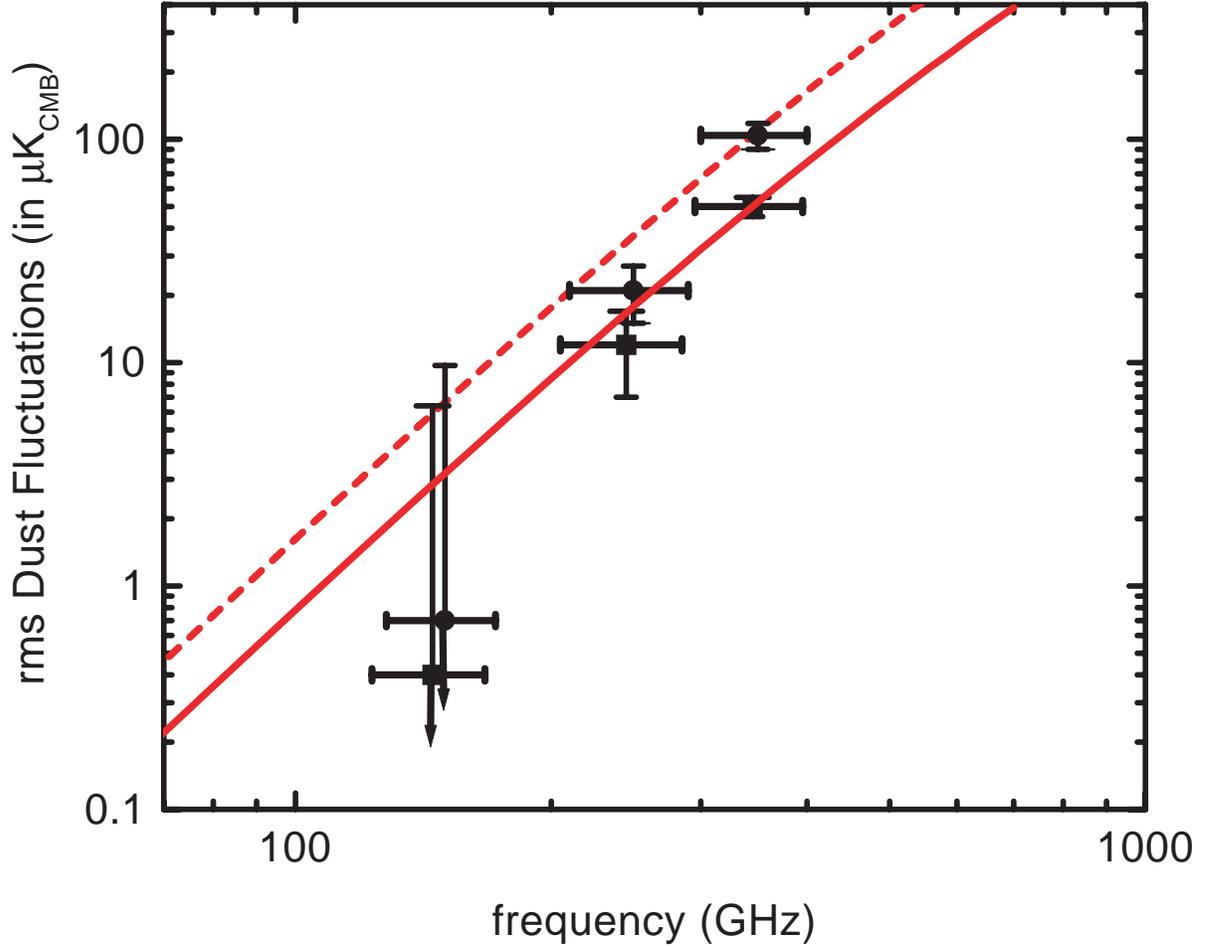} \caption{Data Points:
IRAS-correlated dust fluctuations detected by \bk: the squares
refer to the Deep Survey region ($74^o < RA < 90^o$; $-50^o < dec
< -40^o$); the circles refer to the part of the Shallow Survey
closer to the Galactic Plane ($90^o < RA < 105^o$; $-50^o < dec <
-40^o$, 2027 pixels of 15'). The continuous line is Model 8 of
\cite{finkbeiner99}, normalized at 345 GHz for the Deep Region.
The dashed line is the same model, scaled to the shallow region
via the ratio between the IRAS/DIRBE rms in the two regions.
\label{fig:spectrumdust}}
\end{figure}

\cite{finkbeiner99} have studied the extrapolation to longer
wavelengths of dust fluctuations detected by IRAS . Our results
are consistent with their extrapolations. Their model 8 for the
specific brightness of ISD is :
\begin{equation}
B_{dust}(\nu) \propto q f_1 \nu^{\alpha_1} B(\nu, T_1) + (1-f_1)
\nu^{\alpha_2} B(\nu, T_2)
\end{equation}
with $\alpha_1 = 1.67$, $T_1 =9.4 K$, $\alpha_2 = 2.7$, $T_2 =16.2
K$, $f_1=0.0363$, $q=13$. The rms of the extrapolated maps in the
deep survey region are 2, 11 and 47 $\mu K_{CMB}$ in the 145, 245
and 345 GHz bands respectively. So this model fits quite well our
measurements (see also Fig. \ref{fig:spectrumdust}).

With no detection at 145 GHz, we use the spectrum of Model 8 to
extrapolate - with all the necessary caveats - to 145 GHz.
Integrating in our bands (specified in Fig. \ref{fig:specplot}),
and normalizing to our 345 GHz point, we get  $\Delta
T_{dust,rms}=(2.9 \pm 1.2) \mu K_{CMB}$ at 145 GHz. This value is
much smaller than the CMB anisotropy at the same frequency (the
total sky rms at 145 GHz in our window function is $\sim 90 \mu
K$).

The dust fluctuations are non-gaussian. However, only very few
structures deviate significantly from the rms stated above: these
are the clouds visible in the 354-145 difference map at (RA,dec) =
($88.5^o,-48.3^o$) and ($84.6^o,-48.2^o$). Their size is $\simlt
1^o$, and their brightness, extrapolated at 145 GHz with Model 8,
is $\simlt 20 \mu K$.

Since the polarization of the diffuse cirrus at these frequencies
is $\simlt 10 \%$ of the brightness \cite[]{ponthieu05}, we expect
a polarized contribution from ISD $\simlt 1 \mu K$ rms, which is
small with respect to the expected CMB polarization
($\sqrt{\langle EE \rangle }\sim 3 \mu K$ rms in our beam and with
14$^\prime$ pixelization we are using for the analysis in this
paragraph).

The contamination estimates above depend on the assumption that
ISD is well monitored by the IRAS/DIRBE template. They do not take
into account the possible existence of a colder dust component,
undetected by IRAS, and with a different angular distribution.

Studying the residuals from the best fit of equations
\ref{eq:bestfit} and \ref{eq:bestfit2}, we find that the 145 GHz
residuals are consistent with instrumental noise, while there
are, in the 245 and 345 GHz residuals, structures which could be
either instrumental artifacts or real sky fluctuations, or a
combination of the two. The SFD extrapolation to our frequencies
fails to produce these features (the SFD dust temperature is in
fact fairly constant, at 17-18 K, throughout the deep survey
region). These structures are quite dim: the rms is $\sim
90 \mu K$ at 345 GHz and $\sim 70 \mu K$ at 245 GHz. These numbers
include the effect of instrumental noise and should be considered
 conservative upper limits for the sky fluctuations. Their
extrapolation to 145 GHz, using a reasonable dust spectrum like
Model 8, gives $\Delta T_{rms} \simlt 10 \mu K_{CMB}$ at 145 GHz.

Our deep survey region is representative of a fairly large
fraction of the high latitude sky. To see this, we select
increasing fractions of the sky by requiring that dust brightness,
monitored by the IRAS/DIRBE map at 100 $\mu m$, is equal or lower
than a give threshold. We find that the best $\sim 40 \%$ of the
sky has the same rms fluctuation of dust brightness as in our deep
survey, and that the best 75 $\%$ of the sky has a brightness
fluctuation $\simlt 3$ times larger than the one in our deep
survey.

\subsubsection{Foreground Contamination by Compact Sources}

A few point sources are evident in the I maps. These AGNs have
been used for testing the pointing reconstruction procedures as
explained in \S \ref{subs:attrec} . The effect of the full
population of resolved and unresolved AGNs as a contaminant in CMB
anisotropy measurements is an important topic of discussion, in
view of the forecasted ultra-sensitive surveys of the CMB
anisotropy and polarization. The SED of Blazars is almost flat
($S(\nu) \sim \nu^{-a}$ with $0 \simlt a \simlt 0.3$ between few
GHz and 100 GHz). At our three frequencies, bracketing the
frequency of maximum brightness of CMB fluctuations, the
contamination of the AGNs is thus expected to be minimal.

\cite{giommi04} have searched AGN catalogs for all the sources in
the region observed by \boom . From their list of 54 sources, we
find 8 AGNs in the deep survey region. Once we exclude the three
brighter ones (which are evident in the map) their equivalent CMB
temperature in a 9.5' beam is $< 400 \mu K$ at 145 GHz. Since the
same region is covered by about 5000 independent beams, it is
evident that the presence of mm AGN emission cannot contaminate
the morphology of the I maps. The integrated flux $\langle S
\rangle$ and its fluctuations $\langle (\Delta S)^2 \rangle $
produced by all resolved and unresolved AGNs with differential
LogN-LogS distribution ${dN \over dS}$ are given by
\begin{eqnarray}
\langle S \rangle  &=& \int_{S_{min}}^{S_{max}} S {dN \over dS} dS \\
\langle (\Delta S)^2 \rangle &=& \int_{S_{min}}^{S_{max}} S^2 {dN
\over dS}  dS
\end{eqnarray}
When detected with an instrument with effective solid angle
$\Omega$, these flux fluctuations produce a rms signal equivalent
to a CMB anisotropy $\Delta T_{AGN}$ given by
\begin{equation}
\langle (\Delta T_{AGN})^2 \rangle  = T_{CMB}^2 {\langle (\Delta
S)^2 \rangle \over \Omega  \left[ {x e^x \over e^x -1 } B(\nu,
T_{CMB}) \right]^2}
\end{equation}
Using the ${dN \over dS}$ and the SED of \cite{giommi04}, we get
$\langle (\Delta S)^2 \rangle \sim $ 30, 25, 21 Jy$^2$/sr at 145,
245, 345 GHz respectively. Using the \bk beams to compute
$\Omega$, we obtain $\sqrt{\langle (\Delta T_{AGN})^2 \rangle}=$
5.8, 4.2, 6.0 $\mu K$ rms at 145, 245, 345 GHz respectively. This
is negligible with respect to the measured anisotropy of the map.

AGN emission is usually polarized at a level $p \simlt 5 \%$.
Assuming random polarization directions and neglecting the natural
dispersion in the $p$ values, we get
\begin{equation}
\langle (\Delta Q_{AGN})^2 \rangle  = {p^2 \over 2} \langle
(\Delta T_{AGN})^2 \rangle
\end{equation}
So we find $\sqrt{\langle (\Delta Q_{AGN})^2 \rangle } = $ 0.20,
0.15, 0.21 $\mu K$ rms at 145, 245, 345 GHz respectively.

This is negligible with respect to the amplitude of the E-mode CMB
polarization that we expect in our maps ($\sim 3 \mu K$ rms for
the 14$^\prime$ pixelization, and $\sim 4 \mu K$ rms for the
3.5$^\prime$ pixelization), and to our statistical noise.  Future
experiments that seek to detect the much smaller B-mode
polarization signal - particularly the lensing signal at small
angular scales - will need to carefully account for the
contribution to the B-mode signal by compact sources.

\subsection{Polarization of deep, high latitude maps at 145 GHz}

\begin{figure}[p]
\begin{center}
\includegraphics[angle=0,width=7cm]{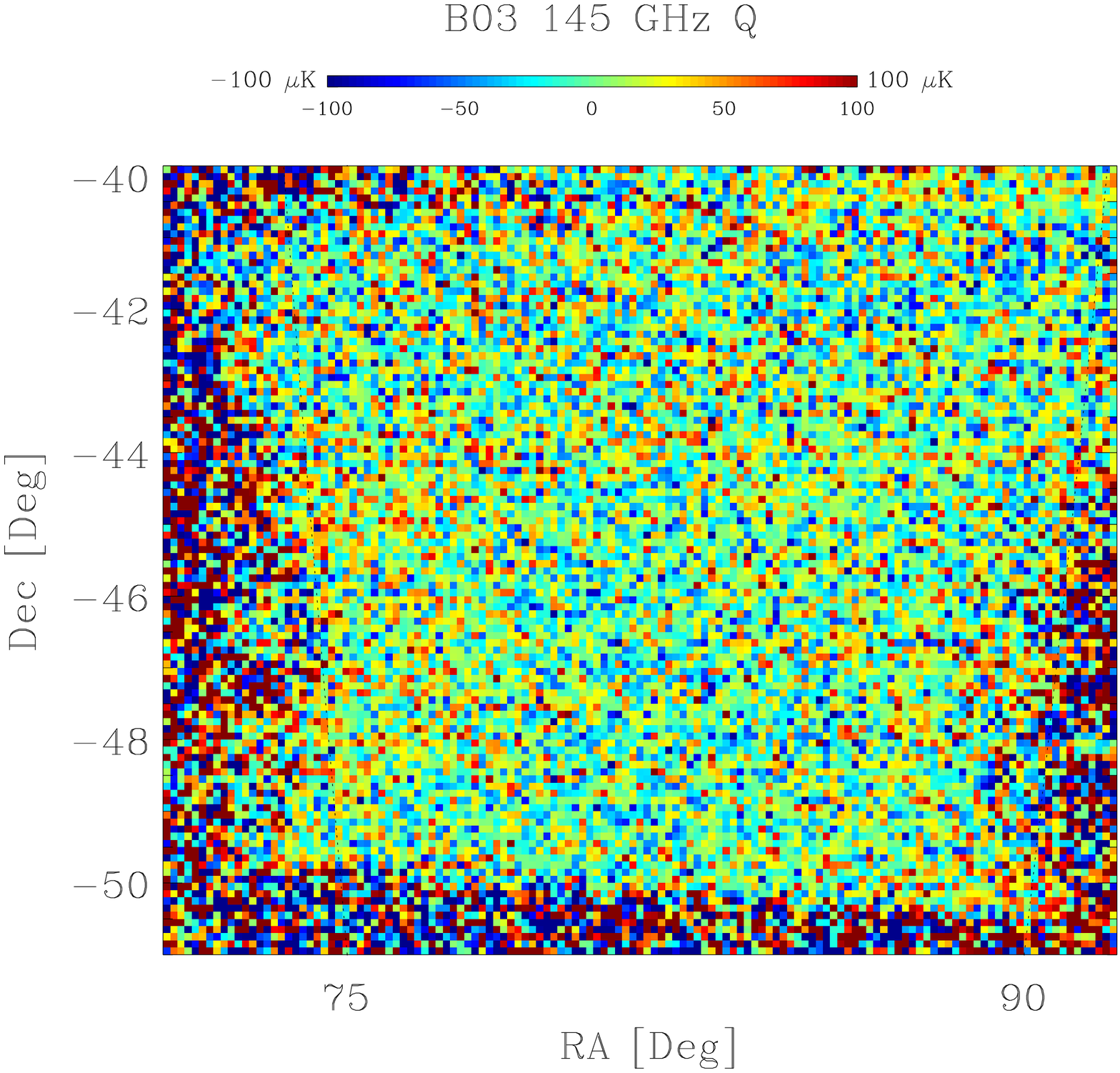}
\includegraphics[angle=0,width=7cm]{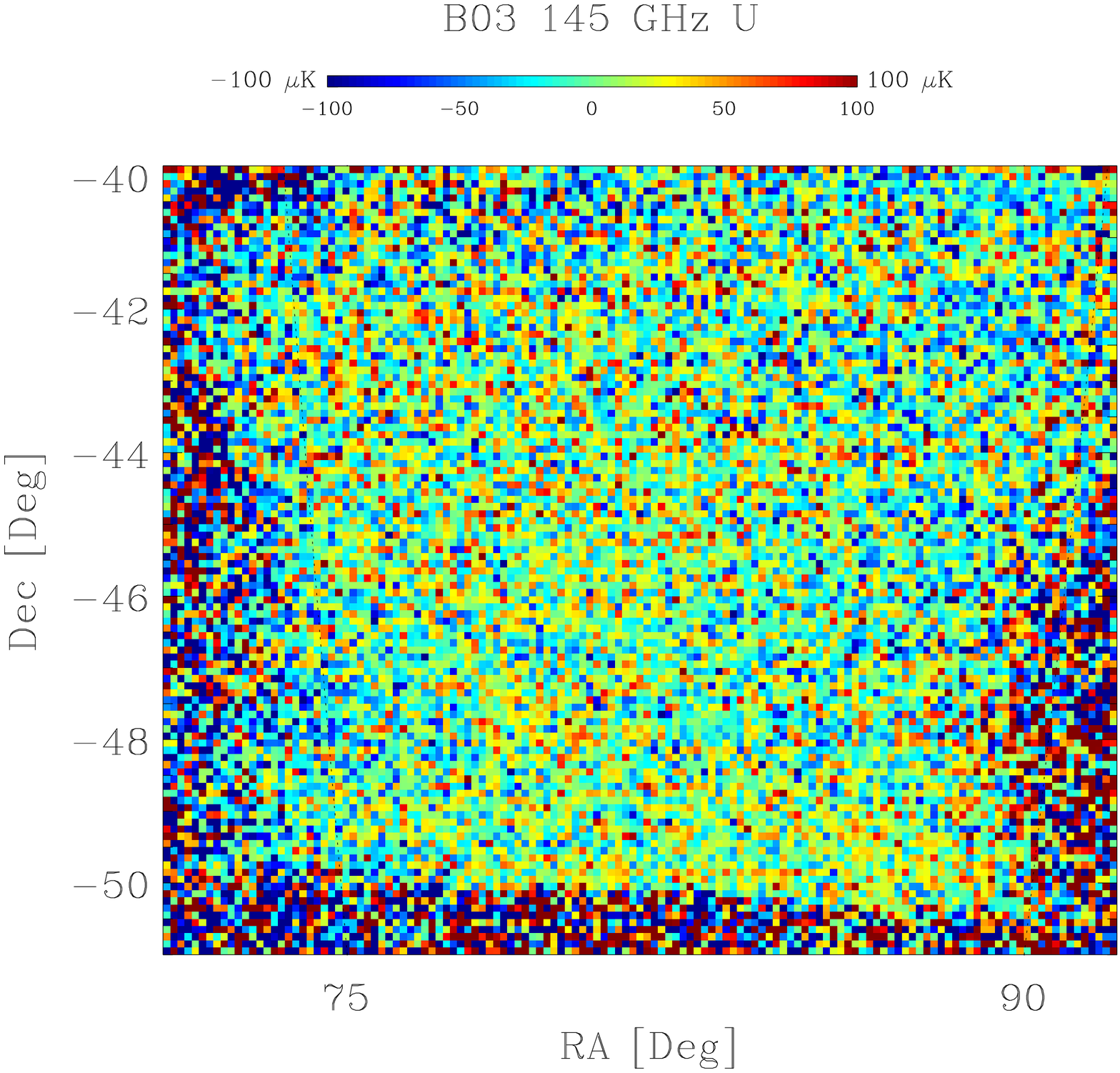}
\caption{Polarization in the deep survey. Left: Q map from the 145
GHz PSB bolometers; right: U map from the same bolometers. Note
that the stretch of the false color scale is a third of the one in
the T maps. These maps are dominated by instrumental noise.
\label{fig:deep_QU}}
\end{center}
\end{figure}

In Fig. \ref{fig:deep_QU} we present maps of the Stokes parameters
Q and U at 145 GHz in the deep survey region.

\begin{figure}[p]
\includegraphics[angle=0,width=13cm]{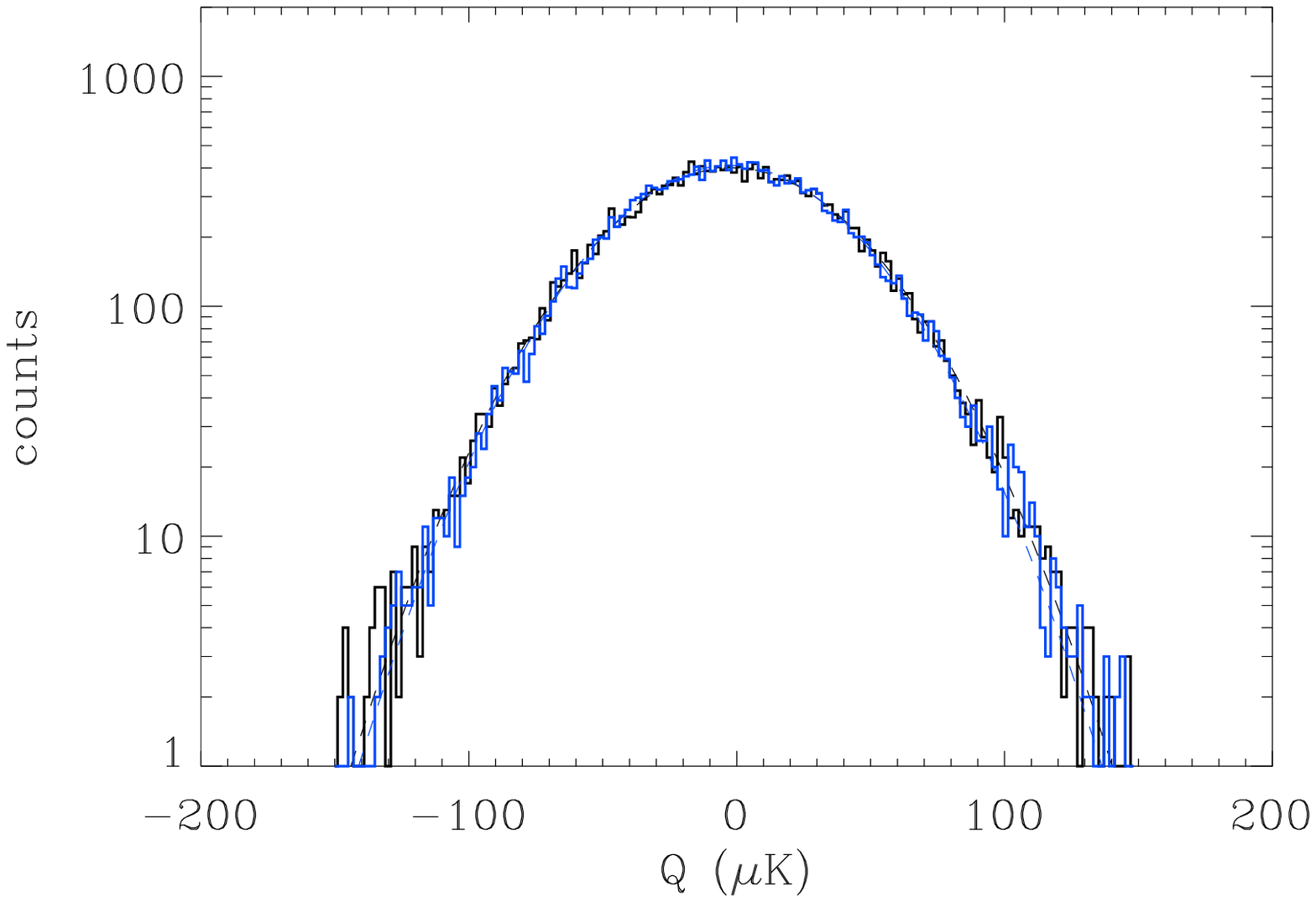}
\includegraphics[angle=0,width=13cm]{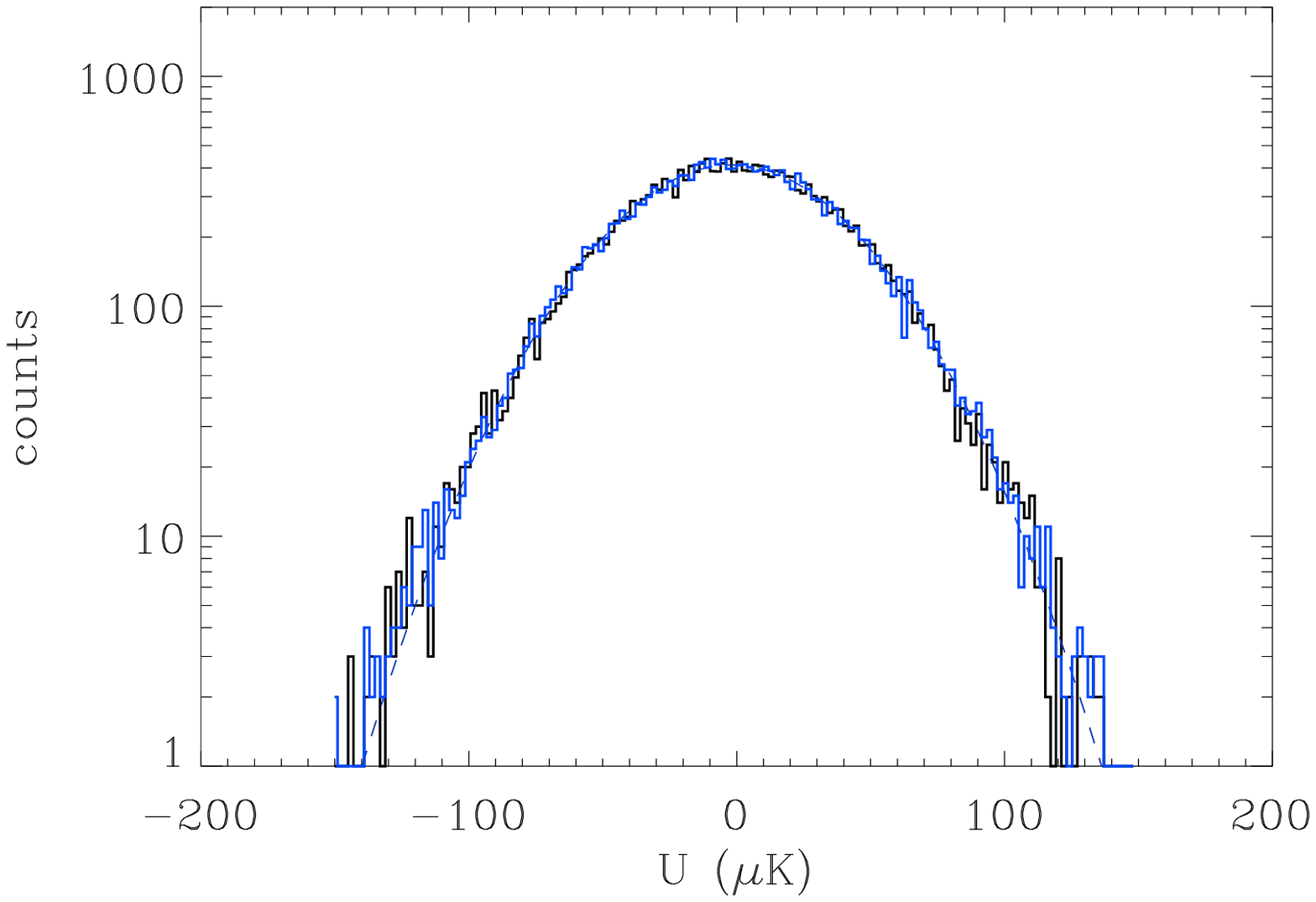}
\caption{ Comparison of the 145 GHz maps of Stokes Q (top) and U
(bottom) taken in the two halves of the deep survey. The
pixelization has NSIDE=1024 (3.5' pixels). We histogram pixel
values resulting from the sum (heavy line) and difference (light
line) maps. The sum and difference histograms are very similar: at
the level visible here there is no indication of systematic
effects that could arise from changes of observation conditions
during the survey. \label{fig:jack_QU}}
\end{figure}

Unlike the I maps, here the CMB signal in each pixel is smaller
than the noise. To estimate the signal to noise ratio (S/N) of the
map, we have produced again two separate 145 GHz maps, one from
the first half (D1) of the observations in the deep region, and
another from the second half (D2). In Fig. \ref{fig:jack_QU} we
compare the histogram of the Q (and U) data from the sum map
(D1+D2)/2 to the same histogram from the difference map (D1-D2)/2.
The difference histogram is very similar to the sum histogram,
confirming that the signal is small with respect to the noise. For
this reason we do not expect any CMB polarization structure to be
visible "by eye" in the Q and U maps. The remarkable agreement of
the sum and difference histograms in Fig. \ref{fig:jack_QU}
indicates the stability of the system and the absence of
systematic effects over the 6 days of measurement on the deep
survey.

\begin{figure}[p]
\begin{center}
\includegraphics[width=13cm]{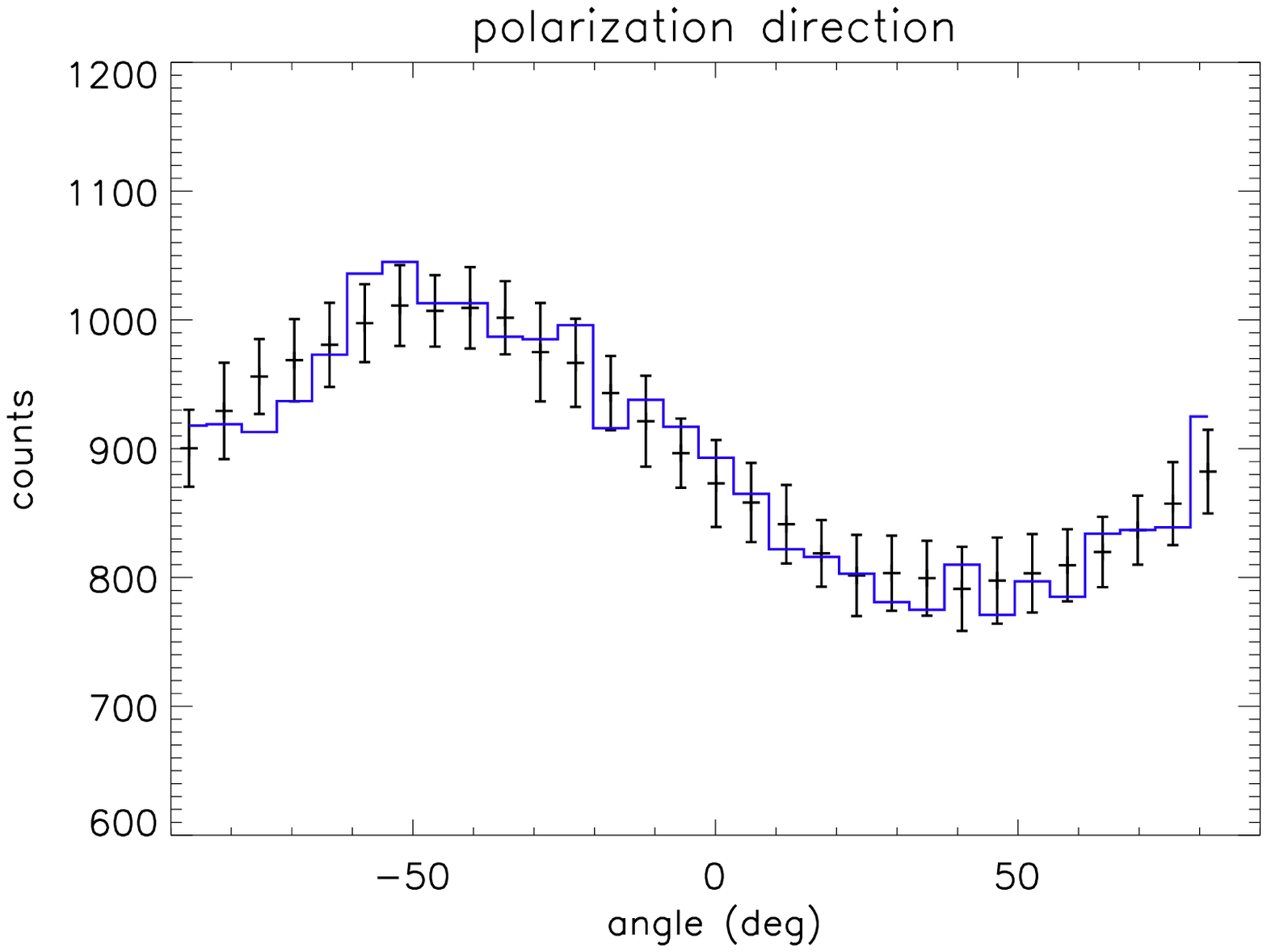}
 \caption{Histogram of the directions
of the polarization vectors, computed as $\alpha = \arctan[U/Q]$,
in the deep survey at 145 GHz (blue line). The $\sim 15 \%$
sinusoidal deviation from uniformity is due to unequalized noise
of the different PSBs. This is confirmed by the analysis of maps
obtained from realistic noise simulations (black data points with
1-$\sigma$ error bars). \label{fig:histoalpha}}
\end{center}
\end{figure}

In Fig. \ref{fig:histoalpha} we plot the histogram of the measured
polarization vector directions, computed as $\alpha_i =
\arctan[U_i/Q_i]$. Since the region observed by \bk is large
compared to the typical correlation scale ($\sim 1^o$)
characteristic of CMB polarization, and since the two signals are
dominated by uncorrelated noise, we naively expect a uniform
distribution of the $\alpha_i$. The data plotted in
Fig.\ref{fig:histoalpha} approximately confirm this expectation.
The sine-like deviation present in the data is due to the
anisotropy of the noise. Even though the 8 polarized detectors
cover all possible directions (in steps of $\pi/8$), the rotation
of the focal plane with respect to the sky during the flight is
small enough that for each direction of polarization only a few
detectors have most of the statistical weight. If one of the
bolometers is noisier than the others (see table \ref{tab:nep}),
the excursions of the noise will be larger in that direction, and
the measured polarization will be preferentially oriented in the
same direction. This has been confirmed by repeating the same
analysis on maps obtained from realistic noise simulations.  The
\bk data are in full agreement with the distribution of the
results of the simulations, which are plotted for comparison in
Fig.\ref{fig:histoalpha}. We can conclude that anisotropy effects
are contained within $\simlt 10\%$ and are well described by our
simulations. These effects can thus be treated in a self
consistent way by the map making and by the spectral estimation
analysis and simulation procedures.

\begin{figure}[p]
\begin{center}
\includegraphics[width=13cm]{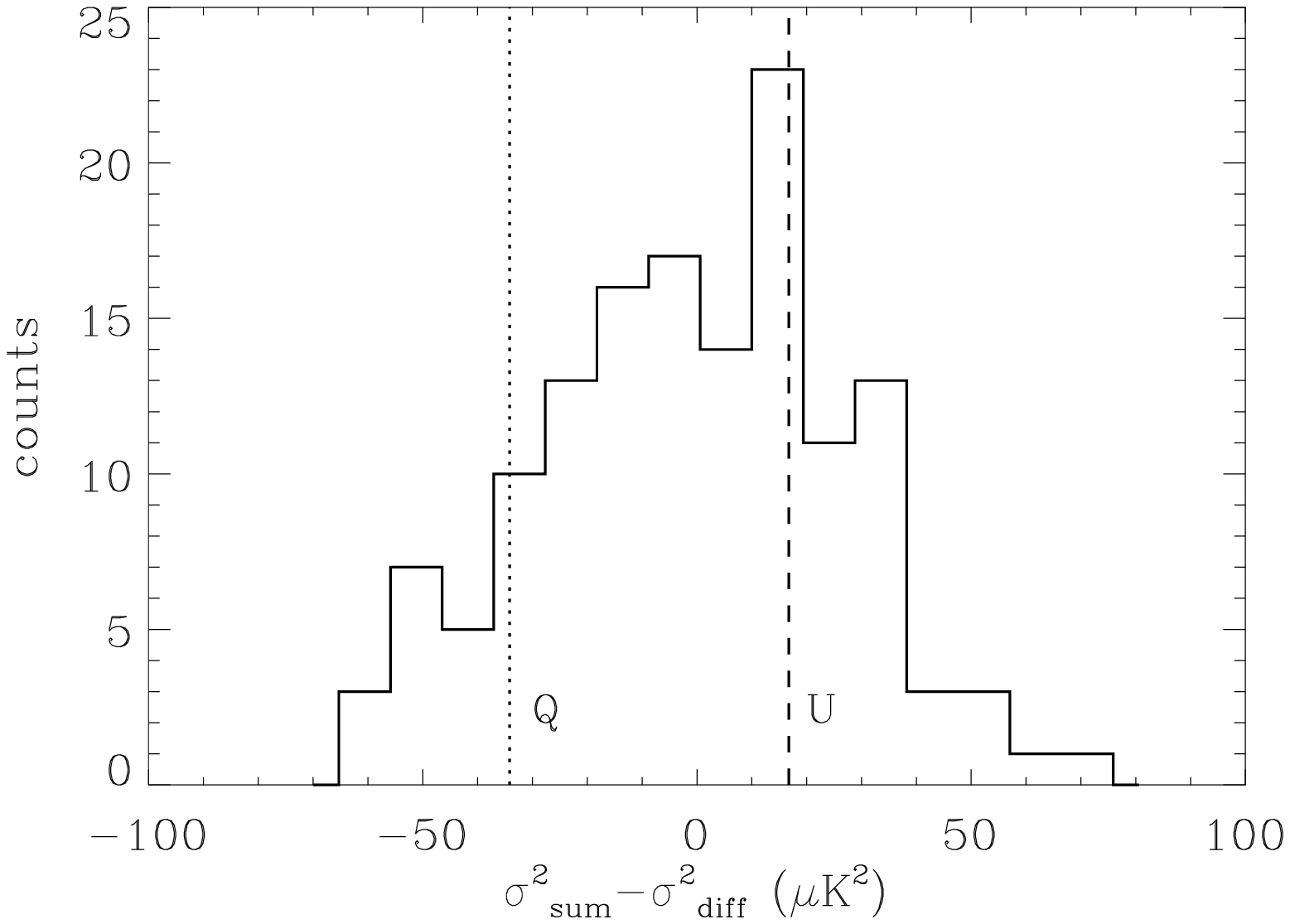}
 \caption{Histogram of the quantity
 $\sigma_{Q,sum}^2-\sigma_{Q,diff}^2$ from simulations of the B2K
 observations of the Deep Region. The simulations used the
 concordance model to generate CMB maps and realistic noise
 simulations for all the 145 GHz detectors. The measured values of
$\sigma_{Q,sum}^2-\sigma_{Q,diff}^2$ and
$\sigma_{U,sum}^2-\sigma_{U,diff}^2$ are plotted as vertical
dashed lines. \label{fig:q_u_histo}}
\end{center}
\end{figure}

From the sum and difference maps we compute the quantities
$\sigma_{Q,sky}^2=\sigma_{Q,sum}^2-\sigma_{Q,diff}^2$ and
$\sigma_{U,sky}^2=\sigma_{U,sum}^2-\sigma_{U,diff}^2$. In the
absence of systematics, these are estimators of the mean square
polarized signal from the sky. From a gaussian fit to the $Q$
histograms we find $\sigma_{Q,sum}=(42.5 \pm 0.2) \mu K$, while
$\sigma_{Q,diff}=(42.9 \pm 0.2) \mu K$. Analogously for $U$ we get
$\sigma_{U,sum}=(42.0 \pm 0.2) \mu K$, while
$\sigma_{U,diff}=(41.8 \pm 0.2) \mu K$. These results do not allow
one to extract the small sky signal (we expect $\sim 4 / \sqrt{2}$
$ \mu K_{rms}$ for $Q$ and $U$ in the concordance model  with
3.5$^\prime$ pixelization), but are fully consistent with the
level of noise of our receivers (see Fig. \ref{fig:q_u_histo}).

In fact the uncertainty in $\sigma_{Q,sky}^2$ and
$\sigma_{U,sky}^2$ obtained from realistic simulations of the
instrumental noise and of the observations is 25 $\mu K^2$ (see
Fig. \ref{fig:q_u_histo}). So we have $\sigma_{Q,sky}^2=(-34 \pm
25) \mu K^2$, and $\sigma_{U,sky}^2=(17 \pm 25) \mu K^2$. For the
polarization degree we have $P^2
=\sigma_{Q,sky}^2+\sigma_{U,sky}^2=(-17 \pm 35)\mu K^2$. This sets
an upper limit to the rms linear polarization of the sky of
$P_{rms} < 7.1 \mu K$ (95$\%$ confidence level). This limit is
close to the $4 \mu K_{rms}$ expected for the CMB polarization
signal in the 3.5$^\prime$ pixelization.

Power spectrum methods provide much more powerful tools to extract
the sky polarization from noisy maps, since it is much easier to
separate different components according to their multipole
content. In particular the effect of low-frequency noise and drift
mostly affects the low multipoles $\ell \simlt 50$. While the
$rms$ analysis above does not separate the different multipoles,
the spectral analysis does; artifacts at low multipoles can be
removed, allowing a much more sensitive probe for signal at higher
multipoles. The application of these methods to \bk is described
in detail in \cite{piacentini05} and \cite{montroy05}, where it is
shown that there is in fact a CMB polarization signal hidden in
the $Q$ and $U$ maps described here.

\section{Conclusions}

The results of this flight of BOOMERanG  demonstrate the
effectiveness of balloon-borne 145 GHz PSBs in the measurement of
CMB anisotropy and polarization, and in the survey of mm-wave
emission from the interstellar medium.

In our three frequency bands the long duration balloon platform
offers an excellent tradeoff between cost and performance. With
bolometers cooled to 0.3K, observing through an uncooled
telescope,  the stratosphere at altitudes $\simgt 25$ Km offers a
very stable environment, and optimal loading conditions for these
measurements.

During the entire flight, the \bk cryogenics and electronics
produced the correct environment (in terms of operating
temperature stability, radiative background level and stability,
and electromagnetic disturbances) to operate the PSBs. These
performed very closely to the theoretical limit, as described in
\S \ref{subs:flightnoise} and \ref{subs:noisest}. The attitude
control system allowed us to reconstruct the pointing of the
telescope with an accuracy of a few arcmin, sufficient for the
main purpose of the experiment, i.e. the measurement of the power
spectra of CMB anisotropy and polarization.

\bk produced maps of the sky at 145, 245 and 345 GHz in two
high-latitude regions and one region at low Galactic latitudes, in
the Southern sky, producing a new survey of more than 1000 square
degrees with $\sim 10^\prime$ resolution. At high latitudes degree
and sub-degree scale anisotropy of the CMB is evident in the 145,
245 and 345 GHz maps. The rms fluctuation of the sky temperature
measured from the 145 GHz map is perfectly consistent with the
current $\Lambda$CDM model for the anisotropy of the CMB. The 145
GHz Stokes Q and U maps presented here allow a statistical (power
spectrum) detection of CMB polarization at 145 GHz, as discussed
in companion papers.

The approach discussed here also shows the effectiveness of
multi-band observations in monitoring the foreground signal. The
Galactic and extragalactic foreground is negligible ($\simlt 10
\%$ compared to the cosmological signal both in anisotropy and in
polarization) at 145 GHz and at high Galactic latitudes. Based on
the SFD maps, about 40 \% of the sky has ISD contamination equal
or lower than in the deep survey region studied here. Future
B-modes searches, seeking a much smaller polarization signal, will
be heavily affected by the foregrounds measured here.

The results presented here are comparable in sensitivity to those
published using interferometric techniques at lower frequencies;
however, this measurement is the first using bolometric detectors,
which can be scaled to large numbers and high sensitivity for
future investigations of CMB polarization.  In the nearer term,
the measurements presented here provide a glimpse at what the
Planck HFI instrument will achieve at 145 GHz;  using the same
technology (PSB detectors), the HFI will reach over the whole sky
roughly the same sensitivity per pixel as was achieved by \bk in
the deep survey.

\bigskip
\acknowledgments We gratefully acknowledge support from the CIAR,
CSA, and NSERC in Canada; Agenzia Spaziale Italiana, University La
Sapienza and Programma Nazionale Ricerche in Antartide in Italy;
PPARC and the Leverhulme Trust in the UK; and NASA (awards
NAG5-9251 and NAG5-12723) and NSF (awards OPP-9980654 and
OPP-0407592) in the USA. Additional support for detector
development was provided by CIT and JPL. CBN acknowledges support
from a Sloan Foundation Fellowship, WCJ and TEM were partially
supported by NASA GSRP Fellowships. Field, logistical, and flight
support were supplied by USAP and NSBF; data recovery was
particularly appreciated. This research used resources at NERSC,
supported by the DOE under Contract No. DE-AC03-76SF00098, and the
MacKenzie cluster at CITA, funded by the Canada Foundation for
Innovation. We also thank the CASPUR (Rome-ITALY) computational
facilities and the Applied Cluster Computing Technologies Group at
the Jet Propulsion Laboratory for computing time and technical
support. Some of the results in this paper have been derived using
the HEALPix package~\cite[]{gorski99}.

\clearpage

\end{document}